\documentclass{article}
\usepackage{epsfig}
\usepackage{amsfonts}
\usepackage{amsmath}
\usepackage{amssymb}

\newcommand{\simge}{\hspace*{0.2em}\raisebox{0.5ex}{$>$}
     \hspace{-0.8em}\raisebox{-0.3em}{$\sim$}\hspace*{0.2em}}
\def\simle{\hspace*{0.2em}\raisebox{0.5ex}{$<$}
     \hspace{-0.8em}\raisebox{-0.3em}{$\sim$}\hspace*{0.2em}}
\newcommand{\beq}{\begin{equation}}
\newcommand{\eeq}{\end{equation}}
\newcommand{\beqa}{\begin{eqnarray}}
\newcommand{\eeqa}{\end{eqnarray}}
\newcommand{\boldtau}{\mbox{\boldmath $\tau$}}

\newcommand{\slashT}{\slash \hspace{-0.4em}T}
\newcommand{\slashpi}{\slash \hspace{-0.4em}\pi}

\newcommand{\slashI}{\slash \hspace{-0.4em}I}

\def\slashchar#1{\setbox0=\hbox{$#1$}           
  \dimen0=\wd0                                    
  \setbox1=\hbox{/} \dimen1=\wd1                  
  \ifdim\dimen0>\dimen1                           
    \rlap{\hbox to \dimen0{\hfil/\hfil}}            
    #1                                             
  \else                                          
    \rlap{\hbox to \dimen1{\hfil$#1$\hfil}}        
    /                                           
 \fi}                                           %

\begin{document}

\begin{titlepage}

\vspace{2.0cm}

\begin{center}
{\Large\bf 
The Time-Reversal- and Parity-Violating 
Nuclear Potential\\
\vspace{.3cm}
in Chiral Effective Theory}

\vspace{1.7cm}

{\large \bf  C.M. Maekawa$^1$, E. Mereghetti$^2$, J. de Vries$^3$,} 
{\large and}
{\large \bf U. van Kolck$^2$}

\vspace{0.5cm}

{\large 
$^1$ 
{\it Instituto de Matem\'atica, Estat\'{i}stica e F\'{i}sica, 
Universidade Federal do Rio Grande}
{\it Campus Carreiros, PO Box 474, 96201-900 Rio Grande, RS, Brazil}
}

\vspace{0.25cm}

{\large 
$^2$ 
\it Department of Physics, University of Arizona, Tucson, AZ 85721, USA}

\vspace{0.25cm}

{\large 
$^3$ 
{\it KVI, Theory Group, University of Groningen,
 9747 AA Groningen, The Netherlands}}

\end{center}

\vspace{1.5cm}

\begin{abstract}
We derive the parity- and time-reversal-violating 
nuclear interactions stemming from
the QCD $\bar\theta$ term
and quark/gluon operators of effective dimension 6:
quark electric dipole moments, 
quark and gluon chromo-electric dipole moments, and two four-quark operators. 
We work in the framework of two-flavor chiral
perturbation theory, where a systematic expansion is possible. 
The different chiral-transformation properties of the sources
of time-reversal violation lead to different hadronic interactions.
For all sources considered the leading-order potential involves
known one-pion exchange, but its specific form 
and the relative importance of short-range interactions depend on the source.
For the $\bar\theta$ term,
 the leading potential is solely given by one-pion exchange,
which does not contribute to the deuteron electric dipole moment. 
In subleading order, a new two-pion-exchange potential is obtained.
Its short-range component is indistinguishable from one of 	 two undetermined
contact interactions that appear at the same order and represent
effects of heavier mesons and other short-range QCD dynamics.
One-pion-exchange corrections at this order are discussed as well.
\end{abstract}

\vfill
\end{titlepage}

\section{Introduction}
\label{intro}

There are three sources of $CP$ violation in the lowest-dimension operators
of the Standard Model.
Due to its flavor-changing properties, the phase of the CKM matrix is 
the best known and investigated source 
(see, for example, Ref. \cite{Hocker+06}). 
The recent discovery of neutrino masses opens up the possibility of
analogous leptonic $CP$ violation \cite{neutrinos}.
The third possible source comes from the $\bar{\theta}$ term in QCD
\cite{thetaorigin}. 
Flavor-conserving quantities, in particular hadronic,
nuclear, and atomic electric dipole moments
(EDMs), are sensitive probes of this qualitatively
different mechanism of $CP$ violation.
EDMs require the simultaneous violation of parity ($P$) and
time-reversal ($T$); for a review, see Ref. \cite{Posp+05}.

Years of careful experimental investigation have 
set stringent bounds on the neutron EDM,
$|d_n| < 2.9 \cdot 10^{-26} \, e$ cm \cite{currentbound}. 
This upper limit points to a tiny value 
for the $\bar\theta$ parameter,
$\bar\theta\simle 10^{-10}$,
despite the naive-dimensional-analysis expectation that it
be of 
${\cal O}(1)$. 
Besides the neutron EDM, there are measurements on atomic EDMs.
The leading contribution in a paramagnetic system comes from the EDM of the
unpaired electron. 
In contrast, a diamagnetic atom has zero total electron
angular momentum, and the main contribution to the atomic EDM comes from 
the nuclear 
Schiff moment (SM), the residual electron-nucleus interaction
generated by the distribution of the EDM throughout the nuclear region.
A less stringent bound on the proton EDM,
$|d_p|< 7.9 \cdot 10^{-25} \, e$ cm, 
is extracted from the EDM of the $^{199}$Hg atom  
\cite{mercury} through a calculation of the nuclear SM
\cite{dmitriev}.
Atomic EDMs can also receive contributions from higher nuclear
moments that violate $T$, such as the magnetic quadrupole moment (MQM).

The unnaturally small value of $\bar\theta$ leaves room for other 
flavor-diagonal sources of $T$ violation, which have their origin in physics 
beyond the Standard Model, at a high-energy scale $M_{\slashT}$.
Well below the new physics scale, these effects manifest themselves in 
interactions between Standard Model fields represented by higher-dimension 
$T$-violating (TV) operators, suppressed by more and more powers of 
$M_{\slashT}$. We expect the most important TV effects to be captured by the 
operators of lowest dimension
\cite{dim6origin,Weinberg:1989dx,Grzadkowski:2010es,RamseyMusolf:2006vr}, 
the dimension-6 quark electric and 
chromo-electric dipole moments (qEDM and qCEDM), 
gluon chromo-electric dipole moment (gCEDM), 
and 
TV four-quark (FQ) operators.

A recent renewal of the longstanding interest in EDMs comes from
a new generation of experiments, which will improve the precision of EDM
observables significantly. 
It is expected that
the current bound on the neutron EDM will be
pushed down to $10^{-27}-10^{-28}\;e$ cm \cite{Snow+05} 
using high-density
ultra-cold neutron sources at 
SNS \cite{SNS} and ILL+PSI \cite{ILL}.
It has also been proposed that the EDM of charged particles
could be measured 
in storage-ring experiments \cite{Far04}.
In addition to the proton, we could see a measurement of the 
deuteron EDM with a projected
precision of $\left\vert d_{d}\right\vert \simle 10^{-29 }\;e$ cm.
A measurement on $^{3}$He is also a possibility.
These measurements probe dimension-6 sources at scales
comparable to the scale reached at the LHC, so 
they could very well turn up positive signals.

Calculating hadronic and nuclear EDMs (and higher moments)
directly from QCD with TV sources is a daunting
task, although progress has been made
for nucleons, in the case of $\bar\theta$, 
using lattice simulations \cite{lattice}.
An alternative is to use a low-energy effective field theory (EFT)
of QCD, chiral perturbation theory (ChPT) \cite{weinberg79,StrongSec} 
(for a review, see for example Ref. \cite{ulfreview}).
In this case, we can describe hadronic and nuclear observables
in a controlled expansion with minimal dynamical assumptions, 
where the symmetries of QCD ---in particular the chiral $SU_L(2)\times SU_R(2)$
symmetry for two quark flavors---
are respected order by order.
While the simplest $T$-violating (TV) interactions
have been known for a long time \cite{Crew+79,su3,scott,BiraEDM+05,dim6NEDFF},
a comprehensive analysis 
---similar to the well-known $T$-conserving (TC) sector---
of the TV chiral Lagrangian from 
the $\bar\theta$ term \cite{emanuele} and the dimension-6 sources 
\cite{dim6}
has only been performed recently.

In the case of the 
nucleon EDM, one finds that, for $T$ violation from the $\bar\theta$ term, 
the leading contributions
come from the pion cloud, where the pion couples to the nucleon
via a non-derivative $P$- and $T$-odd interaction,
and from shorter-range interactions.
The former is purely isovector, and provides an estimate
of the EDM: the characteristic $\ln m_\pi$, with $m_\pi$ the pion mass,
is not expected to be canceled by short-range contributions
and thus provides the bound on $\bar\theta$ mentioned
above \cite{Crew+79,su3,BiraEDM+05}.
One can extend the calculation to the full electric dipole form factor (EDFF)
as well \cite{BiraEDM+05}, where in leading orders the
radius, a contribution to the nucleon SM, is predicted \cite{scott}.
However, because of two possible short-range structures,
isoscalar and isovector,
even measurements of both neutron and proton EDMs
would be insufficient to determine the three unknown parameters
that appear at leading order.

When dimension-6 sources are included in the picture, one finds that the qCEDM 
generates a nucleon EDFF that is not distinguishable from the one stemming 
from $\bar\theta$ \cite{dim6NEDFF}. 
Furthermore, while the momentum dependence of the EDFF is qualitatively 
different if the qEDM or the gCEDM, rather than $\bar\theta$ or the qCEDM, are 
the dominant sources of $T$ violation, a measurement of the neutron and proton 
EDMs alone can be fitted equally well by any of these sources.
It is clear, then, that to pinpoint the dominant mechanism(s) of $T$ violation 
at high energy, more observables are needed, and nuclear EDMs are natural 
candidates. 

Nuclear EDMs and other moments receive various contributions.
There are, of course, 
contributions from the individual nucleons' EDMs.
In the deuteron the isovector component cancels,
while in $^{3}$He one can expect a cancellation between the contributions
of the two nearly anti-aligned protons.
Thus, nuclear EDMs, in particular the deuteron's 
\cite{Pospelov+04, jordysdeut},
are sensitive to a different combination of hadronic
TV parameters than the neutron EDM.
However, these one-nucleon contributions are modified from their ``in-vacuum''
counterparts because the nucleons are not free but bound in the nucleus.
There are many-nucleon effects that are TV.
First, the TV component of the pion cloud can generate 
a TV pion-exchange interaction among nucleons, and no symmetry forbids
interactions of shorter range, either.
These TV nuclear forces will mix-in components of the nuclear wave function
that do not appear in the absence of TV. 
It is a source of polarization effects for the entire nucleus.
Second, there may be multi-nucleon contributions to the 
TV coupling of the photon; such TV currents can be generated
by either pion exchange or shorter-range dynamics.

TV one-pion exchange (OPE) has long been recognized as an important component
of the TV two-nucleon ($NN$) potential, and expressed 
\cite{HaxHenley83,Herczeg} in terms of three 
non-derivative pion-nucleon couplings \cite{Barton},
associated with isospin $I=0, 1, 2$.
So far, the analysis of TV nuclear effects has been based on
tree-level potentials where OPE is 
sometimes supplemented by the single exchange of 
heavier mesons, the eta \cite{GHM93}, rho \cite{TH94}, and omega \cite{TH94}
being most popular. 
Allowing sufficiently many couplings of these mesons 
to nucleons one can produce \cite{Tim+04} the most general
short-range TV $NN$ local interaction with one derivative \cite{H66}.
This is the TV analog of the DDH approach \cite{DDH}
for nuclear TC $P$ violation (PV).

The contributions from such potentials to the deuteron and $^3$He EDMs
have been calculated in the literature under various assumptions.
OPE from the $I=2$ TV pion-nucleon coupling does not contribute
to the $NN$ system at tree level.
It was noticed early on  \cite{Avishai}
that OPE from the $I=0$ TV pion-nucleon coupling
does not contribute to the deuteron EDM, either,
but it does for $^3$He, where it was estimated with a phenomenological
strong-interaction potential  \cite{Avishai2}.
The deuteron EDM that arises from
an $I=1$ TV exchange of either pion- or shorter-range, 
together with a separable
strong-interaction potential, was calculated in Ref.  \cite{Avishai}.
The effects of OPE on the deuteron EDM and MQM
were calculated using both zero-range
and phenomenological strong-interaction potentials
in Ref. \cite{KhriKork00}.
More recent calculations of the deuteron EDM and MQM \cite{Tim+04}
and of the $^3$He EDM \cite{Stet+08} have considered
other TV contributions besides those from the TV potential,
and used more modern, ``realistic'' strong-interaction potentials.
Meson-exchange currents were found small in the deuteron
\cite{Tim+04}, and neglected in $^3$He \cite{Stet+08}.
The TV-potential contributions are consistent
with earlier results; they are
dominated by OPE
from the $I=1$ pion-nucleon coupling 
in the case of the deuteron \cite{Tim+04}, 
and from all three pion-nucleon couplings 
in the case of $^3$He \cite{Stet+08}.

TV moments of heavier nuclei are more difficult to calculate.
It has been argued  \cite{HaxHenley83} that the TV potential can lead to an 
enhancement over the nucleon EDM
thanks to the near-degeneracy of levels of opposite parity,
while meson-exchange currents are comparatively small.
The size of the effect can be estimated through 
the single-particle potential obtained by averaging the $NN$ potential over
a closed nuclear core.
The OPE from the $I=0,2$ pion-nucleon
couplings are proportional to the nuclear $I_3$, $(N-Z)/A$
\cite{HaxHenley83,Herczeg,GV91}, while the OPE from the $I=1$ coupling
does not have such a suppressing factor \cite{Herczeg,GV91}.
It has also been found that the matrix elements from the rho and omega
are small compared to the $I=1$ pion contribution \cite{TH94}.
EDMs and MQMs (for example from $I=0$ OPE \cite{HaxHenley83})
and SMs \cite{crippled} 
of several interesting nuclei have been estimated.
A sample of recent SM calculations can be found in Refs. \cite{dmitriev,engel}.

There are, of course, other nuclear tests of TV, see for example Ref. 
\cite{nukeTV}. The most promising for effects of the TV nuclear interaction
seems to be neutron scattering \cite{gudkov}.
On the proton  \cite{Tim+06} and deuteron \cite{moregud}, 
TV neutron scattering is again dominated by OPE,
but sensitive mostly to the $I=0,2$ and $I=0,1$ couplings, respectively.
For heavy nuclei, one can again obtain estimates
using the single-particle potential \cite{Herczeg,nukeTV}.

For consistency,
we would like to describe nuclear TV observables in the same 
framework used for the calculation of the nucleon EDM.
The non-analytic behavior of the nucleon EDM in $m_\pi$ and
the dominance of OPE in nuclear observables point to 
the need of a framework that can account for both effects simultaneously,
with chiral symmetry playing a central role.
In fact, different sources of $T$ violation have different transformation 
properties under chiral symmetry.
As a consequence, 
the relative importance of various pion-nucleon and short-range interactions 
is not the same for all sources.
Here we use the chiral Lagrangian built in Refs. \cite{emanuele,dim6},
where TV interactions stemming from the 
$\bar\theta$ term and the dimension-6 sources were constructed and 
ordered according 
to the same power counting used to order  
TC interactions in ChPT.

In addition to consistency between one- and few-nucleon TV interactions,
nuclear TV also requires consistency between TV and TC forces,
in order there to be no mismatch in the off-shell behaviors
of the various ingredients. Of course, off-shell effects are 
dependent on the choice of fields, while physical quantities are not,
provided the same choice of fields has been made throughout the 
calculation.
As far as TV nuclear interactions are concerned, 
phenomenological TC models bring additional uncertainties,
such as the choice of zero-range or finite-range interactions and 
the role of heavy mesons.
On the other hand, ChPT has been extended to multi-nucleon systems \cite{nEFT},
leading to the derivation of TC nuclear forces and currents.
This opens up the possibility of describing all necessary 
ingredients in a single framework.

The goal of this paper is to provide the first step in the extension
of TV interactions in the EFT to the multi-nucleon sector
where OPE is treated non-perturbatively.
Some of us have recently looked at the deuteron's PV, TV electromagnetic
moments in an EFT where pions are treated in perturbation theory 
\cite{jordysdeut}.
In this case, the size of uncertainties is set by the relatively
low scale where OPE becomes significant.
Treating OPE non-perturbatively extends the EFT to higher momenta
and improves convergence. 
The TV nuclear potential is the most important ingredient
in this extension,
and here we derive it for the most important TV sources.
This is the TV potential
to be used, for example,
with the TC, parity-conserving (PC) potentials from Refs. 
\cite{TCpot2,TCpot3,isoviolphen,isoviolOPE,isoviolNN,Friar:2004ca,isoviolNNN}.
The construction here is similar to that of the
TC PV potential \cite{PVNN,morePVNN,PV3Npiless}, which extends the EFT
from the TC, PV one-nucleon sector \cite{PVvarious}
to multi-nucleon systems. 
Such a framework provides an alternative to the DDH approach \cite{DDH},
allowing for a model-independent analysis
of nuclear TC PV phenomena  \cite{PVNNpiless,PVNNpiful}.
Our present TV PV EFT framework stands in respect to previous approaches
like this TC PV EFT framework with respect to the DDH approach.

As we are going to see, the form of the TV potential
at a given order in the 
chiral expansion depends on the source. For the dimension-6 sources,
all non-derivative
pion-nucleon couplings appear at leading order,
although these sources differ in the relative strength of 
the $I=2$ pion-nucleon coupling and short-range interactions.
For these sources, the leading-order potential is sufficient 
for most applications we envision.
The situation is different for the $\bar\theta$ term: only the  
$I=0$ pion-nucleon coupling, which is suppressed in some cases
of interest, appears at leading order.
For this source, we thus derive the potential 
up to subleading order. 
We show that new elements appear
with respect to phenomenological treatments,
such as two-pion exchange (TPE) at the same level
as short-range interactions representing heavier-meson exchange.
This richness is a blessing, as it potentially reveals the TV source
\cite{dim6NEDFF,jordysdeut}.

The rest of the paper is organized as follows.
In the next section, Sect. \ref{ChPT},
we present a summary of basic ChPT ideas and give the TC and TV 
chiral Lagrangians needed in the following sections.
For processes involving momenta 
below 
$M_{nuc}\sim 100$ MeV,
pion degrees of freedom can be
integrated out and the dominant TV contact interactions 
(the leading-order TV potential in the so-called pionless EFT)
are obtained in Sect. \ref{pionless}. 
The nuclear potential 
in ChPT,
which should apply beyond $M_{nuc}$,
is then presented in momentum (Sect. \ref{qspace})
and coordinate (Sect. \ref{rspace}) spaces.
(We relegate details of the Fourier transformation
to the Appendix.)
In Sect. \ref{discussion} we discuss the size of different components of the 
potential and compare them with phenomenological forms.
We draw our conclusions in Sect. \ref{conclusion}.

\section{Chiral Perturbation Theory}
\label{ChPT}

QCD is characterized by an intrinsic mass scale $M_{QCD}\sim 1$ GeV.
At momenta $Q$ comparable to the pion mass, $Q \sim m_{\pi} \ll M_{QCD}$, 
interactions among nucleons and pions  are described by the most general 
Lagrangian that involves these degrees of freedom and that has the same 
symmetries as QCD. A particularly important role at low energy is played 
by the approximate symmetry of QCD under the chiral group 
$SU_L(2) \times SU_R(2) \sim SO(4)$. Since it is not manifest in the 
spectrum, which instead exhibits an approximate isospin symmetry, 
chiral symmetry must be spontaneously broken down to the isospin subgroup 
$SU_{L+R}(2) \sim SO(3)$. The corresponding Goldstone bosons can be 
identified with the pions, which provide a non-linear realization of 
chiral symmetry. 

Chiral symmetry and its spontaneous breaking strongly constrain the form of 
the interactions among nucleons and pions.
In particular, in the limit of vanishing quark masses and charges,
when chiral symmetry is exact,
pion interactions proceed through a covariant derivative, 
which in stereographic coordinates $\vec \pi$ for the pions is \cite{Wbook} 
\begin{equation}
D_{\mu} \vec \pi = D^{-1} \partial_{\mu}  \vec \pi
\label{piond}
\end{equation}
with 
\begin{equation}
D = 1 + \vec \pi^{\, 2}/F^2_{\pi}
\end{equation}
and $F_{\pi} \simeq 186$ MeV the pion decay constant.
One can also construct the covariant derivative of this covariant 
derivative,
\begin{equation}
\mathcal D_{\nu} D_{\mu} \vec \pi = 
\partial_{\nu}D_{\mu} \vec \pi 
+ \frac{2}{F^2_{\pi}} \left(\vec \pi \; D_{\nu} \vec \pi \cdot D_{\mu} \vec \pi 
-D_{\nu} \vec \pi \; \vec \pi \cdot D_{\mu} \vec \pi  \right),
\label{eq:pioncov}
\end{equation}
and so on.
Nucleons are described by an isospin-1/2 field $N$, 
and we can define a nucleon covariant derivative
\begin{equation}
\mathcal D_{\mu} N = \left(\partial_{\mu} + \frac{i}{F^2_{\pi}} 
     \vec\tau \cdot \left(\vec\pi \times D_{\mu} \vec \pi \right)\right)N,
\label{nucd}
\end{equation}
where $\tau_i$, $i=1,2,3$, are the Pauli matrices in isospin space.
We also define $\mathcal D^{\dagger}$ through 
$\bar N \mathcal D^{\dagger} = \overline{\mathcal D N}$,
and use the shorthand notation
\begin{equation}
\label{shorthand}
\mathcal D^{\mu}_{\pm} \equiv \mathcal D^{\mu} \pm \mathcal D^{\dagger \mu}, 
\qquad 
\mathcal D^{\mu}_{\pm} \mathcal D^{\nu}_{\pm} 
\equiv \mathcal D^{\mu} \mathcal D^{\nu} 
+\mathcal D^{\dagger \nu} \mathcal D^{\dagger \mu}  
\pm \mathcal D^{\dagger \mu} \mathcal D^{\nu} 
\pm \mathcal D^{\dagger \nu} \mathcal D^{\mu} 
\end{equation}
and
\begin{equation}
\label{shorthandtau}
\tau_i \mathcal D^{\mu}_{\pm} \equiv 
\tau_i \mathcal D^{\mu} 
\pm \mathcal D^{\dagger\, \mu} \tau_i, 
\qquad 
\tau_i \mathcal D^{\mu}_{\pm} \mathcal D^{\nu}_{\pm} \equiv 
\tau_i \mathcal D^{\mu} \mathcal D^{\nu} 
+ \mathcal D^{\dagger \nu} \mathcal D^{\dagger \mu} \tau_i 
\pm \mathcal D^{\dagger \mu} \tau_i \mathcal D^{\nu} 
\pm \mathcal D^{\dagger \nu} \tau_i \mathcal D^{\mu}.
\end{equation}
When acting on a nucleon bilinear of non-zero isospin,
for example $\mathcal D_{\mu} (\bar N \vec\tau N) $,
the covariant derivative is meant to be 
in the adjoint representation, that is, 
the isospin matrix in Eq. \eqref{nucd} 
should be replaced by  $(t^j)_{ik} = i \varepsilon^{ijk}$.

At $Q \sim m_{\pi} \ll m_N$, the nucleon mass,
nucleons are essentially non-relativistic; 
as such the only coordinate with which their fields vary rapidly is 
$v\cdot x$, where $v$ is the nucleon velocity, $v_{\mu} = (1, \vec 0)$ 
in the nucleon rest frame. It is convenient therefore to use a heavy-nucleon 
field from which this fast variation has been removed \cite{Jenkins:1990jv}. 
This simplifies the gamma-matrix algebra, leaving only the spin operator 
$S^{\mu}$, where $S^{\mu} = (0, \vec{\sigma}/2)$ in the nucleon rest frame. 
Below we use the subscript $\perp$ to denote the component of a four-vector 
perpendicular to the velocity, for example 
\begin{equation}
\mathcal D_{\mu \, \perp} \equiv 
\mathcal D_{\mu} - v_{\mu} \,v \cdot \mathcal D.
\end{equation}
Baryon states above the nucleon, such as the delta isobar, 
can be included in EFT along similar lines, 
but for simplicity we do not include them here.

The ChPT Lagrangian in the chiral limit includes all the interactions made out 
of $D_{\mu} \vec \pi$, $N$ and their covariant derivatives that are 
chiral invariant.
Chiral symmetry is explicitly broken, however,
which introduces pion interactions that might not include derivatives,
but are proportional to powers of the symmetry-breaking parameters.
Since the explicit breaking of chiral symmetry  is small, 
chiral-symmetry-breaking operators can be systematically included as 
a perturbation on the chiral-invariant Lagrangian.
Their forms are also not arbitrary, being instead determined
by the chiral transformation properties of their progenitors in
the QCD Lagrangian.
The construction of chiral-symmetry-breaking operators 
from quark masses and ``hard'' electromagnetic interactions
(those from photons with momenta beyond the EFT regime)
is extensively treated in Refs. \cite{Wbook, vanKolck}. 
``Soft'' interactions via an explicit photon field $A_\mu$
appear in gauge-covariant derivatives,
\begin{eqnarray}
\left(D_\mu \pi_a\right) &\to& 
\frac{1}{D} \left(\partial_\mu \delta_{ab}-eA_\mu\varepsilon_{3ab}\right)\pi_b,
\\
{\mathcal D}_\mu N &\to& \left[\partial_\mu +
\frac{i}{F_\pi^2}\vec\tau \cdot \left(\vec\pi \times D_\mu \vec\pi\right) 
-ie A_\mu \frac{1+\tau_3}{2}\right] N,
\end{eqnarray}
where $e$ is the proton electric charge,
and in gauge-invariant interactions built from the 
photon field strength 
\begin{equation}
F_{\mu\nu}=\partial_\mu A_\nu -\partial_\nu A_\mu .
\end{equation}

Approximate chiral symmetry,
together with the  heavy-baryon formalism, allows us to systematically expand 
observables in the mesonic and one-nucleon sectors in powers of $Q/M_{QCD}$, 
where $Q$ is the typical momentum of the process under consideration.
The ChPT Lagrangian contains an infinite number of terms, which can be 
organized using an integer ``chiral index'' $\Delta$ and the number $f$ of 
fermion fields \cite{weinberg79,Wbook}:
\begin{equation}
\mathcal L = \sum_{\Delta  = 0}^{\infty} \sum_{f} \mathcal L_{f}^{(\Delta)},
\label{lagrpc}
\end{equation}
where $\Delta = d + f/2 -2 \ge 0$, with $d$ the number of derivatives,
powers of the pion mass or of the electric charge.
For processes with at most one nucleon, $A= 0,1$,
all momenta and energies are typically $\sim Q$.
The contribution of a diagram 
to the amplitude $T$ can then be estimated by
\begin{equation}\label{ampprc}
T \propto  Q^{\nu} \mathcal F(Q/\mu),
\end{equation}
where $\mathcal F$ is a calculable function, $\mu$ is the renormalization scale, 
and the counting index $\nu$ is 
\begin{equation}
\nu = 4- 2C - A + 2L + \sum_i \Delta_i.
\label{nudelta}
\end{equation}
Here, 
$C=1$ and $L$ are respectively the number of connected pieces and loops
in the diagrams, and 
$i$ counts the number of insertions of vertices from 
$\mathcal L^{(\Delta)}_{f}$.
{}From Eq. \eqref{nudelta} it is apparent that diagrams with increasingly 
higher number of loops and  non-vanishing-index interactions
are increasingly suppressed, 
leading to a perturbative expansion.
Assigning to loops a characteristic factor $Q^2/(4\pi)^2$
and using naive dimensional analysis \cite{weinberg79,NDA,Weinberg:1989dx}
to estimate the EFT parameters,
the suppression scale is $M_{QCD}\sim 2\pi F_\pi$.
Note that in this sector of the theory nucleon recoil
is a subleading effect: the nucleon is nearly static.

The ChPT power counting formula \eqref{nudelta} cannot directly be applied to 
processes with $A\ge 2$ \cite{Weinb90,nEFT}. 
Indeed, in diagrams in which the intermediate state consists purely of 
propagating nucleons
---which are called ``reducible''--- the contour of integration  for 
integrals over the 0th components of loop momenta cannot be deformed in 
way to avoid the poles of the nucleon propagators, thus picking up energies
$\sim Q^2/m_N$ from nucleon recoil, no longer a subleading effect,
rather than $\sim Q$. 
There is also an extra factor of $4\pi$.
These diagrams are therefore enhanced by factors of $4\pi m_N/Q$ with 
respect to 
the ChPT power counting that assigns $Q^2/(4\pi)^2$ to a loop, 
and the need to resum them leads to the appearance 
of shallow bound states in systems with two or more nucleons, nuclei.
Diagrams whose intermediate states contain interacting nucleons and pions 
---``irreducible''--- do not suffer from this infrared enhancement,
and in them nucleon recoil remains a small effect.
Reducible diagrams are thus obtained by patching together irreducible diagrams 
with intermediate states consisting of $A$ free-nucleon propagators. 
Calling $V$ the sum of all irreducible diagrams, the amplitude can be written 
schematically as
\begin{equation}\label{LS}
T = V + V G_0 V + V G_0 V G_0 V + \ldots = V + V G_0 T,
\end{equation}
where $G_0$ is the free-nucleon, non-relativistic Green's function. 
Equation \eqref{LS} is just the Lippmann-Schwinger equation, 
which is formally equivalent to a Schr\"{o}dinger equation with a 
potential $V$. 

Naive dimensional analysis suggests \cite{Weinb90} that irreducible
diagrams follow the ChPT power counting rule \eqref{nudelta} with $C\ge 1$. 
While this is true for pion-exchange diagrams, the situation
is more complicated for contact interactions. 
In fact, it can be shown that the iteration of 
the singular one-pion exchange 
requires for renormalization at the same order a finite number of $f=4$ 
interactions, some of which
are less suppressed than expected on the basis of naive dimensional analysis
\cite{kids,nogga}.
On the other hand, corrections, which should be perturbative, are expected 
to still conform to dimensional analysis \cite{bingwei,pavao}.
Since the TV potential is very small, it should be 
amenable to an expansion in powers of $Q/M_{QCD}$,
with different contributions 
organized according to their chiral index 
$\nu$, or, equivalently, according to the number of inverse powers of 
$M_{QCD}$.

In Sect. \ref{qspace}  we compute the TV 
nuclear potential,
in the case of
the QCD $\bar\theta$ term 
up to order $\nu = 3$,
which means
up to ${\cal O}(Q^2/M^2_{QCD})$ with respect to the leading piece.
Such a calculation 
requires the knowledge of the TC and 
TV ChPT Lagrangians up to $\Delta =2$ 
and $\Delta = 3$, respectively.
In the remainder of this section we present 
the relevant interactions.
Throughout, we use nucleon field redefinitions to eliminate
nucleon time derivatives from subleading interactions.

\subsection{$T$-Conserving Chiral Lagrangian}

The calculation of the TV potential in ChPT requires certain
TC interactions 
with $f=0, 2$, which we list here. (A more complete list
can be found in the literature, for example Refs. \cite{ulfreview,nEFT,Bernard:1992qa}.)
These interactions
stem from the quark (color-gauged) 
kinetic and mass terms in the QCD Lagrangian.

The leading TC chiral Lagrangian has chiral index $\Delta = 0$ and is given 
by 
\begin{equation}
\mathcal{L}_{f \le 2, T}
^{\left( 0\right) }=\frac{1}{2}D_{\mu}\vec{\pi}\cdot
D^{\mu}\vec{\pi}-\frac{m_{\pi}^{2}}{2D}\vec{\pi}^{\,2}
+\bar{N} iv\cdot \mathcal D  N
-\frac{2 g_{A}}{F_{\pi}} D_{\mu} \vec{\pi} \cdot 
 \bar{N} \vec{\tau} S^{\mu} N, 
\label{LagStrong0}
\end{equation}
where $g_A$ is the pion-nucleon axial coupling, $g_A \simeq 1.27$.
The pion mass term originates in explicit chiral-symmetry breaking by the 
average quark mass $\bar m = (m_u +m_d)/2$ 
and, by naive dimensional analysis, $m^2_{\pi} =\mathcal O(\bar m M_{QCD})$.
Neglecting for the moment isospin-breaking operators,
at chiral order $\Delta=1$ 
the relevant Lagrangian consists of 
\begin{equation}
\mathcal{L}_{f \le 2, TI}^{(1)}  
= -\frac{1}{2m_N} \bar{N} \mathcal D_{\perp}^{2}N
+\frac{ g_{A}}{F_{\pi} m_N} \left(i v\cdot D \vec{\pi}\right) \cdot 
\bar{N}\vec{\tau}\, S\cdot \mathcal D_{-} N  
+ \Delta m_N \left(1 - \frac{2\vec\pi^{\, 2}}{F^2_{\pi} D}\right) \bar{N} N.
\label{LagStrong1}
\end{equation}
Here the first two terms  
are the nucleon kinetic energy and a relativistic correction to 
the pion-nucleon coupling,
the coefficients of both operators being fixed by 
Galilean invariance. 
The third term is the nucleon sigma term with a coefficient
$\Delta m_N =\mathcal O(m^2_{\pi}/M_{QCD})$. 
At the next chiral order, $\Delta=2$,
\begin{eqnarray}
\mathcal L^{(2)}_{ f \le 2, TI}
& = &  -\frac{\Delta^{} m^{\, 2}_{\pi}}{2 D^2} \vec \pi^2 
+ \frac{g_A}{4 F_{\pi} m_N^2} D_{\mu} \vec\pi \cdot 
\bar N \vec{\tau} \left(S^{\mu}  \mathcal D^2_{\perp,\, -} 
- 
\mathcal D^{\mu}_{\perp, -} \, S \cdot \mathcal D_{\perp, -} \right) N 
\nonumber\\ 
& &-\frac{2g_A }{F_{\pi}}
 \left[c_A \mathcal D^2_{\perp} D_{\mu} \vec\pi
-d_A
  \left(1 - \frac{2\vec \pi^{\, 2}}{F_{\pi}^2D}\right) D_{\mu} \vec \pi 
  \right]\cdot \bar N \vec \tau S^{\mu} N. 
\label{LagStrong2}
\end{eqnarray}
The first term 
is a correction to the pion mass, 
$\Delta^{} m^2_{\pi} = \mathcal O\left( m^4_{\pi}/M^2_{QCD}\right)$. 
The second term
represents further relativistic corrections to 
the $g_A$ term in Eq. (\ref{LagStrong0}). 
The constraints imposed by Lorentz invariance on Eqs. \eqref{LagStrong1} 
and \eqref{LagStrong2} agree with the results of Ref. \cite{Bernard:1992qa}, 
once a field redefinition is used to eliminate time derivatives acting on 
the nucleon field from the subleading 
$\Delta = 1$ and $\Delta = 2$ Lagrangians.
The operator with coefficient 
$c_A=\mathcal O\left(1/M^2_{QCD}\right)$ 
in Eq. \eqref{LagStrong2} 
is a contribution to the square radius of the pion-nucleon form factor, 
while $d_A=\mathcal O\left(m^2_{\pi}/M^2_{QCD}\right)$ 
is a chiral-symmetry-breaking 
correction to $g_A$ \cite{isoviolphen},
which provide the so-called Goldberger-Treiman discrepancy.

Isospin-breaking operators in the chiral
Lagrangian \cite{vanKolck}
stem from the quark mass difference 
$m_d -m_u = 2 \bar m \varepsilon$ and from 
quark coupling to photons
through the fine-structure constant $\alpha_{\textrm{em}}=e^2/4\pi$.
Here, for simplicity, we count $\varepsilon\sim 1/3$ as ${\cal O}(1)$
and $\alpha_{\textrm{em}}/4\pi$ as
${\cal O}(m_\pi^3/M^3_{QCD})$,
since numerically 
$\alpha_{\textrm{em}}/4\pi \sim \varepsilon m^3_{\pi}/(2\pi F_\pi)^3$.
Isospin-violating terms
first contribute to the $\Delta = 1$ Lagrangian,
\begin{equation}
\mathcal L^{(1)}_{f \le 2, T\slashI} =  
 - \frac{\breve{\delta} m^2_{\pi} }{2 D^2} 
   \left(\vec \pi^{\, 2} - \pi^2_3\right) 
 + \frac{\delta m_{N}}{2}
  \bar N\left(\tau_3-\frac{2 \pi_3}{F^2_{\pi} D}\vec\pi \cdot \vec\tau\right)N,
\label{chiralbreak1}
\end{equation}
while at order $\Delta =2$,  
\begin{equation}
\mathcal L^{(2)}_{f \le 2, T\slashI} 
= -  \frac{\delta^{} m^2_{\pi}}{2 D^2} \pi^2_3
+  \frac{\breve{\delta} m^{}_{N} }{2} 
   \bar N  \left[\tau_3+\frac{2}{F_{\pi}^2 D}  
 \left(\pi_3 \vec \pi \cdot \vec \tau - \vec \pi^{\, 2} \tau_3\right)\right] N 
+ \frac{\beta_1}{F_{\pi}} 
  \left(D_{\mu}\pi_3-\frac{2 \pi_3}{F^2_{\pi} D} 
  \vec\pi\cdot D_{\mu} \vec\pi \right) \bar N S^{\mu} N.
\label{chiralbreak2}
\end{equation}
Here $\breve{\delta} m^2_{\pi} = 
\mathcal O\left(\alpha_{\textrm{em}} M^2_{QCD}/4\pi\right)$
is the leading electromagnetic contribution to 
the pion mass splitting,
while the quark-mass-difference contribution,
$\delta^{} m^2_{\pi}= 
\mathcal O\left(\varepsilon^2 m^4_{\pi}/M^2_{QCD}\right)$, 
is smaller by a power of $\varepsilon m_{\pi}/M_{QCD}$.
The pion mass splitting,
$m^2_{\pi^{\pm}} - m^2_{\pi^0} =\breve\delta m^2_{\pi} -\delta^{} m^2_{\pi}
=(35.5 \; {\rm MeV})^2$ \cite{Nakamura:2010zzi},
is dominated by the electromagnetic contribution.
The nucleon mass splitting, 
$m_n - m_p = \delta m^{}_{N} + \breve{\delta} m^{}_{N}= 1.29$ MeV 
\cite{Nakamura:2010zzi}
also receives contributions 
from electromagnetism and from the quark masses.  In this case, 
the quark-mass contribution  $\delta m_{N}$
is expected to be the largest. 
By dimensional analysis
$\delta m_N = {\cal O}(\varepsilon m_\pi^2/M_{QCD})$,
and lattice simulations estimate it to be 
$\delta m_N=2.26 \pm 0.57 \pm 0.42 \pm 0.10$ MeV 
\cite{latticedeltamN},
which is in agreement with an extraction from charge-symmetry breaking 
in the $pn\to d \pi^0$ reaction \cite{CSBd}.
The electromagnetic contribution is
$\breve{\delta} m_{N}  = \mathcal O\left( \alpha_{\rm{em}}M_{QCD}/4\pi\right)$,
that is, $\mathcal O\left(\varepsilon m^3_{\pi}/M^2_{QCD} \right)$
and about the $20\%$ of $\delta m_N$. 
Using the Cottingham sum rule, 
$\breve{\delta} m_{N}=-(0.76 \pm 0.30)$ MeV \cite{Cott}, 
which is consistent with dimensional analysis.
The operator with coefficient 
$\beta_1=\mathcal O\left(\varepsilon m^2_{\pi}/M^2_{QCD} \right)$ 
is an isospin-violating pion-nucleon coupling.
At present there are only bounds on $\beta_1$ from isospin violation 
in nucleon-nucleon scattering. For example,
a phase-shift analysis of two-nucleon data gives 
$\beta_1=(0 \pm 9) \cdot 10^{-3}$ \cite{isoviolphen,isoviolOPE}, which
is comparable to estimates of $\beta_1$ from $\pi$-$\eta$ mixing.

For the solution of the Lippmann-Schwinger equation, it is convenient to 
eliminate  the nucleon mass difference $m_n - m_p$
from the 
nucleon propagator and from asymptotic states.
This result can be accomplished through a field redefinition, 
defined in Ref. \cite{Friar:2004ca}. 
After the field redefinition, 
Eqs. \eqref{chiralbreak1} and \eqref{chiralbreak2} become
\begin{eqnarray}
\mathcal L^{(1, 2)}_{f \le 2, T\slashI} & = &  
 - \frac{1}{2 D^2} \left( \breve{\delta} m^2_{\pi}  
 - \delta m_{N}^2\right) \left(\vec\pi^2 - \pi^2_3\right) 
    - \frac{\delta^{} m^2_{\pi}}{2 D^2} \pi^2_3
 - (\delta m_{N} + \breve\delta m_N) 
   \left(\vec \pi \times v\cdot D \vec \pi \right)_3 \nonumber\\ 
& &
+ \frac{g_A \delta m_{N}}{F_{\pi} m_N} i \varepsilon_{3 a b} \pi_a   
  \bar N \tau_b S \cdot \mathcal D_{-} N  
+ \frac{\beta_1}{F_{\pi}} \left( D_{\mu} \pi_3 
  - \frac{2 \pi_3}{F^2_{\pi} D} \vec \pi \cdot D_{\mu} \vec \pi \right) 
    \bar N S^{\mu} N .
\label{chiralbreak3}
\end{eqnarray}
We will incorporate isospin-breaking effects in the potential using the 
Lagrangian \eqref{chiralbreak3}.

\subsection{$T$-Violating Chiral Lagrangian From $\bar\theta$}

The lowest-dimension TV operator that can be added to the TC 
QCD  Lagrangian is the dimension-4 
$\bar \theta$ term.
With an appropriate choice of the quark fields $q = (u,d)^T$, the 
$\bar \theta$ term can be expressed as a complex mass term \cite{Baluni},
\begin{equation}
\mathcal L_{\slashT 4}=m_{\star}\bar\theta \; 
\bar{q} i\gamma_{5}q  ,
\label{LtrvQCD}
\end{equation}
where $m_{\star}=m_{u}m_{d}/\left( m_{u}+m_{d}\right) = 
\mathcal O\left(m_{\pi}^{2}/M_{QCD}\right) $ 
and $\bar{\theta}$ is the QCD vacuum angle,
here already assumed to be small as indicated by the 
bound on the neutron EDM, $\bar\theta \lesssim 10^{-10}$.
The $\bar \theta$ term transforms under chiral symmetry as the 
fourth component of an $SO(4)$ vector 
$P = (\bar q \vec{\tau} q, \bar q i \gamma_5 q)$,
whose third component is responsible for 
quark-mass isospin violation \cite{vanKolck}. 
TV from the $\bar \theta$ term and isospin violation from the quark mass 
difference are therefore intrinsically linked;
this link appears in certain relations \cite{Crew+79,emanuele}
between the coefficients of 
TV and isospin-breaking operators in ChPT 
through a coefficient $\rho= (1-\varepsilon^2)\bar \theta/2\varepsilon$.

The pion-nucleon TV Lagrangian from the QCD $\bar\theta$ term was constructed 
in Ref. \cite{emanuele}. 
The leading $f\le 2$ TV interaction generated by the $\bar \theta$ term 
appears at $\Delta =1$ and consists of an isoscalar pion-nucleon coupling, 
\begin{equation}
\mathcal L_{f = 2, \slashT 4} 
^{\left( 1\right)}=  
- \frac{ \bar{g}_{0}}{F_{\pi} D}\vec{\pi}\cdot \bar{N}\vec{\tau}N.  
\label{LagLO}
\end{equation}
The relation to isospin-breaking operators implies
that  $\bar g_0$
can be expressed in terms of the quark-mass contribution to the nucleon mass 
difference,
$\bar g_0 =  \rho \, \delta m_{N} = 
\mathcal O\left(\bar\theta m^2_{\pi}/M_{QCD}\right)$.
Increasing the chiral index by one, 
we find a single term,
\begin{equation}
\mathcal L_{f = 2, \slashT 4}^{\left( 2 \right) }=
- \frac{2\bar{h}^{}_{0}}{F^2_{\pi} D} \; \vec{\pi}\cdot D_{\mu }\vec{\pi}\;
  \bar{N}S^{\mu}N,   
\label{LagSLOTodd}
\end{equation}
where $\bar{h}^{}_{0}$ is an isoscalar two-pion-nucleon coupling,
which is related to the coefficient $\beta_1$ in Eq. \eqref{chiralbreak3} by
$\bar h_0 =  \rho \, \beta_1= 
\mathcal O\left(\bar\theta m_{\pi}^{2}/M_{QCD}^{2}\right) $.
The $\Delta = 3$ 
operators relevant to the calculation of the TV potential at order $\nu = 3$ 
are
\begin{eqnarray}
\mathcal L^{(3)}_{f = 2, \slashT 4}
&=& - \frac{1}{F_{\pi}} \left(\frac{\bar g_1}{D}  
    - 2 \bar g_0 \frac{\Delta m_N}{\delta m_N} 
    \frac{\delta^{} m^2_{\pi}}{m^2_{\pi}}\right)
   \left(1-\frac{2\vec \pi^{\, 2}}{F_{\pi}^2D}\right) \pi_3 
   \bar N N \nonumber\\
& & 
- \frac{1}{F_{\pi}}\left( \frac{\Delta \bar g_0}{D}  
  - \bar g_0 \frac{\delta^{} m^2_{\pi}}{m^2_{\pi}} \right)
 \left(1 - \frac{2\vec \pi^{\, 2}}{F^2_{\pi}D}\right) 
 \vec \pi \cdot \bar N \vec \tau N 
+ 
\frac{\bar\eta}{2F_{\pi}}\left(1-\frac{2\vec\pi^{\, 2}}{ F^2_{\pi}D}\right)
\left( \mathcal D_{\mu\, \perp} D^{\mu}_{\perp}\vec \pi \right) \cdot \bar N \vec\tau N,  
\nonumber\\
&  & +
\frac{\bar g_0}{8 m_N^2 F_{\pi}D} 
\left\{
\vec \pi \cdot \bar N \vec \tau \,\mathcal D^2_{\perp\, -} N
+2\left(1-\frac{\vec\pi^{\, 2}}{ F^2_{\pi}}\right)
D_{\nu} \vec \pi \cdot
\bar N \vec \tau \left[S^{\mu},S^{\nu}\right]  \mathcal D_{\mu\, -}  N 
\right\},
\label{LagT2}
\end{eqnarray}
where once again we eliminated operators containing nucleon time derivatives 
and neglected multi-pion operators. 
Equation \eqref{LagT2} is obtained after rotating away the subleading pion 
tadpole operator  $-(\bar g_0\delta^{} m^2_{\pi}/2 \delta m_N) \pi_3$ 
from the mesonic TV Lagrangian, as detailed in Ref. \cite{emanuele}. 
The first operator in Eq. \eqref{LagT2} is the most important contribution 
of the $\bar \theta$ term to the isospin-breaking TV non-derivative 
coupling $\pi_3 \bar N N$. Its coefficient is of 
$\mathcal O\left(\bar\theta\varepsilon  m^4_{\pi}/M^3_{QCD}\right)$ 
and is  suppressed by two powers of $m_\pi/M_{QCD}$ with respect to 
$\bar g_0$. 
The second term in Eq. \eqref{LagT2} is basically a correction to $\bar g_0$
of the same order. 
Of the TV operators with two derivatives, 
$\bar\eta=\mathcal O\left(\bar\theta   m^2_{\pi}/M^3_{QCD}\right)$ 
represents 
a contribution to the radius of the TV pion-nucleon form factor, 
while the remaining operators relevant at this order are relativistic 
corrections to the $\bar g_0$ term with coefficients fixed by 
Lorentz invariance.
As in the case of the $\Delta = 1, 2$ TV Lagrangians,  
the TV coefficients in Eq. \eqref{LagT2} are related to isospin-violating 
parameters in the $\Delta = 3$ TC Lagrangian, 
which are at present poorly determined 
(see discussion in Ref. \cite{emanuele}).

For the nuclear potential we will need in addition operators involving
more nucleon fields, which are constructed 
in the same way as were the $f=0,2$ interactions in Ref. \cite{emanuele}. 
In the $f=4$ sector of the theory, the first contribution to the 
Lagrangian comes at $\Delta = 2$,
\begin{equation}
\mathcal L^{(2)}_{f= 4, \slashT 4} = 
-\frac{1}{F_{\pi} D} \vec \pi \cdot 
\left(\bar\gamma_{s} \bar N \vec \tau N \,\bar N N
+4\bar\gamma_{\sigma} \bar N \vec \tau S_{\mu} N \,\bar N S^{\mu}N\right),
\label{LTVpiNNNN}
\end{equation}
in terms of two TV parameters $\bar \gamma_{s, \sigma}$.
Just as for $f\le 2$, here too there is a link with isospin-violating 
operators, in this case
\begin{equation}
\mathcal L^{(2)}_{f= 4, T\slashI}
= \frac{\gamma_{s}}{2}  
\bar N\left(\tau_3-\frac{2 \pi_3}{F^2_{\pi} D}\vec\pi \cdot \vec\tau\right)N
\,\bar N N
+2\gamma_{\sigma}
\bar N\left(\tau_3-\frac{2 \pi_3}{F^2_{\pi} D}\vec\pi \cdot \vec\tau\right)
S_{\mu} N
\,\bar N S^{\mu}N,
\end{equation}
which generate the dominant contributions to the short-range
isospin-violating two-nucleon potential \cite{vanKolck,isoviolphen}. 
The isospin-violating coefficients 
$\gamma_{i} = \mathcal O(\varepsilon m^2_{\pi} /F^2_{\pi} M^2_{QCD})$
can be seen as low-energy remnants of $\rho$-$\omega$ mixing 
\cite{isoviolphen} and $a_1$-$f_1$ mixing \cite{coon}.
The TV parameters are related to them by
$\bar\gamma_{i}=\rho\, \gamma_{i} 
=\mathcal O(\bar\theta m^2_{\pi} /F^2_{\pi} M^2_{QCD})$.
The operators in Eq. \eqref{LTVpiNNNN}
are not relevant for the nuclear potential to the order
we work, but contribute 
to the three-nucleon TV potential at next order.

The first operators relevant to the calculation of the TV two-nucleon  
potential are
\begin{equation}
\mathcal L^{(3)}_{f = 4, \slashT 4}
= \left(1 - \frac{2\vec\pi^2}{F^2_{\pi}D}\right) 
\left[\bar C_{1} \bar N N \partial_{\mu} (\bar N S^{\mu} N)  
+ \bar C_{2} \bar N \vec \tau N \cdot 
  \mathcal D_{\mu} (\bar N S^{\mu} \vec\tau N) \right],
\label{4ntodd}
\end{equation}
where the $\bar C_{i}$ are two new TV parameters.
These interactions are related to isospin-breaking operators
\begin{equation}
\mathcal L^{(3)}_{f =4, T\slashI} 
=  \frac{2\pi_3}{F_{\pi}D} 
\left[C_{1} \bar N N \partial_{\mu} (\bar N S^{\mu} N)  
+ C_{2} \bar N \vec \tau N \cdot 
  \mathcal D_{\mu} (\bar N S^{\mu} \vec\tau N) \right], 
\label{4nIV}
\end{equation}
with coefficients 
$C_{i}= \mathcal O(\varepsilon m^2_{\pi}/F^2_{\pi} M^3_{QCD})$.
The TV parameters 
are
$\bar C_{i}= \rho \, C_{i}
=\mathcal O(\bar\theta m^2_{\pi}/F^2_{\pi} M^3_{QCD})$
and, therefore, contribute to the TV potential at order $\nu = 3$.
The coefficients $C_{i}$ could in principle be determined from 
pion production in the two-nucleon system and/or from 
isospin-violating three-nucleon forces. 
However, even lower-order isospin-violating three-nucleon forces
are very small \cite{isoviolNNN}, so prospects 
for extracting $C_{i}$
from TC data are grim.
As before, we do not write in Eq. \eqref{4ntodd} 
other interactions 
that contain more pion fields.

One can continue the construction of the TV Lagrangian not only to higher 
orders, but also to more nucleon fields.
For any given $f$, the dominant TV terms are expected to
be those without derivatives that transform like the fourth component $P_4$,
just as in Eq. \eqref{LTVpiNNNN}.
This type of term will involve an odd number of pions in
addition to the $f$ nucleon fields. It thus contributes at tree
level only to $(f+2)/2$-nucleon forces, or to 
absorption/production/scattering
of pions on $f/2$-nucleon systems.
The first short-range $f/2$-nucleon TV force comes
from four-vectors $P$ that involve one derivative,
as in Eq. \eqref{4ntodd}.
Since nuclear forces tend to become less important
as $f$ increases (see, for example, Ref. \cite{nEFT}),
it is unlikely that terms with $f\ge 6$ need to be constructed.
The first $f=6$ operators that contribute to the three-nucleon TV
potential appear at next chiral order, $\Delta=4$.

\subsection{$T$-Violating Chiral Lagrangian From Dimension-6 Sources}

The smallness of $\bar{\theta}$ leaves room for other sources
of $T$ violation in the strong interactions, which have their origin in an 
ultraviolet-complete 
theory at a high-energy
scale, such as, for example, supersymmetric 
extensions of the Standard Model \cite{RamseyMusolf:2006vr}.
Well below the scale $M_{\slashT}$ characteristic of $T$ violation,
we expect TV effects to be captured by the
lowest-dimension interactions
among Standard Model fields that respect the theory's
$SU_c(3)\times SU_L(2)\times U_Y(1)$ gauge symmetry.
Just above $M_{QCD}$,
strong interactions are described by the
most general Lagrangian with Lorentz, and color and electromagnetic
gauge invariance among 
the lightest quarks, 
gluons,
and photons. 
The effectively dimension-6 TV terms at this scale can be written as 
\cite{dim6origin,Weinberg:1989dx,Grzadkowski:2010es,RamseyMusolf:2006vr}
\begin{eqnarray}
{\cal L}_{\slashT 6}
&=&
-\frac{1}{2}\bar q \left(d_0+d_3 \tau_3\right)\sigma^{\mu \nu} i\gamma^5 q 
\; F_{\mu \nu}
-\frac{1}{2}\bar q \left(\tilde{d}_0+\tilde{d}_3 \tau_3\right)\sigma^{\mu \nu} 
i\gamma^5\lambda^a q 
\; G^a_{\mu \nu}
\nonumber\\
&&+\frac{d_W}{6}\varepsilon^{\mu\nu\lambda\sigma}f^{abc}
G^a_{\mu \rho}G_\nu^{b\,\rho}G^c_{\lambda \sigma}+
 \frac{1}{4} \textrm{Im}{\Sigma_1} \left( \bar q q\, \bar q i \gamma^5 q 
- \bar q \vec\tau q\, \cdot \bar q \vec\tau i \gamma^5 q \right) 
\nonumber \\
 &&+ \frac{1}{4} \textrm{Im}{\Sigma_8} \left( \bar q \lambda^a q\, 
\bar q \lambda^a i \gamma^5 q - \bar q \lambda^a\, \vec\tau q\, \cdot 
\bar q \lambda^a\, \vec\tau i \gamma^5 q \right),
\label{eq:dim6}
\end{eqnarray}
in terms of the photon and gluon field strengths $F_{\mu \nu}$ and 
$G^a_{\mu \nu}$, 
the standard products of gamma matrices $\gamma^5$ and $\sigma^{\mu \nu}$ in 
spin space,
the totally antisymmetric symbol $\varepsilon^{\mu\nu\lambda\sigma}$,
the Pauli matrix $\tau_i$ in isospin space,
the Gell-Mann matrices $\lambda^a$ in color space,
and the associated Gell-Mann coefficients
$f^{abc}$. 

In Eq. (\ref{eq:dim6}) the first (second) term represents
the isoscalar $d_0$ ($\tilde{d}_0$) and isovector $d_3$ ($\tilde{d}_3$)
components of the qEDM (qCEDM). Although these interactions have canonical
dimension 5, they originate just above the Standard Model scale $M_{W}$
from dimension-6 operators  
\cite{dim6origin} involving 
in addition the carrier of electroweak symmetry breaking, 
the Higgs field. 
They are thus proportional to the vacuum expectation value of the Higgs field,
which we can trade for the ratio 
of the quark mass to Yukawa coupling, $m_q/f_q$.
Writing the proportionality constant as 
$e\delta_q f_q/M_{\slashT}^{2}$ ($4\pi \tilde{\delta}_q f_q/M_{\slashT}^{2}$),
\begin{equation}
d_{0,3} \sim \mathcal O\left(e \delta_{0,3}
\frac{\bar m}{M_{\slashT}^2}\right),
\qquad
\tilde d_{0,3}\sim \mathcal O\left(4\pi \tilde \delta_{0,3} 
\frac{\bar m}{M_{\slashT}^2}
\right),
\end{equation}
in terms of the average 
light-quark mass $\bar{m}$ and the dimensionless factors 
$\delta_{0,3}$ and $\tilde{\delta}_{0,3}$ representing typical values 
of $\delta_q$ and $\tilde{\delta}_q$.
The third term in Eq. (\ref{eq:dim6}) 
\cite{Weinberg:1989dx} is the gCEDM, with coefficient
\begin{equation}
d_W \sim \mathcal O\left(\frac{4\pi w}{M^2_{\slashT}}\right)
\label{w}
\end{equation}
in terms of a dimensionless parameter $w$.
The fourth and fifth operators \cite{Grzadkowski:2010es,RamseyMusolf:2006vr}
are 
TV FQ operators, with coefficients
\begin{equation}
\textrm{Im} \Sigma_{1,8} = 
\mathcal O \left(\frac{(4\pi)^2 \sigma_{1,8}}{M^2_{\slashT}}\right)
\label{sigma}
\end{equation}
in terms of further dimensionless parameters $\sigma_{1,8}$.
The sizes of 
$\delta_{0,3}$, $\tilde{\delta}_{0,3}$, $w$, and $\sigma_{1,8}$
depend on the exact mechanisms of electroweak and $T$ breaking
and on the running to the low energies where non-perturbative QCD effects
take over.
The minimal assumption is that they are 
$\mathcal O(1)$, $\mathcal O(g_s/4\pi)$, $\mathcal O((g_s/4\pi)^3)$,
and $\mathcal O(1)$,
respectively, with 
$g_s$ the strong coupling constant.
However they can be 
much smaller (when parameters encoding TV beyond the Standard Model
are small) or much larger (since $f_q$ is unnaturally small).
For discussion and examples, see Refs. \cite{Posp+05,RamseyMusolf:2006vr}.

The dimension-6 operators in Eq. \eqref{eq:dim6} have different transformation 
properties under chiral symmetry.
The isoscalar and isovector qEDM (qCEDM) transform, respectively, as the fourth
and third components of two $SO(4)$ vectors $V$ and $W$ ($\tilde V$ and 
$\tilde W$), with
\begin{eqnarray}\label{eq:2.2.55}
\begin{array}{lcr}
V=\frac{1}{2}\left(\begin{array}{c}\bar{q} \sigma^{\mu \nu} \vec\tau q\\
i\bar{q}\sigma^{\mu \nu} \gamma^5 q\end{array}\right) F_{\mu \nu},& &
W=\frac{1}{2}\left(\begin{array}{c}-i\bar{q} \sigma^{\mu \nu} \gamma^5 
\vec\tau q\\
\bar{q}\sigma^{\mu \nu} q\end{array}\right)F_{\mu \nu},
\end{array}
\end{eqnarray}
and
\begin{eqnarray}\label{eq:2.2.56}
\begin{array}{lcr}
\tilde{V}=\frac{1}{2}\left(\begin{array}{c}\bar{q} \sigma^{\mu \nu} \vec\tau 
\lambda^a q\\
i\bar{q}\sigma^{\mu \nu} \gamma^5 \lambda^a q\end{array}\right)G^a_{\mu \nu},
& &
\tilde{W}=\frac{1}{2}\left(\begin{array}{c}-i\bar{q} \sigma^{\mu \nu} \gamma^5
\vec\tau \lambda^a q\\
\bar{q}\sigma^{\mu \nu} \lambda^a q\end{array}\right)G^a_{\mu \nu}.
\end{array}
\end{eqnarray}
In contrast, the gCEDM and the two TV FQ operators $\Sigma_{1,8}$ 
are singlets of the chiral group, respectively
\begin{equation}
I_W=\frac{1}{6} \varepsilon^{\mu\nu\lambda\sigma}f^{abc} 
G_{\mu\rho}^aG_{\nu}^{b \, \rho} G_{\lambda\sigma}^c
\label{IW}
\end{equation}
and
\begin{eqnarray}
I_{qq1}&=&\frac{1}{4} 
\left( \bar q  q\, \bar q  i \gamma^5 q - \bar q \vec\tau q \cdot \bar q \vec\tau i \gamma^5 q \right), 
\label{Iqq1}
\\
I_{qq8}&=&\frac{1}{4} 
\left( \bar q \lambda^a q\, 
\bar q \lambda^a i \gamma^5 q - \bar q \lambda^a\, \vec\tau q \cdot 
\bar q \lambda^a\, \vec\tau i \gamma^5 q \right).
\label{Iqq8}
\end{eqnarray}
They
do not break chiral symmetry. 

The different chiral properties of various TV sources have profound 
implications for the form and relative importance of nucleon-pion and 
nucleon-nucleon TV couplings in the effective Lagrangian.
Effective interactions are constructed to transform in the same way as
the sources at quark/gluon level. Thus, $\tilde W_3$ 
leads to interactions proportional to $\tilde\delta_3$ that only
appear from $\bar\theta$ in the tensor product between $P_4$ and
$P_3$
from the quark-mass-difference term, and are thus
proportional to $\varepsilon \bar m^2 \bar\theta$ 
(for example, the $\bar g_1$ term in Eq. \eqref{LagT2}).

On the other hand, $\tilde V_4$ generates exactly the same 
interactions as $P_4$,
so $\tilde\delta_0$ and $\bar\theta$ contribute similarly
to low-energy observables.
For qEDM, hadronic interactions arise from integrating
out at least one hard photon, which leads to further breaking of chiral
symmetry in the form of tensor products
of $V$ and $W$ with a antisymmetric chiral tensor \cite{vanKolck}.
The contributions of the qEDM to purely hadronic couplings, 
like pion-nucleon or nucleon-nucleon
couplings, are suppressed by the electromagnetic coupling 
constant, $\alpha_{\textrm{em}}/4\pi\sim \varepsilon m_\pi^3/(2\pi F_\pi)^3$.
electromagnetic currents. We do not explicitly construct the TV potential from the qEDM. 
Because they are chiral invariant, 
the gCEDM and the two TV FQ operators lead to exactly the same
effective interactions, although, of course, with different  strengths.
The interactions from gCEDM are the hadronic matrix elements
of $I_W$ \eqref{IW} and are proportional to $w$ in Eq. \eqref{w},
while those from the TV FQ operators are the hadronic matrix elements
of $I_{qqi}$ in Eqs. \eqref{Iqq1} and \eqref{Iqq8} 
and are proportional to $\sigma_{i}$ in Eq. \eqref{sigma}.
For simplicity of notation, in the following 
$w$ denotes both $w$ and $\sigma_{1,8}$, 
\begin{equation}
\{w,\sigma_1,\sigma_8\} \to w.
\end{equation}
For none of the dimension-6 sources there is an
immediate, useful connection to 
TC 
operators
as for $\bar\theta$.

The pion-nucleon Lagrangian from dimension-6 sources was 
constructed 
in detail in Ref. \cite{dim6}.
The only terms we need in the following belong to the
lowest-order Lagrangian,
\begin{eqnarray}
\mathcal L_{f =  2, \slashT 6}^{(\Delta_{\pi})}&=&  
- \frac{\bar{g}_{0}}{F_{\pi} D}\vec{\pi}\cdot \bar{N}\vec{\tau}N
- \frac{\bar g_{1}}{F_{\pi} D} \pi_3\, \bar N N 
- \frac{\bar g_{2}}{F_{\pi} D} \pi_3\, 
\bar N \left[\tau_3 + \frac{2}{F_\pi^2 D}
     \left(\pi_3\vec{\pi}\cdot \vec{\tau}-\vec{\pi}^2\tau_3 \right) \right]N
\nonumber\\
&&
- \frac{\bar\imath_{0}}{F_{\pi}^2}(v\cdot D\vec\pi \times D_\mu \vec\pi) \cdot 
\bar{N} S^\mu \vec{\tau} N   
- \frac{\bar\imath_{1}}{F_{\pi}}\mathcal D_{\perp} \cdot D_{\perp} \vec\pi
\cdot \bar N \vec \tau N.  
\label{LagLO6}
\end{eqnarray}
In this Lagrangian:

\begin{itemize}

\item 
For qCEDM, $\Delta_{\pi}=-1$. At this order 
the qCEDM contributes only to the non-derivative couplings $\bar g_{0}$ 
and $\bar g_{1}$. 
In contrast to the $\bar\theta$ case (\ref{LagT2}), 
the elimination of the pion 
tadpoles induced by the isovector component $\tilde W_3$ of the qCEDM
generates an interaction of exactly the same chiral form
as other existing interactions, and for simplicity
we absorb the tadpole contribution in the $\bar g_{i}$. 
Then 
$\bar g_{0} = \mathcal O((\tilde\delta_0 + \varepsilon \tilde\delta_3) 
m^2_{\pi} M_{QCD}/M^2_{\slashT})$ and 
$\bar g_{1} = \mathcal O(\tilde\delta_3 m^2_{\pi} M_{QCD}/M^2_{\slashT})$. 

\item 
For qEDM, $\Delta_{\pi}=2$. At this order 
the qEDM contributes also to the non-derivative coupling $\bar g_{2}$,
which arises from the tensor product of $W_3$ with the antisymmetric
chiral tensor generated by a hard photon.
In this case, again after tadpole extermination,
$\bar g_{0} = \mathcal O(\alpha_{\textrm{em}} (\delta_0 + \delta_3) 
(1+\varepsilon) m^2_{\pi} M_{QCD}/4\pi M^2_{\slashT})$,
$\bar g_{1} = \mathcal O(\alpha_{\textrm{em}} (\delta_0 + \delta_3) 
m^2_{\pi} M_{QCD}/4\pi M^2_{\slashT})$,
and
$\bar g_{2} = \mathcal O(\alpha_{\textrm{em}} \delta_3 
m^2_{\pi} M_{QCD}/4\pi M^2_{\slashT})$. 

\item 
For chiral-invariant sources, $\Delta_{\pi}=-1$.
It is not possible to construct  chiral-invariant, TV pion-nucleon couplings 
with zero or one derivatives. 
The leading TV couplings then contain either two derivatives or one insertion 
of the quark mass, and the couplings 
$\bar g_{0,1}$ and $\bar\imath_{0,1}$ 
all appear at the same order. 
More precisely, after the elimination of the pion tadpole, we find
$\bar g_{0}= \mathcal O(w (1+\varepsilon^2) m^2_{\pi}M_{QCD}/M^2_{\slashT})$
and
$\bar g_{1}= \mathcal O(w \varepsilon m^2_{\pi} M_{QCD}/M^2_{\slashT})$, 
while 
$\bar\imath_{0,1}= \mathcal O(w M_{QCD}/M^2_{\slashT})$.

\end{itemize}
In the case of qEDM, we need also the 
photon-nucleon TV interactions \cite{dim6NEDFF,dim6}
\begin{equation}
\mathcal L_{f =  2, \slashT 6}^{(1)}=  
2\bar N \left[\bar{d}_0 \left( 1-\frac{2\vec\pi^2}{F_\pi^2 D}\right)
+\bar{d}_1
\left(\tau_3-\frac{2\pi_3}{F_\pi^2 D}\vec{\pi}\cdot \vec{\tau}\right)\right]
S^\mu N v^\nu F_{\mu\nu},
\label{LagLO6edm}
\end{equation}
with $\bar d_{0,1}= \mathcal O(e \delta_{0,3} m^2_{\pi}/M^2_{\slashT} M_{QCD})$
short-range contributions to the nucleon EDM.

The various dimension-6 sources lead to different interactions
also in the $f\ge 4$ sectors. 
For the chiral-invariant sources, there are further leading-order
interactions with $f= 4$,
\begin{equation}
\mathcal L^{(-1)}_{f = 4, \slashT 6} =  
\bar C_{1}  \bar N N \partial_{\mu} (\bar N S^{\mu} N) 
+ \bar C_{2}\bar N\vec\tau N\cdot\mathcal D_{\mu}(\bar N \vec\tau S^{\mu} N),
\label{eq:3.1.50}
\end{equation}
with coefficients 
$\bar C_{1,2} = \mathcal O(w M_{QCD}/F^2_{\pi} M^2_{\slashT})$.
Note that because of different chiral-transformation properties,
the form of Eq. \eqref{eq:3.1.50} is different from the 
contact interactions
stemming from $\bar\theta$, Eq. \eqref{4ntodd}.
For the remaining dimension-6 sources, the form of contact
interactions can be different still.
They appear from tensor products of symmetry-breaking operators
and are subleading.
The qCEDM and qEDM generate operators at chiral index 
$\Delta = \Delta_{\pi} + 1$, 
which have schematically the same form as 
Eq. \eqref{LTVpiNNNN}, but a richer isospin structure. 
These operators do not contribute to the two-nucleon potential 
at tree level,
but they are relevant for three-nucleon forces 
just beyond the order
we consider.
Contributions from these sources 
to the two-nucleon potential are found 
at $\Delta = \Delta_\pi +2$,
\begin{eqnarray}
\mathcal L^{(\Delta_\pi +2)}_{f =4, \slashT 6} &=& 
\left(1 - \frac{2\vec\pi^2}{F^2_{\pi}D}\right) 
\left[\bar C_{1} \bar N N \, \partial_{\mu} (\bar N S^{\mu} N)  
+ \bar C_{2} \bar N \vec \tau N \cdot 
  \mathcal D_{\mu} (\bar N S^{\mu} \vec\tau N) \right] 
\nonumber\\
& & +
\left(\delta_{3k} - \frac{2\pi_3 }{F^2_{\pi}D} \pi_k \right)
\left[\bar C_{3} \bar N \tau_k N \,\partial_{\mu} \left(\bar N S^{\mu} N\right)
+ \bar C_{4} 
\bar N  N \, \mathcal D_{\mu} \left(\bar N \tau_k S^{\mu} N \right) \right]
\nonumber \\
& & + 
\left\{\left(1 - \frac{2\vec \pi^{\, 2}}{F^2_{\pi} D}\right) 
\left[
\bar C_5 \bar N \tau_l N \, \partial_{\mu} \left(\bar N S^{\mu} N\right) 
+
\bar C_6  \bar N  N\, \mathcal D_{\mu} \left(\bar N \tau_l S^{\mu} N \right) 
\right]
\right.
\nonumber \\
& & \left.
+ \bar C_7 
\left(\delta_{3k} - \frac{2\pi_3 }{F^2_{\pi}D} \pi_k \right)
\bar N \tau_k N \,
\mathcal D_{\mu} \left( \bar N S^{\mu}\tau_l N\right) \right\}
\left[\delta_{l 3} + \frac{2}{F^2_{\pi} D} 
\left(\pi_3 \pi_l - \vec\pi^{\, 2} \delta_{3 l}\right)\right]
,
\label{eq:3.1.48}
\end{eqnarray} 
where the $\bar C_i$ are new coefficients.
The qCEDM only contributes to $\bar C_{1,2,3,4}$ 
in lowest order, with
$\bar C_{1,2}=\mathcal O((\tilde\delta_0 + \varepsilon \tilde{\delta}_3) 
m^2_{\pi}/F^2_{\pi} M^2_{\slashT} M_{QCD})$
and 
$\bar C_{3,4}=\mathcal O(\tilde\delta_3 m^2_{\pi}/F^2_{\pi} M^2_{\slashT}
M_{QCD})$.
The qEDM contributes also to $\bar C_{5,6,7}$,
because of tensor products 
with the antisymmetric tensor generated by a hard photon, 
and we have 
$\bar C_{1,2} = \mathcal O(\alpha_{\textrm{em}}(\delta_0 + \delta_3) 
(1+\varepsilon) m^2_{\pi}/4\pi F^2_{\pi} M^2_{\slashT} M_{QCD})$, 
$\bar C_{3,4} = \mathcal O(\alpha_{\textrm{em}}
(\delta_0 + \delta_3) m^2_{\pi}/4\pi F^2_{\pi} M^2_{\slashT} M_{QCD})$, 
$\bar C_{5,6} = \mathcal O(\alpha_{\textrm{em}}
\delta_0 m^2_{\pi}/4\pi F^2_{\pi} M^2_{\slashT} M_{QCD})$,
and $\bar C_7 = \mathcal O(\alpha_{\textrm{em}} 
\delta_3 m^2_{\pi}/4\pi F^2_{\pi} M^2_{\slashT} M_{QCD}))$.
Comparing Eq. \eqref{eq:3.1.48} to Eq. \eqref{4ntodd}, 
we once again see that, differently from the $\bar\theta$ term,
the qCEDM and qEDM generate TV and isospin-breaking operators 
of the same importance as isoscalar TV operators. 
The four-nucleon operators in Eq. \eqref{eq:3.1.48} 
have, this time just like the $\bar\theta$ term,
chiral index that is two units bigger than that of the leading 
pion-nucleon Lagrangian. 
As a consequence, short-range contributions to the 
two-nucleon potential from these sources arise only at next-to-next-to-leading order.
Such high orders will not be considered explicitly below.

\section{Pionless Theory}
\label{pionless}

Before we discuss the TV potential in ChPT, it is instructive
to consider a much simpler EFT.
At momenta
much smaller than the pion mass, pion degrees of freedom can be 
integrated out and one is left with a pionless EFT, 
in which the interactions are represented by 
operators involving only nucleon fields.
If we denote by $M_{nuc}\sim 100$ MeV the scales associated
with pion physics, 
this EFT applies to processes where all momenta $Q\ll M_{nuc}$.
Power counting in this EFT is 
different from ChPT and is reviewed in Ref. \cite{nEFT}.

The lowest-order TC two-nucleon interactions can be taken as 
\cite{Weinb90}
\begin{equation}
\mathcal{L}_{\slashpi, T}^{\left( 0\right) }= 
- \frac{C_{11}}{2} \bar{N}N\bar{N}N 
- \frac{C_{\tau\tau} }{2} \bar {N} \vec\tau N \cdot \bar{N} \vec\tau N,
\label{StrongNNNN}
\end{equation}
where 
$C_{11,\tau\tau}=\mathcal O(4\pi/ m_N \aleph$), 
with $\aleph< M_{nuc}$ a low-energy scale. 
The corresponding potential in momentum space is simply
\begin{equation}
V_{\slashpi, T}^{\left( 0\right) }  = 
\frac{1}{2} \left[C_{11} 
+ C_{\tau\tau} \vec \tau^{\,(1)} \cdot \vec \tau^{\,(2)}\right],
\label{Vpiless}
\end{equation}
where 
$\vec{\tau}^{\,(i)}/2$  is the isospin of nucleon $i$.
These interactions affect only the two $S$ waves.
Since the effect of free two-nucleon propagation
is $\sim m_N Q/4\pi$, 
for momenta $Q\simge \aleph$ these interactions have to be iterated to 
all orders \cite{aleph,PDS}.
Using dimensional regularization with power divergence subtraction \cite{PDS}
at a scale $\mu$,
\begin{equation}
C_{0s} = C_{11}-3C_{\tau\tau} 
= \frac{4\pi }{m_N} \left(\frac{1}{a_s}-\mu\right)^{-1},
\qquad 
C_{0t} = C_{11}
+C_{\tau\tau} = \frac{4\pi }{m_N} \left(\frac{1}{a_t}-\mu\right)^{-1},
\end{equation}
in terms of the 
isospin-singlet ($^3S_1$) and -triplet ($^1S_0$) scattering lengths, 
$a_s$ and $a_t$.
Because the coefficients 
$C_{11,\tau\tau}$ 
subsume physics at the scale 
of the pion mass, their scaling
is different from the one in the pionful EFT.

In leading order, the 
$\bar \theta$ term
and the dimension-6 sources induce TV
four-nucleon operators similar
to those in Eqs. \eqref{4ntodd}, \eqref{eq:3.1.50} and \eqref{eq:3.1.48}:
\begin{eqnarray}
\mathcal L_{\slashpi, \slashT} &=& 
\bar C_{11}  \bar N N \; \partial_{\mu} (\bar N S^{\mu} N) 
+ \bar C_{\tau\tau} 
\bar N \vec \tau N \cdot \partial_{\mu} (\bar N S^{\mu} \vec\tau N)\nonumber\\
&&+\bar C_{3 1}  \bar N \tau_3 N \; \partial_{\mu} (\bar N S^{\mu} N) 
+ \bar C_{1 3} 
\bar N   N \cdot \partial_{\mu} (\bar N S^{\mu} \tau_3 N)
+ \bar C_{3 3} \bar N \tau_3 N  \partial_{\mu} (\bar N S^{\mu} \tau_3 N),
\label{4ntoddpiless}
\end{eqnarray}
where $\bar C_{11, \tau\tau, 1 3, 3 1, 3 3}$ are new short-range parameters.
In  momentum space the interaction Hamiltonian is given by
\begin{eqnarray}
V_{\slashpi, \slashT}(\vec q) & = & 
-\frac{i}{2} 
\left[\bar C_{11}  
 + \bar C_{\tau\tau} \vec\tau^{\,(1)} \cdot \vec\tau^{\,(2)}
+\bar C_{33}\, \tau^{(1)}_3  \tau^{(2)}_3\right]
\left(\vec\sigma^{\,(1)} - \vec\sigma^{\,(2)}\right) \cdot \vec q 
\nonumber\\
&&- \frac{i}{4}\left(\bar C_{1 3} + \bar C_{3 1}\right)   
\left(\tau_3^{(1)} + \tau_3^{(2)}\right)
 \left(\vec\sigma^{\,(1)} - \vec\sigma^{\,(2)}\right) \cdot \vec q 
\nonumber\\
&&
- \frac{i}{4} \left(\bar C_{1 3} - \bar C_{3 1}\right)
 \left(\tau_3^{(1)} - \tau_3^{(2)}\right)
 \left(\vec\sigma^{\,(1)} + \vec\sigma^{\,(2)}\right) \cdot \vec q ,
\label{Vnnnn}
\end{eqnarray}
where $\vec{\sigma}^{\,(i)}/2$ is
the spin of nucleon $i$
and 
$\vec{q}=\vec{p}_{1}-\vec{p}^{\,\prime}_{1} =\vec{p}_2^{\,\prime}-\vec{p}_{2}$
is the momentum transfer.

In nucleon-nucleon scattering, operators that break $P$ and $T$ 
induce mixing between waves of different parity.
At low energy, the most relevant effect is the mixing between $S$ and $P$ 
waves, and indeed the single 
momentum in Eq. \eqref{Vnnnn} 
can only connect an $S$ to a $P$ wave.
At leading order,
the $P$ wave is free.
Since the short-range TV potential involves one $S$ wave,
we expect \cite{nEFT} that in the pionless EFT 
the coefficients $\bar C_{ij}$ scale as $1/\aleph$. 
Indeed, 
the amplitude for a nucleon-nucleon transition can be computed from
Eq. \eqref{Vnnnn} as done in the PV case in Refs. \cite{PVNN,PVNNpiless}.
In leading order, it involves one insertion of the TV 
operators $\bar C_{ij}$, dressed by the all-order iteration  
of the appropriate $S$-wave operator, $ C_{0s}$
or $C_{0t}$.
The renormalization-group invariance of the amplitude implies 
that the $\bar C_{ij}$ follow a renormalization-group equation 
of the form $d(\bar C_{ij}/C_{0})/d\ln \mu  = 0$, 
which is satisfied if the five independent parameters are taken to be
\begin{eqnarray}
\bar C_{1s} = \bar C_{11} -3\bar C_{\tau\tau} 
=\frac{4\pi \bar c_{s}}{m_N} \left(\frac{1}{a_s}-\mu\right)^{-1},
\; 
\bar C_{1t} = \bar C_{11} +\bar C_{\tau\tau} 
=\frac{4\pi \bar c_{t}}{m_N} \left(\frac{1}{a_t}-\mu\right)^{-1},\nonumber\\
\bar C_{3s} = \bar C_{13} - \bar C_{31} 
=\frac{4\pi \bar c_{3s}}{m_N} \left(\frac{1}{a_s}-\mu\right)^{-1},
\; 
\bar C_{3t} =  \bar C_{13} +  \bar C_{31}
=\frac{4\pi \bar c_{3t}}{m_N} \left(\frac{1}{a_t}-\mu\right)^{-1},
\nonumber\\
\bar C_{33} = \frac{4\pi \bar c_{3 3 t}}{m_N}
\left(\frac{1}{a_t}-\mu\right)^{-1},
\end{eqnarray}
in terms of five
$\mu$-independent coefficients 
$\bar c_{s,t,3s,3t,33t}$. 
As in the TC sector, the scaling of the short-range parameters is different 
in the pionless EFT than in ChPT.
We can write
\begin{equation}
\bar C_{11,\tau\tau} = \mathcal O\left(\frac{4\pi}{m_N\aleph} \bar c_{s,t} 
\right), \qquad
\bar C_{13,31} = \mathcal O\left( \frac{4\pi}{m_N\aleph} \bar c_{3s, 3t}
\right) ,\qquad
\bar C_{3 3} = \mathcal O\left( \frac{4\pi}{m_N\aleph} \bar c_{3 3 t}
\right) .
\end{equation}

In order to estimate the coefficients $\bar c_{s,t,3s,3t,33t}$,
we use naive dimensional analysis \cite{weinberg79,NDA,Weinberg:1989dx}
with the pionful EFT as the underlying theory.
We then find that at leading order
the isoscalar $\bar c_{s,t}$ 
receive contributions from all the sources,
\begin{equation}
\bar c_{s,t}=\mathcal O\left(
\frac{\bar\theta}{M_{QCD}},
(\tilde\delta_0 + \varepsilon \tilde\delta_3 )\frac{M_{QCD}}{M^2_{\slashT}}, 
\frac{\alpha_{\textrm{em}}}{4\pi}(\delta_0 + \delta_3) 
\frac{M_{QCD}}{M^2_{\slashT}}, 
w \frac{M_{QCD}}{M^2_{\slashT}}\right),
\label{cst}
\end{equation}
while the isospin-breaking $\bar c_{3s,3t}$  
only from the dimension-6 sources,
\begin{equation}
\bar c_{3s,3t}=\mathcal O\left(
\tilde\delta_3 \frac{M_{QCD}}{M^2_{\slashT}}, 
\frac{\alpha_{\textrm{em}}}{4\pi}
(\delta_0 + \delta_3)\frac{M_{QCD}}{M^2_{\slashT}},
\varepsilon w \frac{M_{QCD}}{M^2_{\slashT}}\right),
\label{c3s3t}
\end{equation}
and $\bar c_{3 3 t}$ only from the qEDM,
\begin{equation}
\bar c_{3 3 t}=\mathcal O\left(
\frac{\alpha_{\textrm{em}}}{4\pi}\delta_3 \frac{M_{QCD}}{M^2_{\slashT}}\right).
\label{c33t}
\end{equation}

In general, one would expect five possible amplitudes 
connecting $S$ to $P$ waves \cite{Tim+04,Tim+06}: 
three ---one for each possible value of $I_3 = 1,0,-1$--- to describe 
the mixing of the isotriplet $^1S_0$ and $^3P_0$ waves, one for 
the mixing of the isosinglet $^3S_1$ and $^1P_1$ states,
and one for the mixing of nucleons in the  $^3S_1$ configuration 
with the isotriplet $^3P_1$ wave. 
The $\bar \theta$ term yields a short-range potential in the form of the 
isospin-conserving terms of Ref. \cite{Tim+04}.
Because the TV operator in Eq. \eqref{LtrvQCD} is isoscalar
and isospin violation is a subleading effect in ChPT,
for which the pionless EFT is the low-energy limit,
the $\bar \theta$ term does not contribute at leading order 
to quantities that violate both $T$ and isospin.
The two terms contribute to $^3S_1$--$^1P_1$ mixing and 
to $^1S_0$--$^3P_0$ mixing, in equal way for the three $I_3$ configurations.
The $^3S_1$--$^3P_1$ mixing vanishes at leading order, a 
fact that has important consequences for the estimate of the 
deuteron EDM \cite{jordysdeut}.
If $\tilde\delta_3$ and $\varepsilon$ are different from zero, 
the qCEDM and the chiral-invariant TV sources also contribute 
to isospin-breaking TV observables at leading order. 
The operator $\bar C_{3t}$ is proportional to the 
third component of the total isospin of the two-nucleon pair, 
and thus it does contribute to $^1S_0$--$^3P_0$ mixing, 
but only for $I_3 = \pm 1$. $\bar C_{3 s}$ is instead 
proportional to the total spin of the two nucleons, and 
it is relevant to $^3S_1$--$^3P_1$ mixing, and, 
consequently,  to the deuteron EDM.  
Only the qEDM produces full isospin breaking in leading order.

A potential with just these five short-range terms
was considered recently \cite{moregud}.
Five low-energy quantities ---such as the spin rotation of a
polarized beam 
and the longitudinal polarization of an unpolarized incident
beam in neutron scattering on a proton target \cite{Tim+06} 
at different energies--- 
are needed to determine the parameters $\bar c_{s,t}$,
$\bar c_{3s,3t}$, and $\bar c_{33t}$.

\begin{figure}[t]
\begin{center}
\includegraphics[width=2.7cm]{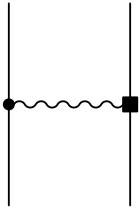} 
\end{center}
\caption{One-photon-exchange diagram 
contributing to the long-range TV 
two-nucleon potential. 
The solid and wavy lines represent nucleon and photon, 
respectively; 
a square stands for the TV photon-nucleon couplings
in $\mathcal L^{(1)}_{f = 2, \slashT 6}$ 
\eqref{LagLO6edm}, 
while the filled circle represents an interaction from 
$\mathcal L^{(0)}_{f \le 2, T}$
\eqref{LagStrong0}. Only one possible ordering is shown.
}
\label{loOphotonE}
\end{figure}

There is, however, an important extra ingredient that needs to be 
added at low energies: one-photon exchange where one of
the vertices originates in the nucleon EDM, as in Eq. \eqref{LagLO6edm}, see Fig. \ref{loOphotonE}.
This long-range potential is particularly important for
qEDM, since for this source the short-range interactions
have suppression by $\alpha_{\textrm{em}}/4\pi$ 
from the hard photon, as can be seen in Eqs. \eqref{cst},
\eqref{c3s3t}, and \eqref{c33t}.
We find
\begin{eqnarray}
V_{\gamma,\slashT}(\vec q ) & =& 
- \frac{ie}{2} 
\left[ \bar d_0 
+ \bar d_1 \tau^{(1)}_3 \tau^{(2)}_3  \right] 
\left( \vec\sigma^{\, (1)} - \vec\sigma^{\,(2)} \right) \cdot 
\frac{\vec q}{\vec q^{\, 2}}
\nonumber \\
& & 
- \frac{ie}{4} 
\left[ (\bar d_0+ \bar d_1) \left(\tau_3^{(1)} + \tau_3^{(2)}\right)
\left( \vec\sigma^{\, (1)} - \vec\sigma^{\,(2)} \right)  
+ (\bar d_1 - \bar d_0)\left(\tau_3^{(1)} -\tau^{(2)}_3 \right) 
\left( \vec \sigma^{(1)} + \vec \sigma^{(2)}\right) 
\right\} \cdot \frac{\vec q}{\vec q^{\, 2}},
\nonumber \\
\label{Vgammapionless}
\end{eqnarray}
where 
\begin{equation}
\bar d_{0,1}=
\mathcal O\left(e \delta_{0,3}\frac{m_\pi^2}{M^2_{\slashT}M_{QCD}}\right)
\label{dst}
\end{equation}
are the isoscalar and isovector components of the nucleon EDM.
In principle, the coefficients $\bar d_{0,1}$ in the pionless theory are different 
from the short-range contributions to the nucleon EDM in the pionful theory introduced in Eq. \eqref{LagLO6edm}. However, since in the case of TV from the qEDM 
pion-loop contributions to the nucleon EDM are suppressed \cite{dim6NEDFF},
at leading order they exactly match.

The photon-exchange potential has the same spin/isospin components of the
short-range potential \eqref{Vnnnn}, but it acts on
all partial waves.
While for the other sources 
this type of potential is important only for very low momenta,
for qEDM it is enhanced with respect to
the short-range potential (47) by a factor of $M_{nuc}/\aleph$ for $Q \sim \aleph$, and it dominates throughout the regime
of the pionless EFT.

In subleading orders more derivatives appear in two-nucleon
contact interactions and photon exchange. 
An important issue is the order where
few-nucleon forces first appear. On
the basis of naive dimensional analysis, we
expect them to be also of subleading order.
In the PV TC case, this is confirmed
for the three-nucleon force 
by a more detailed analysis based on renormalization-group
invariance \cite{PV3Npiless}.
Since the powers of momenta involved here are the same,
the same conclusion should hold.
The potentials \eqref{Vnnnn} 
and \eqref{Vgammapionless} should then be sufficient for most
TV applications of the pionless EFT.

\section{The TV 
Potential in Momentum Space}
\label{qspace}

In processes involving momenta $Q\sim M_{nuc}$, which presumably
comprise the bound states of most nuclei, pion effects are important
and pion degrees of freedom should be included explicitly in the theory. 
In this section we use the interactions
given in Sect. \ref{ChPT} to compute the TV 
nuclear potential in momentum space for the TV sources of dimension up to 6.

In the lowest orders, the TV nuclear potential involves only two nucleons.
We write the two incoming momenta as
$\vec p_1 = \vec P/2 + \vec p$ and $\vec p_2= \vec P/2 -\vec p$,
and the two outgoing momenta as 
$\vec p^{\, \prime}_1 = \vec P/2 + \vec p^{\,\prime}$
and $\vec p^{\,\prime}_2 = \vec P/2 -\vec p^{\, \prime}$.
The TV potential in momentum space can be expressed as 
function not only of the 
momentum transfer $\vec q = \vec p^{} - \vec p^{\, \prime}$, 
but also of the center-of-mass (CM) momentum $\vec P$ and of the variable 
$\vec K = (\vec p + \vec p^{\, \prime})/2$:
$V_{\slashT} = V_{\slashT}(\vec q, \vec K, \vec P)$.
Expressions for the potential in the 
CM frame are obtained 
by setting $\vec P =0$. 
Notice 
that although some of the terms below vanish  
in the 
CM frame, 
they can be relevant to
the calculation of the TV electromagnetic form factors of deuteron,
or for calculations of $T$ violation in nuclei with $A > 2$,
where the interaction with the photon or other nucleons
changes the 
CM momentum of the nucleon pair.

\subsection{$\bar\theta$ Term}
\label{qspacetheta}

\begin{figure}[t]
\begin{center}
\includegraphics[width = 2.7cm]{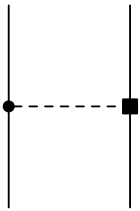} 
\end{center}
\caption{OPE diagram 
contributing to the leading TV 
two-nucleon potential. 
The solid and dashed lines represent nucleon and pion, respectively; 
a square stands for the TV pion-nucleon couplings 
$\bar g_0$ in $\mathcal L^{(1)}_{f = 2, \slashT 4}$ 
\eqref{LagLO}
or $\bar g_{0,1,2}$ and $\bar\imath_1$ in 
$\mathcal L^{(\Delta_\pi)}_{f= 2, \slashT 6}$ \eqref{LagLO6}, 
while the filled circle represents an interaction from 
$\mathcal L^{(0)}_{f \le 2, T}$
\eqref{LagStrong0}. Only one possible ordering is shown.
}
\label{loOPE}
\end{figure}

In leading order, 
the $\bar\theta$-term nuclear potential 
comes from the OPE diagrams of Fig. \ref{loOPE}, 
with TC and TV  pion-nucleon interactions 
taken from
$\mathcal L^{(0)}_{f \le 2, T}$ and $\mathcal L^{(1)}_{f = 2, \slashT 4}$
in Eqs. \eqref{LagStrong0} and \eqref{LagLO}, respectively. 
The strong-interaction vertex introduces a factor of $g_A
Q/F_\pi$, while the TV vertex brings in a factor 
$\bar{g}_{0}\propto m^2_{\pi}/M_{QCD}$. As a
result this contribution goes as $M_{QCD}^{-1}$ and it is of 
order $\nu =1$.
In momentum space, the expression for the potential is simply
\begin{equation}
V^{(1)}_{\bar\theta}
= i \frac{g_{A}\bar{g}_{0}}{ F_\pi^2}
\vec{\tau}^{\, (1)}\cdot \vec{\tau}^{\, (2)}
\left( \vec{\sigma}^{(1)}-\vec{\sigma}^{(2)}\right) \cdot 
\frac{\vec{q}}{\vec{q}^{\,2}+m_{\pi}^{2}},
\label{onepion}
\end{equation}
which agrees with Ref. \cite{HaxHenley83}.
Just like the $\bar\theta$ term contribution to the potential \eqref{Vnnnn}
in the pionless theory, this OPE potential 
contributes to $^1S_0$--$^3P_0$ and $^3S_1$--$^1P_1$ mixing,
but not to the isospin-violating $^3S_1$--$^3P_1$ mixing.
At this order, there is a single unknown TV parameter, $\bar{g}_0$.
Contrary to the PC, TC case \cite{nEFT} 
and more like the PV, TC potential \cite{PVNN}, pion physics is
enhanced relative to short-range physics due to 
the absence of a derivative in the simplest pion-nucleon TV interaction
and the presence of one in the simplest TV two-nucleon contact interaction.

Since in nuclei the OPE from the $I=0$ pion-nucleon coupling $\bar g_0$ is 
suppressed by the factor $(N-Z)/A$,
it is interesting to pursue higher orders,
up until the $I=1$ pion-nucleon coupling $\bar g_1$, whose OPE is not affected 
by such suppression, appears.
This means up to 
$\nu =3$, that is, including corrections 
of 
${\cal O}(Q^2/M^2_{QCD})$ with respect to the leading TV potential.
According to Eq. \eqref{nudelta}, corrections at 
orders $\nu = 2,3$ come from 
one-loop diagrams involving 
$\mathcal L^{(0)}_{f \le 2, T}$ and $\mathcal L^{(1)}_{f = 2, \slashT 4}$ 
only,
and from tree diagrams with insertions of higher-order terms.
The tree contributions come 
from the four-nucleon TV operators in $\mathcal L^{(3)}_{f = 4, \slashT 4}$,
Eq. \eqref{4ntodd}, 
and from OPE diagrams in which either the 
TC 
or the TV vertices originate in the power-suppressed $f\le 2$ Lagrangians.

\begin{figure}[t]
\begin{center}
\includegraphics[width = 2.35cm]{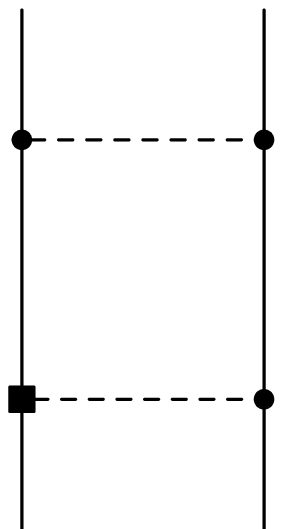} \hspace{0.75cm}
\includegraphics[width = 2.35cm]{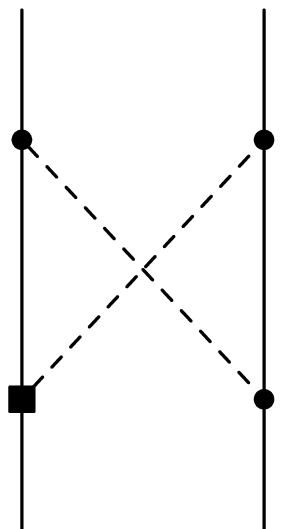} \hspace{0.75cm}
\includegraphics[width = 2.35cm]{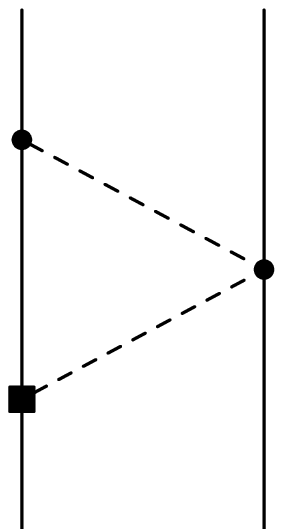} 
\end{center}
\caption{Box, crossed, and triangle TPE diagrams 
contributing to the subleading 
TV two-nucleon potential. 
Notation as in Fig. \ref{loOPE}. Only one possible ordering per topology is shown.}
\label{box}
\end{figure}

The most important loop diagrams are from TPE,
depicted in Fig. \ref{box}. 
The $T$-odd pion-nucleon coupling $\bar g_0$ and 
one of the strong-interaction vertices bring in a factor of 
$\bar g_0 g_A/F^2_{\pi}$. 
The other two vertices of the box and
crossed diagrams of Fig. \ref{box} are 
strong-interaction pion-nucleon
vertices from Eq. (\ref{LagStrong0}), 
and combined with the  $(4\pi)^2$ from the loop integration, 
they yield the suppression factor 
$g_A^2/(4\pi F_{\pi})^2 \sim 1/M^2_{QCD}$.  
For the triangle diagrams, the seagull vertex is the Weinberg-Tomozawa term
also from Eq. (\ref{LagStrong0}), 
which brings in a factor of $1/F^2_{\pi}$ that, combined with the  $(4\pi)^2$ 
from the loop, also leads to a suppression of 
$1/(4\pi F_{\pi})^2 \sim 1/M^2_{QCD}$.
All these 
diagrams are thus of order  $M_{QCD}^{-3}$.
Care of course has to be taken with the subtraction
from the box diagrams in Fig. \ref{box} 
of the iterated static OPE, which is infrared enhanced
and already included in the computation of wave functions at lower order.  
Following the procedure described
for instance in Ref. \cite{PVNN}, 
the subtraction is accomplished by exploiting the identity 
\begin{equation}\label{distrib}
\frac{i}{-v\cdot k + i\varepsilon} = 
-\frac{i}{v\cdot k + i\varepsilon} + 2\pi \delta(v\cdot k).
\end{equation}
When Eq. \eqref{distrib} is used in place of one of the nucleon propagators 
in the box diagrams, the first term on the right-hand side leads to 
a contour integral over the 0th component of the loop momentum, 
which can be performed without picking up the nucleon poles
and is free of the infrared enhancement discussed in Sect. \ref{ChPT}, 
while the delta function corresponds to the two-nucleon pole
and must be discarded in the calculation of the potential.
For the crossed-box and triangle diagrams, instead, it is always possible to 
avoid the nucleon poles, and these diagrams only contribute to the potential.

The TPE diagrams in Fig. \ref{box} are 
ultraviolet divergent. We regulate them in dimensional regularization
in $d$ spacetime dimensions, where divergences get encoded in the
factor
\begin{equation}
L = \frac{2}{4-d} - \gamma_E + \ln 4\pi ,
\label{L}
\end{equation}
where $\gamma_E$ is the Euler constant.
We denote by $\mu$ the renormalization scale.
Proper renormalization requires that sufficiently many
counterterms appear at the same order to compensate for the $L$ and $\mu$ 
dependence of the loops.
Indeed, here
this dependence can be absorbed by the renormalization of the 
contact interaction $\bar C_{2}$ from ${\cal L}_{f = 4, \slashT 4}^{(3)}$,
Eq. \eqref{4ntodd},
which we do by redefining it through
\begin{equation}
\bar C_{2} \to
\bar C_{2} + \frac{2 g_A \bar g_0}{F^2_{\pi}} \frac{1}{(2\pi F_{\pi})^2} 
\left[ \left(3g_A^2- 1\right) 
\left(L + \ln \frac{\mu^2}{m^2_{\pi}}\right) + 2\left( g_A^2 -1	\right)  
\right].
\label{C2ren}
\end{equation}
Note that we chose to 
absorb in $\bar C_{2}$ some finite constant pieces.
TPE does not renormalize the coupling $\bar C_{1}$ at this order.
With this redefinition,
the contact interactions yield the short-range potential
\begin{equation}
V^{(3)}_{\bar\theta, \textrm{SR}}(\vec q) = 
-\frac{i}{2} 
\left[\bar C_{1} 
+\bar C_{2}\, \vec\tau^{\,(1)} \cdot \vec \tau^{\,(2)} \right]
\left(\vec \sigma^{\,(1)} - \vec \sigma^{\,(2)}\right) \cdot \vec q ,  
\label{Vnnnnpion}
\end{equation}
which is formally identical to the leading 
$\bar\theta$ potential in the pionless EFT,
Eq. \eqref{Vnnnn}. 
The couplings, however, are different.
We can see from  Eq. \eqref{C2ren} that the natural size
of the coefficients $\bar C_{i}$ is,
as advertised, $\bar\theta m_\pi^2/F_\pi^2 M_{QCD}^3$,
implying a suppression of $Q^2/M_{QCD}^{2}$ with respect to TV OPE.

Once the divergent, short-range part of TPE has been lumped with
the contact terms, we are left with
the non-analytic contributions of medium range,
\begin{equation}
 V^{(3)}_{\bar\theta, \textrm{MR}} = 
-i  \frac{2 \bar g_0 g_A}{F^2_{\pi}} \frac{1}{(2\pi F_{\pi})^2} 
\left[2g_A^2 \; B\left(\frac{\vec q^{\,2}}{4m_{\pi}^{2}}\right) 
- T\left(\frac{\vec q^{\,2}}{4m_{\pi}^{2}}\right)\right]
\vec \tau^{\, (1)}\cdot \vec \tau^{\, (2)}  
\left(\vec\sigma^{(1)} - \vec\sigma^{(2)}\right) \cdot \vec q  , 
\label{TPEP}
\end{equation}
in terms of the functions
\begin{equation}
T(x) =
\sqrt{\frac{1+x}{x}}
\ln\left(\sqrt{x}+\sqrt{1+x}\right)=\frac{1+x}{1+\frac{3}{2}x} B(x).
\end{equation}
As the leading OPE potential, Eq. \eqref{onepion},
the TPE potential is a function only of the momentum transfer $\vec q$.
The scale of momentum variation is, as one would expect,
$2m_\pi$.
TPE and leading OPE share the same
spin-isospin structure, which means they can only be separated
if we probe their different momentum dependences.

\begin{figure}[tb]
\center
\includegraphics[width=2.5cm]{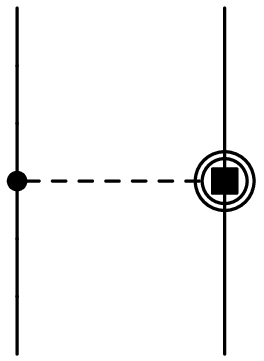}
\includegraphics[width=2.5cm]{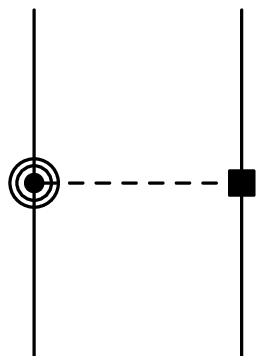}
\includegraphics[width=2.5cm]{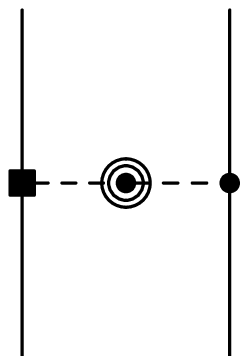}
\caption{OPE corrections to the  
$\bar\theta$-term 
two-nucleon potential up 
to order  $\mathcal O(Q^3/M_{QCD}^{3})$. 
The double circles denote vertices in the $\Delta =1,2$ 
TC chiral Lagrangians, $\mathcal{L}_{f \le 2, TI}^{(1)}$
\eqref{LagStrong1}, 
$\mathcal L^{(2)}_{f \le 2, TI}$ \eqref{LagStrong2},
and $\mathcal L^{(1, 2)}_{f \le 2, T\slashI}$ \eqref{chiralbreak3}. 
The doubly-circled square denotes vertices from the $\Delta = 3$ TV Lagrangian, 
$\mathcal L^{(3)}_{f = 2, \slashT 4}$
\eqref{LagT2}. Other notation as in Fig. \ref{loOPE}. 
Only one possible ordering is shown.
}\label{opesub}
\end{figure}

A much richer structure arises from 
the remaining $\nu \le 3$ contributions to the two-nucleon TV potential,
which come from the OPE diagrams depicted in Fig. \ref{opesub}. 
Doubly-circled 
vertices in the first two diagrams denote  
$\mathcal O(Q^2/M^2_{QCD})$ corrections to the TC 
and TV pion-nucleon couplings,
given 
by the operators in the 
Lagrangians $\mathcal L^{(1, 2)}_{f \le 2, TI}$,
$\mathcal L^{(1, 2)}_{f\le 2, T\slashI}$, 
and $\mathcal L^{(3)}_{f = 2, \slashT 4}$ found in  
Eqs. \eqref{LagStrong1}, \eqref{LagStrong2}, \eqref{chiralbreak3},
and \eqref{LagT2}.
The last diagram
is proportional to  corrections to the pion mass in 
$\mathcal{L}_{f \le 2, TI}^{(2)}$ and 
$\mathcal L^{(1, 2)}_{f \le 2, T\slashI}$,
and to the nucleon mass difference
in $\mathcal L^{(1, 2)}_{f \le 2, T\slashI}$.
Note that, as we argue shortly, there are no further loop diagrams
to consider explicitly.

Corrections that originate in the pion mass are 
closely connected to
the leading OPE, Eq. \eqref{onepion}.
At the order we are considering,
the pion mass 
receives corrections from one-loop diagrams, 
which we absorb \cite{StrongSec} in the renormalization of the coupling 
$\Delta^{} m^2_{\pi}$ in Eq. \eqref{LagStrong2}. 
With the definitions of Eqs. \eqref{LagStrong0}, \eqref{LagStrong2}, 
\eqref{chiralbreak1}, and \eqref{chiralbreak2},
the physical masses of the neutral and charged pions are, respectively,  
$m^2_{\pi^0} = m^2_{\pi} + \Delta^{} m^2_{\pi} + \delta^{} m^2_{\pi} 
= (135 \; {\rm MeV})^2$ and 
$m^2_{\pi^{\pm}} = m^2_{\pi} +\Delta^{} m^2_{\pi} +\breve\delta^{} m^2_{\pi}
= (139.6 \; {\rm MeV})^2$ \cite{Nakamura:2010zzi}. 
The isospin-symmetric correction to the pion mass 
can be accounted at $\nu =3$
by substituting $m^2_{\pi} \rightarrow m^2_{\pi} +\Delta^{} m^2_{\pi}$
in the leading-order TV potential.
Isospin-breaking corrections come
from the different masses of the neutral and charged pions.
With the assumption 
$\alpha_{\textrm{em}}/4\pi \sim \varepsilon m^3_{\pi}/M^3_{QCD}$, 
which is numerically reasonable, 
the pion mass splitting is dominated by the electromagnetic contribution 
$\breve \delta m^2_{\pi}$, which gives rise to a potential of order $\nu = 2$. 
The quark-mass-difference contribution $\delta^{} m^2_{\pi}$ appears at $\nu =3$, 
when one should also consider diagrams with two insertions of 
$\breve \delta m^2_{\pi}$. 
The sum of these components generates two structures,
an isoscalar
\begin{equation}
V^{(2 + 3)}_{\bar\theta, \textrm{a}}(\vec q)  =  
- i \frac{\bar g_0 g_A}{3 F^2_{\pi}} 
\left( 2 \breve{\delta} m^2_{\pi} +  \delta^{} m^2_{\pi}     
- 2 \frac{(\breve{\delta} m^2_{\pi})^2}{q^2 + m^2_{\pi}} \right) 
\vec \tau^{\, (1)} \cdot \vec\tau^{\, (2)}
\, \left(\vec\sigma^{(1)} - \vec\sigma^{(2)}\right) \cdot 
\frac{\vec q  }{(\vec q^{\, 2} + m^2_{\pi})^2} ,
\label{pionmass2}
\end{equation}
and an isotensor
\begin{equation}
V^{(2+3)}_{\bar\theta, \textrm{b}}(\vec q)  =   
i \frac{\bar g_0 g_A}{3 F^2_{\pi}} 
\left(\breve{\delta} m^2_{\pi}  - \delta^{} m^2_{\pi}  
- \frac{(\breve{\delta} m^2_{\pi})^2}{q^2 + m^2_{\pi}}  \right)
\left(3\tau^{(1)}_3\tau^{(2)}_3 -\vec\tau^{\, (1)}\cdot\vec\tau^{\, (2)}\right)
\left(\vec\sigma^{(1)} - \vec\sigma^{(2)}\right) \cdot 
\frac{\vec q}{(\vec q^{\, 2} + m^2_{\pi})^2} .  
\label{pionmass1}
\end{equation}
These isoscalar and isotensor components can be rewritten 
using the physical pion masses.
The isoscalar component, Eq. \eqref{pionmass2},
can be obtained by using the physical values of the neutral and 
charged pion mass in the leading potential, Eq. \eqref{onepion}. 
We can write 
\begin{equation}
V^{(1)}_{\bar\theta}(\vec q)
+ V^{(2 + 3)}_{\bar\theta,\textrm{a}}(\vec q)
 = i \frac{\bar g_0 g_A}{3 F^2_{\pi}} 
\left(\frac{2}{\vec q^{\, 2} + m^2_{\pi^{\pm}}} 
+ \frac{1}{\vec q^{\, 2} + m^2_{\pi^{0}}}\right)
\vec \tau^{\, (1)} \cdot \vec \tau^{\, (2)}  \,  
\left(\vec\sigma^{(1)} - \vec\sigma^{(2)}\right) \cdot \vec q  ,
\label{pionmass4}
\end{equation}
which, expanding in $\breve\delta m^2_{\pi}$ and $\delta^{} m^2_{\pi}$, 
reproduces Eq. \eqref{pionmass2}. The combination of neutral and charged pion 
propagators in Eq. \eqref{pionmass4} represents an ``average'' 
pion static propagator, which naturally appears in the isoscalar contribution. 
Similarly, we can rewrite the tensor component
as 
\begin{equation}
V^{(2+3)}_{\bar\theta, \textrm{b}}(\vec q)  =  -
i \frac{\bar g_0 g_A}{3 F^2_{\pi}} 
\left( \frac{1}{\vec q^{\, 2} + m^2_{\pi^{\pm}}} 
-  \frac{1}{\vec q^{\, 2} + m^2_{\pi^{0}}}\right)
\left(3\tau^{(1)}_3\tau^{(2)}_3 -\vec\tau^{\, (1)}\cdot\vec\tau^{\, (2)}\right)
\left(\vec\sigma^{(1)} - \vec\sigma^{(2)}\right) \cdot \vec q .   
\label{pionmass5}
\end{equation}
In applications to nucleon-nucleon scattering, 
this tensor component would contribute 
at low energies to $^1S_0$--$^3P_0$ mixing, 
affecting proton-proton and neutron-neutron ($I_3 = \pm 1$)
and neutron-proton ($I_3 = 0$) scattering differently. 
It is worth stressing that, in contrast to phenomenological approaches, 
the isotensor component coming from  the ${\bar \theta}$ term is not 
a leading contribution.

Corrections from the nucleon mass come in several guises.
The use of a heavy-nucleon field ensures that the large scale $m_N$
appears always in denominators.
In the isospin-symmetric limit the first effects of $m_N$ 
enter in the $\Delta=1,2$ TC Lagrangians, 
Eqs. \eqref{LagStrong1} and \eqref{LagStrong2}, the $\Delta=3$ TV Lagrangian, 
Eq. \eqref{LagT2}, and, via the on-shell condition for the nucleons, 
the energy of the potential-pion propagator.
They yield relativistic corrections to the leading OPE 
with the same spin-isospin structure,
\begin{equation}
V_{\bar\theta,\textrm{c}}^{(3)}(\vec q, \vec K, \vec P) = 
- i \frac{ g_A \bar g_0}{F_{\pi}^2 m_N^2} 
\left(\vec K^2 + \frac{\vec P^2}{4}
- \frac{1}{4 } \frac{(\vec P \cdot \vec q\,)^2}{\vec q^{\, 2} + m^2_{\pi}}
\right)
\vec \tau^{\, (1)} \cdot \vec \tau^{\, (2)} \, 
\left(\vec\sigma^{(1)} - \vec\sigma^{(2)}\right) \cdot 
\frac{\vec q}{\vec q^{\, 2} + m^2_{\pi}} ,
\label{Va}
\end{equation}
and with new structures,
\begin{eqnarray}
V_{\bar\theta, \textrm{d}}^{(3)}(\vec q, \vec K, \vec P) &=& 
- i \frac{ g_A \bar g_0}{4 F_{\pi}^2 m_N^2} 
\vec \tau^{\,(1)} \cdot \vec \tau^{\,(2)} \, 
\frac{1}{\vec q^{\, 2} + m^2_{\pi}} 
\left\{
\vec P \cdot \vec q
\left[ \left(\vec\sigma^{(1)}-\vec\sigma^{(2)}\right)\cdot \frac{\vec P}{2} 
+  \left(\vec\sigma^{(1)}+\vec\sigma^{(2)}\right)\cdot \vec K \right] 
\right.
\nonumber
\\
&& \left.
+i\vec\sigma^{(1)} \cdot 
\left[\vec q\times\left(\frac{\vec P}{2}+\vec K\right)\right]\, 
\vec\sigma^{(2)} \cdot\vec q  
+i\vec\sigma^{(1)} \cdot \vec q \, 
\vec\sigma^{(2)}\cdot\left[\vec q \times\left(\frac{\vec P}{2}-\vec K\right)
\right]
\right\}.
\label{Vc}
\end{eqnarray}
Note that in writing Eqs. \eqref{Va} and \eqref{Vc} 
we have omitted pieces that vanish due to 
energy-momentum conservation for on-shell nucleons,
which implies $\vec K \cdot \vec q = \vec p^2 - \vec p^{\, \prime 2} = 0$.
This potential includes the contribution of the $1/m_N$ correction to $g_A$ 
in Eq. \eqref{LagStrong1}.
Naively, one would expect this correction to contribute at order $\nu =2$; 
however the interaction brings in a factor of $v\cdot q$, which, 
for on-shell nucleons, becomes 
$v\cdot q = \vec q \cdot \vec P/2 m_N$,
suppressing 
the potential by a further factor of $1/m_N$.
There is a subtlety in this argument. 
When one performs the integral involving 
both OPE with a pion energy in the numerator and another interaction 
in the potential, and picks the pion pole, one gets a one-loop contribution
to the potential. However, by power counting, such diagrams are suppressed
by a further $Q/M_{QCD}$. 
(This of course does not preclude enhancements
by factors of $\pi$ that in principle might affect any ChPT loop,
but are hard to incorporate in power counting.)

Corrections to the nucleon mass can be removed from nucleon propagators
by redefinitions of the nucleon field.
The chiral-symmetry-breaking correction to the nucleon mass,
$\Delta m_N$ in Eq. \eqref{LagStrong1}, can be absorbed
in $m_N$, $m_N\to m_N-\Delta m_N$, by a redefinition of the nucleon field
of the same type of that which eliminates the mass from Eq. \eqref{LagStrong0} 
in the first place.
The isospin-violating nucleon mass splittings 
$\delta m_N$ and $\breve\delta m_N$ can be dealt with
the field redefinition of Ref. \cite{Friar:2004ca}, which leads to
Eq. \eqref{chiralbreak3}.
The corresponding potential linear in $\delta m_N$ 
has a $1/m_N$ factor,
\begin{eqnarray}
V^{(3)}_{\bar\theta, \textrm{e}}(\vec q, \vec K,\vec P )  
&=&   \frac{\bar g_0 g_A}{F^2_{\pi}} \frac{\delta m_N}{m_N} 
\frac{1}{\vec q^{\, 2} + m^2_{\pi}} 
\left( \vec \tau^{\, (1)} \times \vec \tau^{\, (2)}\right)_3
\nonumber \\
& & 
\left[\left(\vec\sigma^{(1)} + \vec\sigma^{(2)} \right) \cdot \vec K
+\left(\vec\sigma^{(1)} - \vec\sigma^{(2)} \right) \cdot 
\left(\frac{\vec P}{2} 
+ 
\frac{\vec q}{\vec q^{\, 2} + m^2_{\pi}} 
\vec P \cdot \vec q \right)
\right]. 
\label{pionmass3}
\end{eqnarray}
This potential has the right quantum numbers to produce 
$^3S_1$--$\,^3P_1$ mixing, and therefore  must be included in a 
calculation of the deuteron EDM.
In addition, there are terms quadratic in $\delta m_N$, which 
generate an additional contribution to the isoscalar 
and tensor 
potentials
in Eqs. \eqref{pionmass2} and \eqref{pionmass1}, 
\begin{equation}
V^{(3)}_{\bar\theta, \textrm{f}}(\vec q)  =   
i  \frac{\bar g_0 g_A}{3F^2_{\pi}} \delta m_N^2  
\left[2\vec \tau^{\, (1)} \cdot \vec \tau^{\, (2)}  \,  
- 
\left( 3\tau^{(1)}_3\tau^{(2)}_3 -\vec \tau^{\, (1)} \cdot\vec\tau^{\, (2)}
\right) \right] 
\left(\vec\sigma^{(1)} - \vec\sigma^{(2)}\right) \cdot 
\frac{\vec q}{(\vec q^{\, 2} + m^2_{\pi})^2}.
\label{pionmass7}
\end{equation}
Note that the electromagnetic correction to the nucleon mass,
$\breve\delta m_N$, does not appear in the equations above.
The reason is that in Eq. \eqref{chiralbreak3} it appears
with a $v\cdot \partial \vec\pi$ factor,
which again brings a $v\cdot q$ and consequently an extra suppression
$\sim Q/m_N$. The contribution from $\breve\delta m_N$ is thus next
order. Up to such higher-order terms, we can make 
$\delta m_N\to \delta m_N+\breve\delta m_N=m_n-m_p$ in the expressions above.

Finally we arrive at contributions from $\nu=3$ effects in
the pion-nucleon vertices.
At this order, several contributions can be absorbed into 
redefinitions of 
the couplings $g_A$, $\bar g_0$, and $\bar C_{2}$. 
One-loop corrections to $g_A$ do not introduce 
any non-analytic contribution and, for an on-shell nucleon, 
they  renormalize the coupling 
$d_A$ in Eq. \eqref{LagStrong2} \cite{Park:1993jf}. 
The operator with coefficient 
$d_A$ 
gives rise to a potential like Eq. \eqref{onepion}, with $g_A$ replaced
by 
$- g_A d_A$. For simplicity we absorb 
$d_A$ in $g_A$,
$g_A\to g_A (1+d_A)$.
Similarly, 
the calculation of the pion-nucleon TV form factor in 
Ref. \cite{emanuele} shows that the one-loop corrections to $\bar g_0$ 
do not introduce any non-trivial momentum dependence, so they 
simply renormalize  the coupling $\Delta \bar g_0$ in Eq. \eqref{LagT2}. 
These $m_\pi^2$ corrections to $\bar g_0$ 
can be absorbed in it,
$\bar g_0\to \bar g_0 - \Delta \bar g_0 + \bar g_0 \delta m_\pi^2/m_\pi^2$.
As for the operators with coefficients 
$c_A$ in Eq. \eqref{LagStrong2} and
$\bar \eta$ in Eq. \eqref{LagT2},
they give potentials of the form
\begin{equation}
\vec\tau^{(1)} \cdot \vec\tau^{(2)} 
\left(\vec{\sigma}^{(1)} - \vec\sigma^{(2)}\right) \cdot 
\vec q \, \frac{\vec q^{\, 2}}{\vec q^{\, 2} + m^2_{\pi}} = 
- \vec\tau^{(1)} \cdot 
\vec\tau^{(2)} \left(\vec{\sigma}^{(1)} - \vec\sigma^{(2)}\right) \cdot 
\vec q \frac{m^2_{\pi}}{\vec q^{\, 2} + m^2_{\pi}}
+ \vec\tau^{(1)} \cdot \vec\tau^{(2)} 
\left(\vec{\sigma}^{(1)} - \vec\sigma^{(1)}\right) \cdot \vec q ,
\label{OPEtoOPE+con}
\end{equation}
which cannot be distinguished from those of 
$g_A \bar g_0$ and 
$\bar C_{2}$, 
and can therefore be absorbed into further
redefinitions of $g_A$, $\bar g_0$, and $\bar C_{2}$.
Now the Goldberger-Treiman relation 
for the strong pion-nucleon constant,
$g_{\pi N N} = 2 m_N g_A/F_{\pi}$, applies without an explicit correction.
If for the pion-nucleon coupling constant we use  
$g_{\pi N N} = 13.07$ \cite{piNcoupling}, then in 
the leading-order TV potential we should use $g_A=1.29$.

The remaining contributions come from vertex corrections, both
in the TV sector via 
the TV pion-nucleon coupling $\pi_3 \bar N N$ in Eq. \eqref{LagT2},
and in the TC sector via the isospin-breaking
pion-nucleon axial-vector coupling 
$\partial_\mu\pi_3 \bar N S^\mu N$ in Eq. \eqref{chiralbreak3}.
We find
\begin{eqnarray}
V_{\bar\theta, \textrm{g}}^{(3)}(\vec q) &=& 
\frac{i}{2F^2_{\pi}}\frac{1}{\vec q^{\, 2} + m^2_{\pi}} 
\left\{\left(g_A  \bar g_1  - \frac{\bar g_0 \beta_1}{2} \right) 
\left(\tau_3^{(1)} + \tau_3^{(2)} \right) \,
\left( \vec\sigma^{(1)} -  \vec\sigma^{(2)}\right)\cdot \vec q 
\right.
\nonumber\\
&&\left.+ 
\left(g_A  \bar g_1
+ \frac{\bar g_0 \beta_1}{2 } \right)
\left(\tau_3^{(1)} - \tau_3^{(2)} \right)\,
\left( \vec\sigma^{(1)}  +  \vec\sigma^{(2)}\right)\cdot \vec q \right\},
\label{pi3prime}
\end{eqnarray}
where we redefined $\bar g_1$ to absorb the tadpole contribution
$\bar g_1 \rightarrow \bar g_1 
+ 2 \bar g_0 \Delta m_N\delta m^2_{\pi}/\delta m_N m^2_{\pi}$. 
The first structure contributes to $^1S_0$--$^3P_0$ mixing. 
Being proportional to 
$I_3$, the contribution vanishes in the case of neutron-proton scattering, 
and is only relevant for proton-proton or neutron-neutron scattering.
Because of its isospin structure, it
does not affect the $^3S_1$--$^1P_1$ and $^3S_1$--$^3P_1$ channels, and, 
in particular, it is not relevant for the calculation of the deuteron EDM.
The second structure, in contrast, contributes to $^3S_1$--$^3P_1$ mixing, 
and, consequently, to the deuteron EDM.
Its contribution vanishes in the other low-energy channels.

Note that loop diagrams involving the leading $S$-wave TC four-nucleon 
operators and a TV pion exchange all vanish. 
The analysis of Refs. \cite{kids,nogga} showed that some higher-wave 
TC four-nucleon operators are less suppressed than expected on the grounds 
of naive dimensional analysis, and they must be included in the leading-order 
$f=4$ TC Lagrangian. 
Loop diagrams with $P$-wave operators and a TV pion exchange 
do not vanish.
However, these diagrams do not depend on the momentum transfer $\vec q$ and 
they simply renormalize the couplings $\bar C_1$ and $\bar C_2$.

One can proceed in the same manner to construct higher-order potentials.
At next order there are further OPE and TPE, 
and also one-photon-exchange, contributions
to the two-nucleon potential.
There is also the appearance
of the lowest-order three-nucleon TV  
potential, which arises from essentially three mechanisms:
{\it (i)} a TPE component $\propto g_A \bar{g}_0/m_N F_\pi^4$ 
involving a pion energy in a Weinberg-Tomozawa seagull vertex;
{\it (ii)} a TPE component $\propto g_A^2 \bar{h}_0/F_\pi^4$ 
involving the seagull vertex from 
${\cal L}_{f =2, \slashT 4}^{\left( 2 \right)}$, Eq. \eqref{LagSLOTodd};
and 
{\it (iii)} a one-pion/short-range component 
$\propto g_A \bar{\gamma}_i/F_\pi^2$
involving the short-range pion-two-nucleon interactions from 
${\cal L}^{(2)}_{f =4, \slashT 4}$, Eq. \eqref{LTVpiNNNN}.
The fact that, in the absence of an explicit delta isobar, the 
three-nucleon potential first shows up three orders beyond leading
is completely analogous to the TC PC case \cite{TCpot3}.
An important difference is that, because of the relative enhancement of 
pion exchange compared to short-range physics, the leading
TV PV three-nucleon force does not include a purely
short-range component. Thus, this TV PV three-nucleon force 
is in principle determined by one- and two-nucleon physics.

\subsection{Dimension-6 Sources}
\label{qspacedim6}

Our attention has been focused so far on the TV potential from the 
$\bar \theta$ term, 
in which case the vanishing of $\bar g_1$ at leading order makes it important to 
consider subleading contributions.
We now briefly turn our attention to the TV potential from dimension-6 
sources of $T$ violation.

For all dimension-6 sources, the leading-order potential contains
OPE of the form in Fig. \ref{loOPE}.
Since these sources all generate  $I=0$ and $I=1$ pion-nucleon couplings 
of the same size, we have no motivation to go to subleading order 
in the potential. The leading-order potential, which is a function
of the transfer momentum  $\vec q$ only,
should be sufficient for most phenomenological applications.
Of course, if needed, subleading orders can be derived just as we
have done for $\bar\theta$.

In the case of the qCEDM, the TV couplings $\bar g_{0}$ and $\bar g_{1}$ 
both appear 
in the $\Delta = -1$ Lagrangian, Eq. \eqref{LagLO6}. 
As a consequence, the leading potential from the qCEDM has chiral index 
$\nu = -1$, and it has 
both an isospin-conserving part, which is identical to 
Eq. \eqref{onepion}, and an isospin-breaking one.
The leading potential is
\begin{eqnarray}
V^{(-1)}_{\textrm{qCEDM}}(\vec q ) & =& 
i \frac{g_A \bar g_{0}}{F^2_{\pi}} \vec\tau^{(1)} \cdot \vec\tau^{(2)}
\left(\vec\sigma^{\,(1)} - \vec\sigma^{\,(2)}  \right) \cdot 
\frac{\vec q}{\vec q^{\, 2} + m^2_{\pi}} 
\nonumber \\
& & 
+ i\frac{g_A \bar g_{1}}{2 F^2_{\pi}}  
\left[ \left(\tau_3^{(1)} + \tau_3^{(2)}\right)
\left( \vec\sigma^{\, (1)} - \vec\sigma^{\,(2)} \right)  
+ \left(\tau_3^{(1)} -\tau^{(2)}_3 \right) 
\left( \vec \sigma^{(1)} + \vec \sigma^{(2)}\right) 
 \right] \cdot \frac{\vec q}{\vec q^{\, 2} + m^2_{\pi}}.
\end{eqnarray}

For the TV chiral-invariant (CI) sources, 
that is, gCEDM and TV FQ, the $\Delta = -1$
TV  Lagrangian contains both 
pion-nucleon couplings and 
four-nucleon operators,
see Eqs. \eqref{LagLO6} and  \eqref{eq:3.1.50}.
The additional TV pion-nucleon coupling $\bar\imath_{1}$ produces a potential
of the type \eqref{OPEtoOPE+con},
and thus can be absorbed in redefinitions 
$\bar g_{0}\to \bar g_{0}+m^2_{\pi} \bar\imath_{1}$
and $\bar C_{2}\to \bar C_{2} + 2 g_A \bar\imath_{1}/F^2_{\pi}$.
As a consequence, the leading two-nucleon potential consists 
of an isospin-conserving and an isospin-breaking 
one-pion-exchange contribution and a short-distance piece,    
\begin{eqnarray}
V^{(-1)}_{\textrm{TVCI}}(\vec q ) & = & 
i \frac{g_A \bar g_{0}}{F^2_{\pi}} \vec\tau^{\,(1)} \cdot \vec\tau^{\,(2)}
\left(\vec\sigma^{\,(1)} - \vec\sigma^{\,(2)}  \right) \cdot 
\frac{\vec q}{\vec q^{\, 2} + m^2_{\pi}} 
\nonumber \\
& & 
+ i\frac{g_A \bar g_{1}}{2 F^2_{\pi}}  
\left[ \left(\tau_3^{(1)} + \tau_3^{(2)}\right)
\left( \vec\sigma^{\, (1)} - \vec\sigma^{\,(2)} \right)  
+ \left(\tau_3^{(1)} -\tau^{(2)}_3 \right) 
\left( \vec \sigma^{(1)} + \vec \sigma^{(2)}\right)\right] \cdot 
\frac{\vec q}{\vec q^{\, 2} + m^2_{\pi}} 
\nonumber \\
 && 
-\frac{i}{2} \left[\bar C_{1} 
+ \bar C_{2} \vec\tau^{\,(1)} \cdot \vec\tau^{\,(2)}\right] 
\left(\vec\sigma^{\,(1)} - \vec\sigma^{\,(2)}  \right) \cdot \vec q.
\label{eq:6.3.20}
\end{eqnarray}

The qEDM leading-order potential displays further new structures: in addition
to the $I=2$ pion-nucleon coupling $\bar g_2$ in Eq. \eqref{LagLO6}, 
there is also the 
one-photon-exchange contribution shown in Fig. \ref{loOphotonE},
where one vertex is the
short-distance contribution \eqref{LagLO6edm} to the nucleon EDM: 
\begin{eqnarray}
V^{(2)}_{\textrm{qEDM}}(\vec q ) & =& 
i \frac{g_A  }{F^2_{\pi}} \left(
\bar g_{0}\vec\tau^{(1)} \cdot \vec\tau^{(2)} 
+ \bar g_2 \tau^{(1)}_3 \, \tau^{(2)}_3\right)
\left(\vec\sigma^{\,(1)} - \vec\sigma^{\,(2)}  \right) 
\cdot \frac{\vec q}{\vec q^{\, 2} + m^2_{\pi}} 
\nonumber \\
& & 
+ i\frac{g_A \bar g_{1 }}{2 F^2_{\pi}}  
\left[ \left(\tau_3^{(1)} + \tau_3^{(2)}\right)
\left( \vec\sigma^{\, (1)} - \vec\sigma^{\,(2)} \right)  
+ \left(\tau_3^{(1)} -\tau^{(2)}_3 \right) 
\left( \vec \sigma^{(1)} + \vec \sigma^{(2)}\right) 
\right] \cdot \frac{\vec q}{\vec q^{\, 2} + m^2_{\pi}} 
\nonumber \\
& & 
- i\frac{e}{2} 
\left(\bar d_0 + \bar d_1 \tau^{(1)}_3 \tau^{(2)}_3\right)
\left( \vec\sigma^{\, (1)} - \vec\sigma^{\,(2)} \right) \cdot 
\frac{\vec q}{\vec q^{\, 2}} \nonumber \\
& & - i\frac{e}{4}  
\left[ (\bar d_0+ \bar d_1) \left(\tau_3^{(1)} + \tau_3^{(2)}\right)
\left( \vec\sigma^{\, (1)} - \vec\sigma^{\,(2)} \right)  
+ (\bar d_1 - \bar d_0)\left(\tau_3^{(1)} -\tau^{(2)}_3 \right) 
\left( \vec \sigma^{(1)} + \vec \sigma^{(2)}\right) \right] 
\cdot \frac{\vec q}{\vec q^{\, 2}}.
\nonumber \\
\end{eqnarray}

As for $\bar\theta$, at subleading orders the potential receives corrections from 
various mechanisms.
For all dimension-6 sources there are one-loop diagrams 
with the same topology as in Fig. \ref{box}, the square now denoting 
$\bar g_{0}$, $\bar g_{1}$ or $\bar g_{2}$. 
There are also 
tree-level OPE diagrams of the type in Fig. \ref{opesub}, 
with subleading TC and TV pion-nucleon vertices.
For qCEDM and qEDM, the one-derivative four-nucleon interactions
in the Lagrangian \eqref{eq:3.1.48}
have to be taken into account,
while for TV CI sources one has to include all the possible TV 
four-nucleon operators with three derivatives.
For qEDM, one also needs to include corrections involving
photon exchange.
This calculation proceeds along lines that are very similar to 
Sect. \ref{qspacetheta}.

In the case of the 
$\bar\theta$ term, TV three-nucleon forces only appear at NNNLO,
one order higher than the accuracy of our analysis. 
For qCEDM and qEDM the situation is similar,
but one might wonder whether for 
CI sources,
which appear to be more sensitive to 
short-distance physics, TV three-nucleon forces are more relevant. 
However, also in this case it turns out that the 
three-nucleon potential is a 
NNNLO 
effect.
The lowest-order three-nucleon potential receives various contributions:
$(i)$ a TPE component  $\propto g_A \bar g_{0,1}/ m_N F_{\pi}^4$ involving 
a pion energy in the Weinberg-Tomozawa vertex;
$(ii)$ a TPE component $\propto g_A^2 \bar \imath_{0}/m_N F_{\pi}^4$ 
involving a pion energy in the leading  TV seagull $\bar\imath_{0}$;
$(iii)$ TPE components from TV seagulls in the $\Delta = 0$ pion-nucleon 
Lagrangian, which we did not explicitly construct;
$(iv)$ a one-pion/short-range component $\propto g_A \bar\gamma_i/F^2_{\pi}$, 
with four-nucleon operators 
that contain at least one pion field 
in the $\Delta = 0$ Lagrangian; 
and 
$(v)$ short-range six-nucleon operators, 
which also appear in the $\Delta = 0$ Lagrangian. 
{}From the power counting formula \eqref{nudelta}, 
all these three-nucleon contributions are suppressed 
by three powers of $Q/M_{QCD}$ with respect to the 
effects of the two-nucleon potential \eqref{eq:6.3.20}
in the three-body system.

\section{The TV 
Potential  in Configuration Space}
\label{rspace}

The evaluation of $T$-odd observables in nuclear and atomic systems is 
often more easily carried out in configuration space. 
In this section we give the TV nuclear potential derived in Sect. \ref{qspace}
in coordinate space.
We compare this potential with the literature in the next section.

In the two-body case, it is convenient to introduce the relative position of 
the two nucleons $\vec r = \vec x_1 - \vec x_2$, 
their 
CM coordinate $\vec X = (\vec x_1 + \vec x_2)/2$, 
and the conjugate variables 
$- i \vec \nabla_r \equiv - i \partial/\partial \vec r$ and 
$- i\vec\nabla_X \equiv - i \partial/\partial \vec X$. 
Translation invariance constrains the potential to commute with $\vec \nabla_X$
and, therefore, not to depend on $\vec X$, so that in general the potential 
is a function of $\vec r$ and of the nucleons' relative 
and 
CM momenta,
$V_{\slashT} = V_{\slashT}(\vec r, \vec \nabla_r, \vec \nabla_X)$.
The relations between the potential in momentum space and in coordinate space 
are detailed in the Appendix.
Some care must be taken, and a regularization scheme has to be defined, 
when computing the Fourier transform of functions that blow up as 
$|\vec q\,|$ goes to infinity, as is the case of the subleading TV potential.
As described in the Appendix,
we follow Ref. \cite{Friar96} and 
define the  Fourier transform in $d$ dimensions.
We apply the $d$-dimensional Fourier
integration of the momentum-space potential before setting $d=4$.
This method 
eliminates naturally the divergent factor $\Gamma (2-d/2)$ 
arising from loops and
yields a finite result. As we will see, the divergent behavior at 
large momentum translates in a singular $\sim 1/r^4$ potential at 
short distances. 
Expressions in configuration space
obtained with this method are equivalent to the procedure based on
old-fashioned 
perturbation theory \cite{Friar96}.

In order to write the results of the Fourier transform
we introduce a few functions of the magnitude $r = |\vec r\,|$
of the radial coordinate:
\begin{equation}
U(r) = \frac{1}{12\pi r} 
\left[2\exp\left(-m_{\pi^{\pm}}r\right)+\exp\left(-m_{\pi^{0}}r\right)\right],
\label{Yukawa}
\end{equation}
which reduces to the usual Yukawa function $U(r) = \exp(-m_{\pi} r)/4\pi r$ 
when we ignore the pion mass difference;
\begin{equation}
W(r) =  \frac{1}{4\pi r} 
\left[\exp{(-m_{\pi^{\pm} } r)}- \exp{(-m_{\pi^0} r)}\right],
\label{Yukawadiff}
\end{equation}
which is entirely a consequence of isospin breaking; and
the TPE functions
\begin{eqnarray}
X(r)&=& \frac{1}{4\pi(2 \pi F_{\pi})^2 r^3} 
\int_0^1 dx \left(3 + 3\beta r +\beta^2 r^2\right)
\, \exp(-\beta r),
\label{Xfunk}\\
Y(r)&=&\frac{1}{2\pi(2 \pi F_{\pi})^2 r^3} 
\int_0^1 dx \, (1 + \beta r)\, \exp(-\beta r),
\label{Yfunk}
\end{eqnarray}
with $\beta^2 = m^2_{\pi}/x(1-x)$.

\subsection{$\bar\theta$ Term}
\label{rspacetheta}

The Fourier transform of the leading 
OPE potential
including the corrections from the pion mass \eqref{pionmass4}
and 
from the nucleon kinetic energy \eqref{Va},
is 
\begin{eqnarray}
&&V^{(1)}_{\bar\theta}(\vec r) 
+ V^{(2 + 3)}_{\bar\theta, \textrm{a}}(\vec r)
+ V^{(3)}_{\bar\theta, \textrm{c}}(\vec r, \vec\nabla_{r}, \vec\nabla_{X}) 
=  -  \frac{\bar g_0 g_A}{F^2_{\pi}} 
\vec \tau^{\, (1)} \cdot \vec \tau^{\, (2)}\,  
\left( \vec{\sigma}^{\, (1) }-\vec{\sigma}^{\,  (2) }\right) 
 \cdot
\nonumber \\ 
& &
\left[\left(\vec{\nabla}_r\, U(r)\right) 
\left(1 + \frac{\vec \nabla_X^{\, 2}}{4 m_N^2}\right)  
+ \left\{\frac{\nabla^i_r}{2 m_N} , \left\{ \frac{\nabla^i_r}{2 m_N}, 
\left(\vec\nabla_r  U(r)\right) \right\} \right\} 
+ \frac{(\vec\nabla_r \cdot \vec \nabla_X)^2}{8 m_{\pi}m_N^2} 
\left(\vec\nabla_r \, r U(r) \right) \right],
\label{onepionconfig} 
\end{eqnarray}
where $\{\cdots,\cdots\}$ denotes the anticommutator.
The remaining pion-mass correction, Eq. \eqref{pionmass5},
is
\begin{equation}
V^{(2+3)}_{\bar\theta, \textrm{b}}(\vec r) =    
\frac{ \bar g_0 g_A}{3 F^2_{\pi}} \left( 3 \tau^{(1)}_3 \tau^{(2)}_3 
- \vec \tau^{\, (1)} \cdot \vec \tau^{\, (2)}\right)\, 
\left(\vec\sigma^{(1)} - \vec\sigma^{(2)}\right) \cdot 
\left(\vec \nabla_r  W(r)  \right), 
\label{pimassconfig0}
\end{equation}
while 
the Fourier transform of the other relativistic corrections to the 
leading OPE,
Eq. \eqref{Vc}, is
\begin{eqnarray}
V^{(3)}_{\bar\theta, \textrm{d}}(\vec r, \vec \nabla_r, \vec \nabla_X) & = & 
- \frac{\bar g_0 g_A}{8 F^2_{\pi} m_N^2} 
\vec \tau^{\,(1)} \cdot \vec \tau^{\,(2)}  
\left[ 
\left( \vec \sigma^{\, (1)} - \vec\sigma^{\, (2)}\right)\cdot \vec \nabla_X \, 
\left( \vec\nabla_r U(r)\right)\cdot \vec \nabla_X  \right.
\nonumber
\\
& & \left. + \left(\vec \sigma^{\, (1)} + \vec\sigma^{\, (2)}\right)\cdot 
   \left\{ \vec\nabla_r, \left( \vec\nabla_r U(r) \right)\cdot \vec\nabla_X   
\right\} \right. 
\nonumber  
\\
&& \left.
+2i \varepsilon^{i j k}
\left(\sigma^{(1)\, i} \sigma^{(2)\, l} - \sigma^{(1)\, l} \sigma^{(2)\, i} 
\right)  
\left(\nabla^l_r \nabla^k_r U(r)\right) \nabla^j_r  
\right.
\nonumber \\
&& \left.
+i \varepsilon^{i j k}
\left(\sigma^{(1)\, i} \sigma^{(2)\, l} + \sigma^{(1)\, l} \sigma^{(2)\, i} 
\right)  
\left(\nabla^l_r \nabla^k_r U(r)\right) \nabla^j_X
\right].
\label{relcorr1config}
\end{eqnarray}
The nucleon mass-splitting corrections in
Eq. \eqref{pionmass3} become
\begin{eqnarray}
V^{(3)}_{\bar\theta, \textrm{e}}(\vec r, \vec \nabla_r, \vec \nabla_X )
&=&- i  \frac{ \bar g_0 g_A}{2F^2_{\pi}} \frac{\delta m_N}{m_N} 
\left( \vec \tau^{\, (1)} \times \vec \tau^{\, (2)}\right)_3 
\left[ \left(\vec\sigma^{(1)} + \vec\sigma^{(2)} \right) \cdot 
\left\{\vec \nabla_r, U(r)\right\}
\right.
\nonumber\\
&&\left.
+  \left(\vec\sigma^{(1)} - \vec\sigma^{(2)} \right) \cdot 
 \left( U(r) \vec \nabla_X
- \frac{1}{m_\pi} 
\left(\vec \nabla_r \,\nabla^i_r \; r  U(r) \right) \nabla^i_X \right)\right],
\label{deltamnconfig}
\end{eqnarray}
while those in 
Eq. \eqref{pionmass7} read
\begin{equation}
V^{(3)}_{\bar\theta, \textrm{f}}(\vec r) =   
- \frac{ \bar g_0 g_A}{3 F^2_{\pi}} \frac{ \delta m_N^2  }{m_{\pi}}   
\left[\vec \tau^{\, (1)} \cdot \vec \tau^{\, (2)}
-\frac{1}{2}
\left(3\tau^{(1)}_3\tau^{(2)}_3 -\vec\tau^{\, (1)}\cdot\vec\tau^{\, (2)}\right)
\right] 
\left(\vec\sigma^{(1)} - \vec\sigma^{(2)}\right) \cdot 
\left(\vec \nabla_r \; r U(r) \right).
\label{pimassconfig}
\end{equation}
The last OPE terms, Eq. 
\eqref{pi3prime},
are
\begin{eqnarray}
V_{\bar\theta, \textrm{g}}^{(3)}(\vec r)  & = &    
-\frac{\bar g_0 g_A}{2 F^2_{\pi}}
\left[\left(\frac{\bar g_1}{\bar g_0} 
- \frac{\beta_1}{2 g_A}\right)  
\left(\tau_3^{(1)} + \tau_3^{(2)} \right) \,
\left( \vec\sigma^{(1)}  -  \vec\sigma^{(2)}\right) 
\right.
\nonumber \\
&&\left. + 
\left(\frac{ \bar g_1}{\bar g_0} + \frac{\beta_1}{2 g_A}\right) 
\left(\tau_3^{(1)} - \tau_3^{(2)} \right) \,
\left( \vec\sigma^{(1)} +  \vec\sigma^{(2)}\right) 
 \right] \cdot \left(\vec  \nabla_r U(r) \right). 
\label{isoconfig2}
\end{eqnarray}

Finally,
the Fourier transform of the TPE potential in Eq. \eqref{TPEP} reads
\begin{equation}
V^{(3)}_{\bar\theta,\,\textrm{MR}}(\vec r) =  
- \frac{\bar g_0 g_A}{F^2_{\pi}}
\vec \tau^{\, (1)} \cdot \vec \tau^{(2)} \, 
\left( \vec\sigma^{(1)} - \vec\sigma^{(2)} \right)\cdot 
\left[\vec \nabla_r \left(2 g_A^2 X(r)-Y(r)\right)\right].
\label{unsub}
\end{equation}
The potential in Eq. \eqref{unsub} is singular, 
and the cutoff dependence it introduces in the evaluation of matrix elements 
and observables is absorbed by the renormalization of 
$\bar C_{2}$ in the short-distance potential \eqref{Vnnnnpion},
\begin{equation}\label{deltaconfig}
V^{(3)}_{\bar\theta, \textrm{SR}}(\vec r) = 
\frac{1}{2} \left[\bar C_{1} 
+\bar C_{2} \vec \tau^{\, (1)} \cdot \vec \tau^{\, (2)}\right]
\,  \left( \vec{\sigma}^{\, (1) }-\vec{\sigma}^{\,  (2) }\right) 
\cdot \left(\vec \nabla_r\, \delta^{(3)}(\vec r\,)\right)  .
\end{equation}

\subsection{Dimension-6 Sources}
\label{rspacedim6}

At leading order, the potential from dimension-6 sources is local, and 
depends only on the relative position $\vec r$.

In the case of the qCEDM,  
since the potential arises from one-pion exchange,
its radial dependence is encoded in the Yukawa function
$U(r)$,
\begin{eqnarray}
V^{(-1)}_{\textrm{qCEDM}}(\vec r) 
&=&  -  \frac{\bar g_0 g_A}{F^2_{\pi}} 
\vec \tau^{\, (1)} \cdot \vec \tau^{\, (2)}  
\left( \vec{\sigma}^{\, (1) }-\vec{\sigma}^{\,  (2) }\right) \cdot 
\left(\vec{\nabla}_r\, U(r)\right) 
\nonumber \\ 
& &
-\frac{\bar g_1 g_A}{2 F^2_{\pi}}
\left[ \left(\tau_3^{(1)} + \tau_3^{(2)} \right) 
\left( \vec\sigma^{(1)}  -  \vec\sigma^{(2)}\right) 
+\left(\tau_3^{(1)} - \tau_3^{(2)} \right)
\left( \vec\sigma^{(1)} +  \vec\sigma^{(2)}\right) \right] 
\cdot \left(\vec  \nabla_r U(r) \right).
\end{eqnarray}
For CI sources, there are additional short-range contributions,
\begin{eqnarray}
V^{(-1)}_{\textrm{TVCI}}(\vec r) 
&=&  -  \frac{\bar g_0 g_A}{F^2_{\pi}} 
\vec \tau^{\, (1)} \cdot \vec \tau^{\, (2)} 
\left( \vec{\sigma}^{\, (1) }-\vec{\sigma}^{\,  (2) }\right) \cdot 
\left(\vec{\nabla}_r\, U(r)\right) 
\nonumber \\ 
& &
-\frac{\bar g_1 g_A}{2 F^2_{\pi}}
\left[ \left(\tau_3^{(1)} + \tau_3^{(2)} \right)
\left( \vec\sigma^{(1)}  -  \vec\sigma^{(2)}\right) 
+\left(\tau_3^{(1)} - \tau_3^{(2)} \right) 
\left( \vec\sigma^{(1)} +  \vec\sigma^{(2)}\right) \right] \cdot 
\left(\vec  \nabla_r U(r) \right)
\nonumber \\
& & + \frac{1}{2} 
\left[\bar C_{1}+\bar C_{2} \vec \tau^{\, (1)} \cdot \vec \tau^{\, (2)}\right]
\left( \vec{\sigma}^{\, (1) }-\vec{\sigma}^{\,  (2) }\right) 
\cdot \left(\vec \nabla_r\, \delta^{(3)}(\vec r\,)\right).
\end{eqnarray}
The potential from the qEDM is purely long-distance: 
in addition to pion exchange of range $\sim 1/m_\pi$,
there is a longer-range component from photon exchange,
\begin{eqnarray}
V^{(2)}_{\textrm{qEDM}}(\vec r) 
&=&  -  \frac{ g_A}{F^2_{\pi}} 
\left(\bar g_0 \vec \tau^{\, (1)} \cdot \vec \tau^{\, (2)} 
+ \bar g_2 \tau^{(1)}_3 \tau^{(2)}_3 \right)
\left( \vec{\sigma}^{\, (1) }-\vec{\sigma}^{\,  (2) }\right) \cdot 
\left(\vec{\nabla}_r\, U(r)\right) 
\nonumber \\ 
& &
-\frac{\bar g_1 g_A}{2 F^2_{\pi}}
\left[ \left(\tau_3^{(1)} + \tau_3^{(2)} \right)
\left( \vec\sigma^{(1)}  -  \vec\sigma^{(2)}\right) 
+\left(\tau_3^{(1)} - \tau_3^{(2)} \right) 
\left( \vec\sigma^{(1)} +  \vec\sigma^{(2)}\right)\right] \cdot 
\left(\vec  \nabla_r U(r) \right)
\nonumber \\
& &+  \frac{e}{2} 
\left[\bar d_0 + \bar d_1 \tau^{(1)}_3 \tau^{(2)}_3  \right]  
\left( \vec{\sigma}^{\, (1) }-\vec{\sigma}^{\,  (2) }\right) \cdot 
\left(\vec{\nabla}_r\, \frac{1}{4\pi r}\right) 
 \nonumber \\ 
&&
+\frac{e}{4}
\left[ (\bar d_0 + \bar d_1)\left(\tau_3^{(1)} + \tau_3^{(2)} \right) 
\left( \vec\sigma^{(1)}  -  \vec\sigma^{(2)}\right) 
+ (\bar d_1 - \bar d_0)\left(\tau_3^{(1)} - \tau_3^{(2)} \right)
\left( \vec\sigma^{(1)} +  \vec\sigma^{(2)}\right) \right] 
\nonumber\\
&&\qquad 
\cdot \left(\vec  \nabla_r \frac{1}{4\pi r} \right).
\end{eqnarray}

The various sources thus involve different spin, isospin, and radial
dependences. We discuss some implications in the next section.

\section{Discussion}
\label{discussion}

Traditionally  the study on $T$ violation  in nuclear physics has been carried 
out by considering the most general pion-nucleon 
{\it non}-derivative couplings 
in a phenomenological TV Lagrangian \cite{Barton}, 
which we write in our notation as
\begin{equation}
\mathcal L_{\slashT, \textrm{non}} = 
-\frac{\bar g_0}{F_{\pi}} \bar N \vec \tau \cdot \vec \pi N 
- \frac{\bar g_1}{F_{\pi}} \pi_3 \bar N N 
- \frac{\bar g_2}{F_{\pi}} \pi_3 \bar N \tau_3 N, 
\label{pheno}
\end{equation}
and by inferring from it
the TV two-nucleon potential \cite{Herczeg},
\begin{eqnarray}
V_{\slashT, \textrm{non}} (\vec r)  &= & 
- \frac{g_A}{F^2_{\pi}} 
\left\{\left[\left(\bar g_0 +\frac{\bar g_2}{3}\right) 
\vec \tau^{\,(1)} \cdot \vec \tau^{\,(2)} 
+ \frac{\bar g_1}{2} \left(\tau^{(1)}_3 + \tau^{(2)}_3 \right)
+\frac{\bar g_2}{3}
\left(3\tau^{(1)}_3 \,\tau^{(2)}_3 -\vec \tau^{\,(1)} \cdot\vec \tau^{\,(2)}
\right) \right] 
\right.
\nonumber\\
&& \left.  \qquad\quad 
\left(\vec \sigma^{(1)} - \vec \sigma^{(2)} \right)
+ \frac{\bar g_1}{2} \left(\tau^{(1)}_3 - \tau^{(2)}_3 \right) 
\left(\vec \sigma^{(1)} + \vec \sigma^{(2)} \right) 
\right\} \cdot \left( \vec \nabla U(r) \right), 
\end{eqnarray}
with $U(r)$ defined in Eq. \eqref{Yukawa}.
When short-distance contributions are included in the model, the most general  
TV two-nucleon local potential 
with the {\it minimum} number of derivatives assumes the 
form \cite{H66}
\begin{eqnarray}
V_{\slashT, \textrm{min}} (\vec r) & =&
\left(\vec \sigma^{(1)} - \vec \sigma^{(2)} \right) \cdot \vec\nabla
\left[
\mathcal U_0(r)+ 
\vec \tau^{\,(1)} \cdot \vec \tau^{\,(2)} \mathcal V_0(r)
+\frac{1}{2}\left(\tau^{(1)}_3 + \tau^{(2)}_3 \right) \mathcal U_1(r)
\right.
\nonumber\\
&&\left.
+ \left( 3 \tau^{(1)}_3 \,\tau^{(2)}_3 -\vec\tau^{\,(1)} \cdot\vec\tau^{\,(2)} 
  \right)\mathcal V_2(r)
\right]
+
\frac{1}{2}\left(\tau^{(1)}_3 - \tau^{(2)}_3 \right) 
\left(\vec \sigma^{(1)} + \vec \sigma^{(2)} \right) \cdot 
\vec \nabla \mathcal V_1^{}(r)
\label{sumupnd}
\end{eqnarray}
in terms of five radial functions $\mathcal U_{0,1}(r)$ and
$\mathcal V_{0,1,2}(r)$.
These functions are assumed to originate in one-boson exchange 
\cite{GHM93,TH94,Tim+04}:
pion exchange is taken to give long-range contributions 
to $\mathcal V_0$, $(\mathcal V_1 + \mathcal U_1)/2$ and 
$\mathcal V_2$, 
while eta, rho and omega mesons give shorter-range contributions to the same
quantities, and to 
$\mathcal U_0$ and $\mathcal V_1 - \mathcal U_1$. 
The five momentum-independent 
potentials in Eq. \eqref{sumupnd} are 
treated on the same footing, and they provide enough information to describe 
the five $S$--$P$ mixing amplitudes discussed in Sect. \ref{pionless}.

For TV stemming from the QCD $\bar \theta$ term, 
the proper account of chiral symmetry radically changes the picture. 
As noticed in Ref. \cite{Crew+79}, at leading order the $\bar \theta$ term 
generates only the isoscalar pion-nucleon $T$-odd coupling $\bar g_0$, 
and thus contributes at tree level only to the $I=0$ potential 
\cite{HaxHenley83}.
A coupling of $\bar g_1$ form appears two orders down in the 
ChPT expansion,
and the one of $\bar g_2$ form is even more suppressed \cite{emanuele}. 
To evaluate the effects of the $\bar \theta$ term on observables which, 
like the deuteron EDM, are mostly sensitive to the $I=1$ components, 
it is necessary to consider the TV two-nucleon potential to 
next-to-next-to-leading order in ChPT. 
As described in Sect. \ref{qspace}, this implies the  consideration 
not only of the non-derivative pion-nucleon TV couplings, 
but also of subleading TV  derivative couplings, 
of power-suppressed TC interactions (with particular care for isospin-breaking 
operators, which contribute to the $I=1$ and $I=2$ potentials), 
and of one-loop and short-range contributions to the two-nucleon potential.
When all these elements are considered, 
the potential has a much richer structure than Eq. \eqref{sumupnd}:
{\it (i)} a hierarchy emerges between the five spin-isospin structures 
already present in 
Eq. \eqref{sumupnd}, and 
{\it (ii)} momentum-dependent potentials appear, 
with the same importance as most of the momentum-independent ones.

We first analyze the implications of our $\bar \theta$ results to  
$V^{}_{\slashT, \textrm{min}} (\vec r)$.
Using the chiral index $\nu$, as defined in Eq. \eqref{nudelta}, 
to keep track of the size of different pieces, 
the $\bar \theta$ contributions to Eq. \eqref{sumupnd} are
\begin{eqnarray} 
 \mathcal V^{(1)}_0 (r) & = & 
-  \frac{g_A \bar g_0}{F^2_{\pi}} \, U(r),
\label{summ1}\\
 \mathcal V^{(3)}_0 (r) & = &
- \frac{g_A \bar g_0}{F^2_{\pi}}  
  \left[2g_A^2 X(r)-Y(r)
  +\frac{ (m_n - m_p)^2}{3m_{\pi}} r\, U(r) \right]
+ \bar C_{2}\frac{\delta (r)}{8\pi r^2},
\label{summ2}\\
 \mathcal U^{(3)}_0(r) & = & 
+ \bar C_{1} \frac{\delta (r)}{8\pi r^2}, 
\label{summ3}\\
 \mathcal V^{(3)}_1(r) & = &  
- \frac{g_A \bar g_0}{F^2_{\pi}}
\left(\frac{\bar g_1}{\bar g_0} + \frac{\beta_1}{2 g_A} \right) U(r),
\label{summ4}\\
 \mathcal U^{(3)}_1(r) & = &  
- \frac{g_A \bar g_0}{F^2_{\pi}}
\left(\frac{ \bar g_1}{\bar g_0} - \frac{\beta_1}{2 g_A} \right) U(r),
\label{summ5}\\
 \mathcal V^{(2+3)}_2(r) & = &   
+\frac{g_A \bar g_0}{3 F^2_{\pi}}  
\left[\frac{(m_n - m_p)^2}{2 m_{\pi}} r U(r) +  W(r) \right],
\label{summ6}
  \end{eqnarray}
where $U(r)$, $W(r)$, $X(r)$, and $Y(r)$  are defined in 
Eqs. \eqref{Yukawa}, \eqref{Yukawadiff}, \eqref{Xfunk}, and \eqref{Yfunk},
respectively. 
In Eq. \eqref{summ1} 
the use of the definition \eqref{Yukawa} to express 
$\mathcal V^{(1)}_0$ in terms of the physical pion masses introduces 
subleading corrections in the $\nu = 1$ term,
which strictly speaking would use a function 
$U(r)$ that only depends on a common pion mass, 
say the neutral one,
$U_0(r)= \exp(-m_{\pi^0} r)/4\pi r$.
In Eqs. \eqref{summ2}--\eqref{summ6} we can neglect the pion mass difference 
in $U(r)$, and use $U_0(r)$,
the error thus introduced being 
at higher orders 
in the ChPT power counting.
Similarly, in Eq. \eqref{summ1} the use of 
$g_A$ and $\bar g_0$ with their $m_\pi^2$ corrections included
accounts for some $\nu = 3$ corrections, while 
whether or not such $m_\pi^2$ corrections are included 
in Eqs. \eqref{summ2}--\eqref{summ6}
is beyond the order we consider.
In Eqs. \eqref{summ2} and \eqref{summ6} we replaced $\delta m_N$, 
the quark-mass-difference contribution to the nucleon mass splitting, 
with the physical value of the nucleon mass splitting itself, $m_n - m_p$, 
the difference again being a higher-order contribution in ChPT.

\begin{figure}[tb]
\begin{center}
\epsfxsize=12cm \epsffile{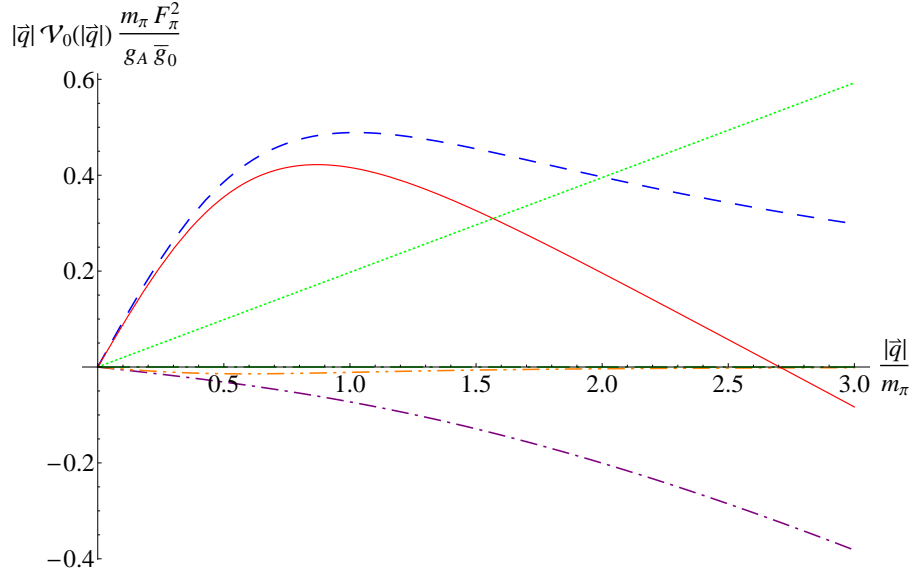}
\end{center}
\caption{Components of the 
$\bar \theta$-term two-nucleon potential $|\vec q\,| \mathcal V_0$, 
in units of $g_A \bar g_0/F_{\pi}^2m_{\pi}$, 
as a function of the transferred momentum $|\vec q\,|$, in units of $m_{\pi}$. 
The (blue) dashed line denotes the leading-order OPE contribution
with physical pion masses;
the (orange) dashed-double-dotted line shows the effect of the 
pion mass difference on the leading OPE contribution;
the (dark green) long-dashed-dotted line accounts for the even smaller
effect of the nucleon mass splitting;
the (purple) dashed-dotted line is the non-analytic TPE contribution; 
and the (green) dotted line presents an 
estimate of the short-range component of the potential.
The (red) solid line  
is the sum of all 
contributions up to next-to-next-to-leading order,
except for the short-range component.
}
\label{plotmomentum}
\end{figure}

\begin{figure}[tb]
\begin{center}
\includegraphics[width=12cm]{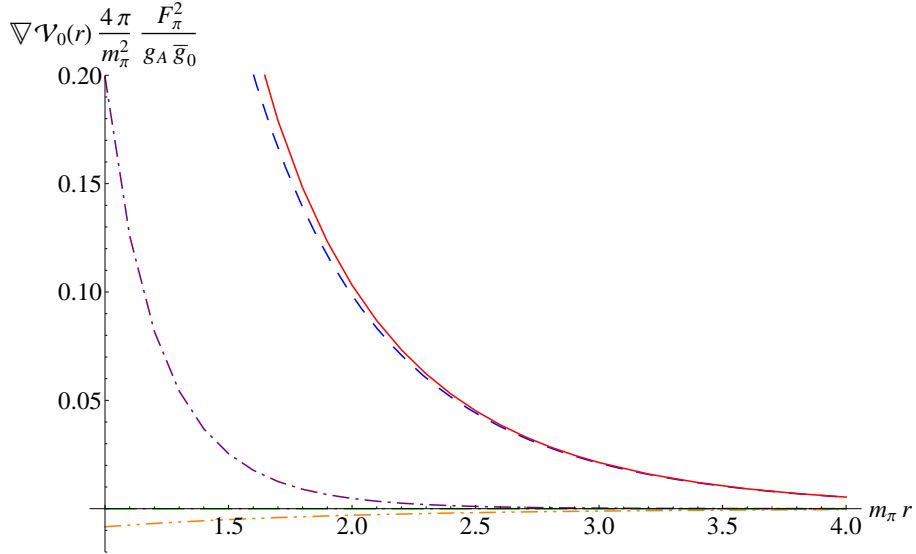}
\end{center}
\caption{Components of the 
$\bar \theta$-term two-nucleon potential $\nabla \mathcal V_0$ 
in units of $g_A \bar g_0 m^2_{\pi}/4\pi  F^2_{\pi}$,
as functions of the distance between the two nucleons $r = |\vec r|$, 
in units of $1/m_{\pi}$. 
Curves as in Fig. \ref{plotmomentum}, except that 
the short-range component of the potential is not shown.
}
\label{plotconfig}
\end{figure}

As one can see, at order $\nu =3$ in ChPT all the possible spin-isospin 
structures considered in Refs. \cite{H66,GHM93,TH94,Tim+04} appear.
The dominant component is the isoscalar $\mathcal V_0$ \cite{HaxHenley83}.
In Figs. \ref{plotmomentum} and \ref{plotconfig} we plot, respectively, 
the momentum-space and configuration-space expressions 
for $\nabla \mathcal V_0$. 
The dashed line 
represents the leading-order $\nabla \mathcal V^{(1)}_0$,
Eqs. \eqref{pionmass4} and \eqref{onepionconfig}
with the use of the definition \eqref{Yukawa} for $U(r)$ to express 
$\mathcal V^{(1)}_0$ in terms of the physical pion masses.
The dashed-double-dotted line illustrates the effect of 
the difference 
between the 
leading OPE potential computed with the function $U(r)$ 
and with $U_0(r)$.
Other isospin-breaking corrections, which come from the nucleon mass 
splitting in $\mathcal V^{(3)}_0$,
Eqs. \eqref{pionmass7} and \eqref{pimassconfig},
are very small, as indicated by the long-dashed-dotted line
barely distinguishable from the $x$-axis.
At next-to-next-to-leading order, 
$\mathcal V_0$ also exhibits a medium-range component originating 
in TPE diagrams and a short-range component.
The dashed-dotted line depicts the non-analytic piece of the TPE diagrams,
Eqs. \eqref{TPEP} 
and \eqref{unsub}.
We estimate the short-range potential
by assuming the coefficient 
$\bar C_{2}$ in Eq. \eqref{C2ren} 
to be dominated by the $\ln \mu^2/m^2_{\pi}$ term with $\mu=m_N$. 
The rationale is that there is no obvious reason to expect that 
such a contribution, non-analytic in $m_{\pi}$, should get 
exactly canceled by $m_\pi$-independent short-distance contributions.
However, the sign cannot be guessed reliably and our choice is purely
arbitrary, for illustration only.
Equation \eqref{Vnnnnpion}
gives rise to the straight dotted line in Fig. \ref{plotmomentum}
but Eq. \eqref{deltaconfig} does not appear in Fig. \ref{plotconfig}
since it is concentrated at $r=0$.
The solid line in 
both figures
is the sum of the long and medium range
contributions to $\nabla \mathcal V_0$.

{}From Fig. \ref{plotmomentum}, we can appreciate that, as expected from 
the ChPT power counting, the medium and short-range corrections to the 
TV potential have comparable size in the momentum range we are considering, 
and for momenta $q \simge m_{\pi}$ they noticeably affect the leading order. 
At momenta of order $300$--$400$ MeV the medium and short-range contributions 
have roughly the same size as the leading potential. 
In this region, degrees of freedom which we have not explicitly included 
in the EFT, like the $\Delta$ isobar,  become relevant, and the 
convergence of the perturbative expansion can be improved by 
extending the EFT to incorporate them. 
Isospin-breaking, long-range corrections, although of formally the same
order as TPE and contact terms, are much smaller, at least in part
because of factors of $\varepsilon$, except at very
small momenta where their longer range compensates.
In Fig. \ref{plotconfig} we 
focus our attention on the long-distance region,
$r \ge 1/m_{\pi}$. 
At distances of up to $ r \lesssim  2/m_{\pi}$ TPE is still 
the dominant correction to the potential, but it is overcome
at longer distances, 
$r \simge 2/m_{\pi}$, by the long-range effects of pion mass splitting in OPE.

\begin{figure}[tb]
\begin{center}
\includegraphics[width=11cm]{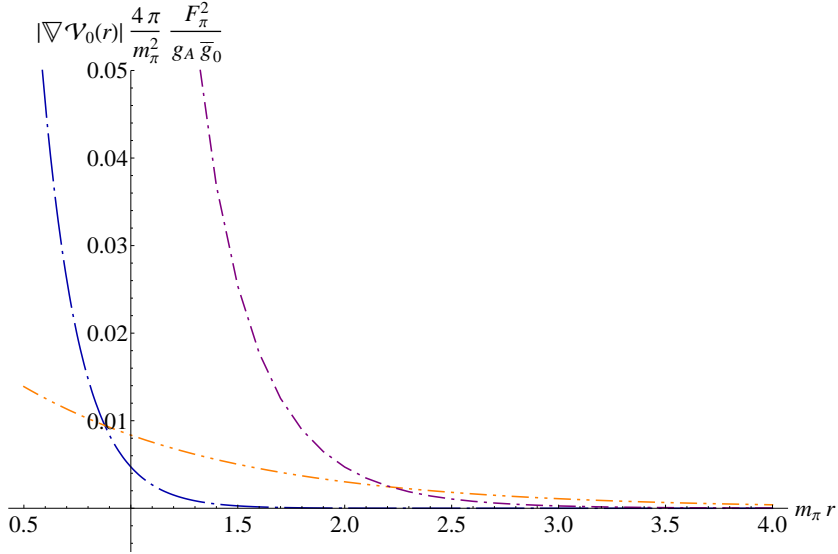}
\end{center}
\caption{Comparison between one-rho-exchange and EFT contributions 
to the magnitude of the 
$\bar\theta$ potential $|\nabla \mathcal V_0|$ 
in units of $g_A \bar g_0 m^2_{\pi} /4\pi  F^2_{\pi}$,
as functions of the distance $r$, 
in units of $1/m_{\pi}$. 
The rho-exchange contribution is depicted as a 
(blue) long-dashed-dotted line, 
while TPE  
and pion mass splitting in OPE 
are as in Fig. \ref{plotmomentum}.}
\label{plotrho}
\end{figure}

It is instructive to compare our results for $\mathcal V_0$ 
to the corresponding TV potential obtained in a one-boson-exchange model.
In such a 
model, $T$ violation in the coupling of a rho meson 
to the nucleon generates  corrections to 
 $\mathcal V_0$ of the form \cite{TH94,Tim+04}
\begin{equation}
\mathcal V^{(\rho)}_{0}(r) = 
- \frac{g_A \bar g_0}{F^2_{\pi}} \frac{g_{\rho N N}}{g_{\pi N N}} 
\frac{\bar g_{0\rho} F_{\pi}}{\bar g_0} \frac{e^{- m_{\rho} r}}{4\pi r},
\label{rhopot}
\end{equation}
where $g_{\rho N N}$ is the TC rho-nucleon vector coupling 
and $\bar g_{0\rho}$ is an isoscalar, TV,
one-derivative rho-nucleon coupling, 
defined, for example, in Ref. \cite{Tim+04}.
In the limit where the rho mass is large, 
$m_{\rho} \rightarrow \infty$, 
$\mathcal V^{(\rho)}_0(r)$ approximates a delta function and 
the effect of TV rho-exchange amounts to a contribution to 
$\bar C_{2}$ of the form
\begin{equation}
\bar C^{(\rho)}_{2} = - 2 \frac{g_A\bar g_0}{F^2_{\pi}} 
\frac{g_{\rho NN}}{g_{\pi NN}} \frac{\bar g_{0\rho} F_{\pi}}{\bar g_0} 
\frac{1}{m^2_{\rho}}.
\label{rhodelta}
\end{equation}
Since $m_{\rho} \sim M_{QCD}$ and there is no reason
for $\bar g_{0\rho} F_{\pi}/\bar g_0$ to be particularly big or small, 
the size of rho-meson contribution is 
comparable to the power-counting expectation,
$\bar C_{2} = \mathcal O(\bar\theta m^2_{\pi}/F_\pi^2 M^3_{QCD})$,
with some suppression coming from the numerical smallness of the 
TC rho-nucleon vector coupling compared to the pion-nucleon coupling. 
Assuming 
the TV pion-nucleon and rho-nucleon couplings to have the 
same strength, $\bar g_{0\rho} F_{\pi}/\bar g_0 = 1$,
and using 
for the rho-nucleon vector coupling 
the value determined in modern high-precision two-nucleon potentials, 
$g_{\rho NN} = 3.2$ 
\cite{rhoNNNij},
in Fig. \ref{plotrho} we compare the rho-meson contribution to 
$\nabla \mathcal V_0$ to the pion-mass-splitting and TPE medium-range 
corrections
discussed above.
For $r \simge 1/m_{\pi}$, the contribution of the rho meson is 
numerically small 
compared to both pion-mass-splitting and TPE corrections. 
At shorter ranges, $r \lesssim 1/m_{\pi}$, 
rho exchange overcomes the effect of pion mass splitting, 
but it always remains smaller than TPE.  
Of course we can make one-rho exchange more important by
jacking up $\bar g_{0\rho} F_{\pi}/\bar g_0$,
but we cannot compensate for the different ranges
of the two contributions, $m_\rho$ {\it versus} $2m_\pi$.
We see little justification for 
neglecting TPE in the 
$\bar\theta$ potential.

We now turn to the other spin-isospin structures in Eq. \eqref{sumupnd},
which in EFT are all suppressed by one power of $Q^2/M^2_{QCD}$
with respect to the leading OPE TV potential.
The function $\mathcal U^{(3)}_0$ only receives contributions from 
short-range physics. Again, in a one-boson-exchange scenario, 
contributions of exactly the size of $\bar C_{1}$ come from 
eta and omega exchanges \cite{GHM93,TH94,Tim+04},
\begin{equation}
\bar C^{(\eta,\omega)}_{1} = 
2 \frac{g_A \bar g_0}{F^2_{\pi}} 
\left(\frac{g_{\eta N N}}{g_{\pi N N}} 
\frac{\bar g_{0\eta} F_{\pi}}{\bar g_0} \frac{1}{m^2_{\eta}} 
-\frac{g_{\omega N N}}{ g_{\pi N N}} 
\frac{\bar g_{0\omega} F_{\pi}}{\bar g_0}\frac{1}{m^2_{\omega}} \right),
\label{etaomega}
\end{equation}
where $m_{\eta}$ ($m_{\omega}$) is the eta (omega) mass,
$g_{\eta N N}$ ($g_{\omega N N}$) is the 
TC eta-nucleon axial (omega-nucleon vector) coupling,
and $\bar g_{0\eta}$ ($\bar g_{0\omega}$) is an 
isoscalar, TV no-derivative eta-nucleon (one-derivative omega-nucleon) 
coupling.
The eta- and 
omega-meson contributions are  
comparable to the power-counting 
expectation 
$\bar C_1 = \mathcal O(\bar\theta m^2_{\pi}/F^2_{\pi} M_{QCD}^3)$. 
For the eta meson, the enhancement due to the relatively light mass is 
offset by the smallness of eta-nucleon TC coupling, $g_{\eta N N} = 2.24$ 
\cite{Tiator:1994et}. The ratio $g_{\omega N N}/g_{\pi N N}$ is instead 
close to one \cite{rhoNNNij}, and, therefore, we have no reason to expect 
the omega-meson contribution to $\bar C_1$ to differ much from the
power-counting estimate.

In contrast, $\mathcal V_1^{(3)}$, $\mathcal U_1^{(3)}$, 
and $\mathcal V_2^{(2+3)}$ sprout entirely from OPE. 
The TV, isospin-breaking coupling $\bar g_1$ 
contributes equally to $\mathcal V_1$ and  $\mathcal U_1$,
as expected \cite{Herczeg}
from the identification at the Lagrangian level, {\it cf.} Eqs. \eqref{LagT2}
and \eqref{pheno}.
However, we expect a comparable
long-range piece 
in $\,\mathcal U_1 - \mathcal V_1$, 
which stems from the combination of the isospin-violating vertex 
$\beta_1$ and the TV vertex $\bar g_0$. 
As discussed in Ref. \cite{emanuele}, strong-dynamics contributions 
to the coefficients of these $I=1$ potentials are in principle determined 
by measurement of TC, isospin-breaking observables. 
For example $\bar g_1/ \bar g_0$  could be extracted from a detailed 
analysis of isospin-breaking effects in pion-nucleon scattering. 
At present, however, even the very sophisticated,  state-of-the art  analysis  
of Ref. \cite{Hoferichter:2009gn} stops one order shy of the accuracy required 
for such extraction.
Similarly, the ratio $\beta_1/g_A$ affects isospin violation in 
nucleon-nucleon scattering, but at present phase-shift analyses 
of two-nucleon data can only provide a bound on  $\beta_1$, 
which is in accordance with the power-counting expectation 
\cite{isoviolphen,isoviolOPE}.
In the absence of better constraints on the parameters
in Eqs. \eqref{summ4} and \eqref{summ5}, 
the ratios $\mathcal V^{(3)}_1/\mathcal V^{(1)}_0$
and $\mathcal U^{(3)}_1/\mathcal V^{(1)}_0$ can only be estimated by 
power counting, as
$\mathcal O(\varepsilon m^2_{\pi}/M^2_{QCD}) \sim 1 \%$.
As for the last component of the phenomenological
potential, $\mathcal V_2^{(2+3)}$
originates entirely from the isospin-violating corrections to the pion 
and nucleon masses, and it is also relatively small.
Note that to this order this component has nothing to do
with the coupling ${\bar g}_2$ of a phenomenological Lagrangian:
because of the isoscalar character of the $\bar\theta$ term, 
Eq. \eqref{LtrvQCD},
${\bar g}_2$ arises in EFT only at higher order.

In one-boson-exchange models the $I=1,2$ potentials
are assumed to arise from
pion, eta, rho, and omega isovector and tensor TV couplings
to the nucleon \cite{GHM93,TH94,Tim+04}. 
In the ChPT power counting, short-range contributions to these
potentials are suppressed with respect to the long-range pieces,
again because of the isoscalar character of the $\bar\theta$ term. 
(Of course, 
because of the factors $\varepsilon$ in the long-range contributions
of this order, short-range terms might not be entirely negligible.)
This is consistent with the argument that the dominant 
meson-exchange contributions are from the pion and the eta \cite{GHM93}.

There are, therefore, a few points of contact between the local
part of our $\nu\le3$ potential and the phenomenological potential
$V^{}_{\slashT, \textrm{min}} (\vec r)$ \eqref{sumupnd}.
However, as we have seen in Sects. \ref{qspace} and \ref{rspace},
at this order EFT yields also momentum-dependent interactions,
which in coordinate space appear as non-local potentials
and corrections that account for 
CM motion
of the nucleon pair.
They can be found in the relativistic and isospin-breaking corrections
to OPE in Eqs. \eqref{onepionconfig}, \eqref{relcorr1config}, 
and \eqref{deltamnconfig}.

At $\nu =3$, the 
$\bar\theta$ two-nucleon potential contains 
in the 
CM frame, $\vec P = 0$,
four spin-isospin structures that are
momentum-dependent,
\begin{eqnarray}
V_{\slashT, \mathrm{more}}(\vec r, \vec p_r) & = &  
\frac{g_A \bar g_0}{4m_N^2 F^2_{\pi}} \vec \tau^{\,(1)} \cdot \vec \tau^{\,(2)}
\left[\left(\vec \sigma^{(1)} - \vec \sigma^{(2)} \right) \cdot  
\left\{p^i_r,\left\{p^i_r , \vec\nabla_r U(r) \right\}\right\}
\right.  
\nonumber \\
&  & \left. 
- \frac{2}{3} \left(\vec \nabla^2_r U(r)\right) 
\left(\vec \sigma^{(1)} \times \vec \sigma^{(2)} \right) \cdot \vec p_r   
+  \left(\vec \sigma^{(1)} \times \vec \sigma^{(2)} \right)^m 
\left(\nabla^{m}_r\nabla^{l}_r U(r) 
-\frac{1}{3}\delta^{lm}\vec\nabla^2_r U(r) \right) p_r^l  
\right]
\nonumber\\
&  & +
\frac{g_A\bar g_0 \, \delta m_N }{2 m_NF^2_{\pi}} 
\left(\vec \tau^{\, (1)} \times \vec\tau^{\, (2)}\right)_3 
\left(\vec\sigma^{(1)} + \vec\sigma^{(2)} \right) \cdot 
\left\{\vec p_r, U(r)\right\}, 
\label{summp4} 
\end{eqnarray}
where $\vec p_r= - i \vec \nabla_r$ 
denotes the quantum-mechanical relative momentum operator.
 
The 
structure of the momentum-dependent TV potentials was considered 
previously in Ref. \cite{H66}, where all possible Hermitian operators
were constructed, which
violate time-reversal and parity,
and contain up to one power of momentum $\vec p_r$.
The momentum-dependent TV potential was parameterized with eleven unknown 
functions $d_i(r)$, $i =1,2 \ldots, 11$. 
The first  term in Eq. \eqref{summp4} is
quadratic in the momentum operator and was not considered in Ref. \cite{H66}. 
The second and third spin-isospin structures correspond, respectively, to the 
isoscalar functions $d_2(r)$ and $d_6(r)$. 
For TV from the QCD $\bar \theta$ term, 
these two functions are 
therefore  dominated by pion-exchange, and their coefficients are fixed 
by Lorentz invariance and do not contain any new TV parameter.
Isospin-breaking effects in the strong interaction  give rise to  
the last term in Eq. \eqref{summp4}, which is proportional to the 
nucleon mass difference, and it is the first contribution of the 
$\bar\theta$ term  to $d_{10}(r)$. Once again, $d_{10}$ is dominated by OPE 
diagrams, and
the only TV parameter intervening is $\bar g_0$.
We find that, at order $\nu =3$ in ChPT, the other isospin-conserving 
(the isoscalar $d_1$ and $d_5$) and 
isospin-breaking ($d_{3} $, $d_4$, $d_7$, $d_8$, $d_9$ and $d_{11}$) 
functions do not receive contributions from $\bar\theta$. 

\begin{figure}[tb]
\begin{center}
\includegraphics[width=12cm]{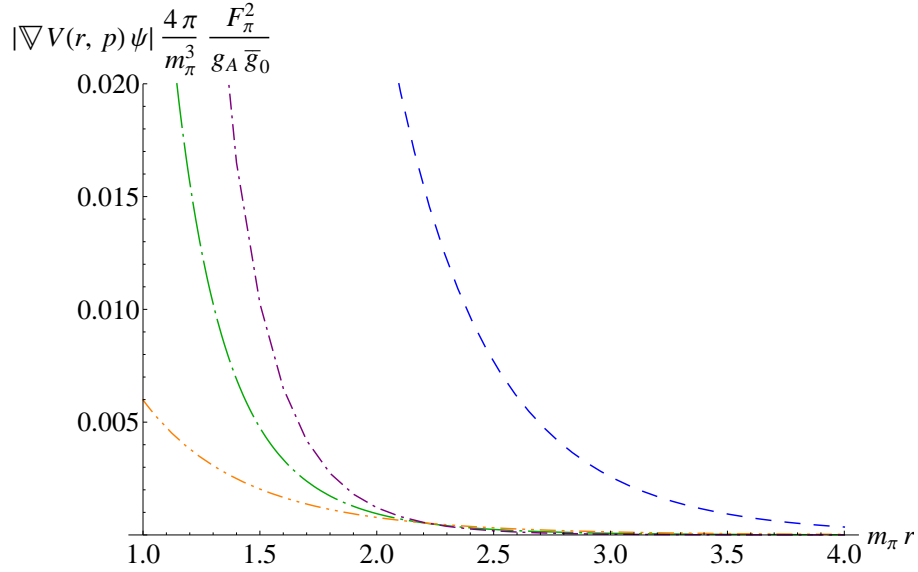}
\end{center}
\caption{Comparison between a relativistic correction to OPE
and local components of the  
$\bar\theta$ two-nucleon 
potential $\nabla \mathcal V_0$ applied to 
an illustrative bound-state
wave function $\psi$, 
in units of $g_A \bar g_0 m^3_{\pi}/4\pi  F^2_{\pi}$,
as functions of the distance $r$, 
in units of $1/m_{\pi}$. 
The  (dark green) long-dashed-dotted line represents the 
term in the
potential 
that is quadratic in momentum.
Other curves are as in Fig. \ref{plotmomentum}.}
\label{plotrel}
\end{figure}

In order to get a sense of the 
importance of the momentum-dependent contributions,
we consider the effect of the relativistic correction in Eq. \eqref{summp4}
that is quadratic in $\vec p_r$. 
In Fig. \ref{plotrel} we compare it (long-dashed-dotted line) to
leading OPE (dashed line), medium-range TPE (dashed-dotted line), 
and pion mass-splitting corrections (double-dotted-dashed line),
all applied to
a simple bound-state wave function with the scale present in the 
$^1S_0$ channel, $a_s = - 23.714$ fm:
\begin{equation}\label{wf}
\psi(r) = \frac{\exp(- r/a_s)}{r}. 
\end{equation}
In this qualitative example the relativistic correction
cannot be neglected with respect to the other $\nu =3$ corrections, 
and we take Fig. \ref{plotrel} as an indication  
that also in actual calculations of TV observables 
it would not be 
advisable to drop the potential in
Eq. \eqref{summp4}, when the 
$\bar\theta$ potential is needed 
to next-to-next-to-leading-order accuracy. 

Finally, to the same order we find 
contributions proportional to the 
CM momentum of the nucleon pair, 
\begin{eqnarray}
V_{\slashT, \mathrm{CM}}(\vec r, \vec p_r, \vec P) & = & 
\frac{g_A\bar g_0}{8 m_N^2 F^2_{\pi}}  \vec\tau^{\,(1)} \cdot \vec\tau^{\,(2)} 
\left\{\left(\vec \sigma^{\,(1)} + \vec \sigma^{\,(2)} \right) \cdot  
\left\{ \vec p_r, \,  \left(\vec \nabla_r U(r)\right)\cdot \vec P  \right\}  
\right.
\nonumber \\ 
&& \left.
+ \varepsilon^{ijk} 
\left(\sigma^{(1)\, i}\,\sigma^{(2)\, l} +\sigma^{(1)\, l}\,\sigma^{(2)\, i}
\right)\, \left( \nabla^l_r \nabla^k_r U(r)\right) \, P^j 
\right.
\nonumber \\ 
& & \left.
+ \left(\vec \sigma^{(1)} - \vec \sigma^{(2)} \right) \cdot
\left[2\left(\vec \nabla_r U(r)\right) \vec P^2 
+ \vec P\,  
\left(\vec \nabla_r U(r)\right) \cdot \vec P
+  \frac{1}{m_{\pi}}(\vec P \cdot \vec\nabla_r)^2
\left(\vec \nabla_r r U(r)\right)
\right]
\right\}
\nonumber \\ 
& & 
+ \frac{g_A \bar g_0\, \delta m_N}{2 m_N F^2_{\pi}}  
\left(\vec \tau^{\, (1)} \times \vec \tau^{\, (2)}\right)_3  
\left(\vec\sigma^{(1)} - \vec\sigma^{(2)} \right) \cdot 
\left[U(r)  \vec P 
- \frac{1}{ m_{\pi}} 
\left( \vec \nabla_r \, \nabla^i_r \, r  U(r) \right) P^i
\right],
\label{summP6}
 \end{eqnarray}
where  $\vec P = - i \vec\nabla_X$.
Although the operators 
in Eq. \eqref{summP6} vanish in the two-nucleon 
CM frame 
and are not important for the study of $T$ violation in nucleon-nucleon 
scattering, 
they  impact observables like TV electromagnetic form factors of the deuteron, 
where the recoil against the photon changes the 
CM momentum of 
the nucleon pair, and they have to be considered in nuclear systems with $A>2$.
An example of the effects of recoil on TC deuteron processes
can be found in Compton scattering 
\cite{comptonQ4}.

At leading order, the dimension-6 sources of TV only contribute to 
$V_{\slashT, \textrm{min}}(\vec r)$, the two-nucleon potential with 
minimal number of derivatives, in Eq. \eqref{sumupnd}.
Combining all these sources,
\begin{eqnarray} 
 \mathcal V^{(\nu_\mathrm{min})}_0 (r) & = & 
- \frac{g_A (3\bar g_{0}+\bar g_{2})}{3F^2_{\pi}} \, U(r)
+ \bar C_{2} \frac{\delta (r)}{8\pi r^2}
+ \frac{e \bar d_1}{24\pi r}, 
\label{eq:6.5.16}\\
 \mathcal U^{(\nu_\mathrm{min})}_0(r) & = & 
+ \bar C_{1} \frac{\delta (r)}{8\pi r^2}
+ \frac{e \bar d_0}{8\pi r}, 
\label{eq:6.5.17}\\
 \mathcal V^{(\nu_\mathrm{min})}_1(r) & = &  
- \frac{g_A \bar g_{1}}{F^2_{\pi}} U(r)
+ \frac{}{}\frac{e (\bar d_1-\bar d_0)}{8\pi r}, 
\label{eq:6.5.18}\\
 \mathcal U^{(\nu_\mathrm{min})}_1(r) & = &  
- \frac{g_A \bar g_{1}}{F^2_{\pi}} U(r)
+ \frac{e (\bar d_0+\bar d_1)}{8\pi r}, 
\label{eq:6.5.19}\\
 \mathcal V^{(\nu_\mathrm{min})}_2(r) & = &  
- \frac{g_A \bar g_{2}}{3F^2_{\pi}} U(r)
+ \frac{e \bar d_1}{24\pi r}, 
\label{eq:X.Y.ZW}
  \end{eqnarray}
where 
\begin{itemize}
\item
for qCEDM, $\nu_\mathrm{min}=-1$ and only the $\bar g_{0,1}$ 
terms apply;

\item
for TVCI sources, $\nu_\mathrm{min}=-1$ and only the 
$\bar g_{0,1}$ and $\bar C_{1,2}$
terms apply; and

\item
for qEDM, $\nu_\mathrm{min}=2$ and only the 
$\bar g_{0,1,2}$ and $\bar d_{0,1}$
terms apply.
\end{itemize}
In the case of the qCEDM, the potential is thus dominated by OPE, 
with shorter- or longer-range contributions expected to be small.
For the gCEDM and the TV FQ operators, short-range effects are also 
leading; as for $\bar\theta$, these can be parametrized
by rho (see Eq. \eqref{rhodelta}), and eta and/or omega exchange
(see Eq. \eqref{etaomega}). 
Only for qEDM all components of $V_{\slashT, \textrm{min}}(\vec r)$
appear in leading order
thanks to both the most general non-derivative pion-nucleon
coupling structure, and to 
the long-distance potential stemming from the nucleon EDM.
The latter cannot be well approximated by heavy-meson exchange.

\section{Conclusions}
\label{conclusion}

The power-counting scheme of EFTs allows us to organize the contributions
to the potential in powers of $M_{QCD}^{-1}$.  
The TV PV potential has an ordering that
has similarities with the TC PV potential \cite{PVNN}:
for all TV sources of dimension up to 6,
the leading potential contains one-pion exchange.
For $T$ violation from the $\bar\theta$ term
and from the quark chromo-EDM, 
the relative importance of two-pion exchange and shorter-range interactions
follows the TC PV case closely. 
However, the situation is different for 
the quark EDM, the gluon chromo-EDM
and TV four-quark operators, where contact interactions or one-photon exchange 
are relatively more important.

For $T$ violation from the $\bar\theta$ term, at leading order, 
${\cal O}(Q/M_{QCD})$, 
we find only the well-known
OPE from the $I=0$ pion-nucleon TV coupling \cite{HaxHenley83}.
The OPE from the $I=1$ pion-nucleon TV coupling
is suppressed by two orders in the expansion parameter and is of
${\cal O}(Q^3/M_{QCD}^3)$.
Since the $I=0$ OPE is suppressed in nuclei,
higher orders in the potential could be important.
We have thus also examined the
corrections in the next two orders, 
which are up to ${\cal O}(Q^2/M_{QCD}^2)$ relative to leading.
We have found that the potential is purely two-body, and:

\begin{itemize}

\item
At the longest, one-pion range, there are more general vertex
corrections than usually assumed. We employed 
the results
of Ref. \cite{emanuele} where the TV pion-nucleon vertex was examined
to this order.
In addition to the qualitatively different 
$I=1$ pion-nucleon TV coupling, there are 
corrections to the local potential stemming from isospin breaking
in the pion and nucleon masses, and in the TC  pion-nucleon coupling.
There are also recoil ($\propto 1/m_N$) and relativistic ($\propto 1/m_N^2$)
corrections to leading OPE, 
which make the potential non-local and dependent on the total momentum
of the nucleon pair. 

\item
At 
this order, 
we find, additionally, two-pion exchange from the $I=0$ TV coupling. 
The non-analytic, medium-range part of the TPE potential 
is independent of the choice of fields and regulators.
Like the leading OPE potential, this part of the TPE potential
has as only (so-far) unknown quantity the $I=0$ TV pion-nucleon coupling.
Under reasonable assumptions about the strengths of TV couplings,
this potential is stronger, and has a different radial dependence,
than phenomenological one-meson-exchange potentials.
The main effect of TPE is to modify the potential
in the same channels as the leading OPE potential.

\item
The short-range part of the TPE potential, on the other hand, cannot
be separated from contact interactions,
the most general form of which we also write at this order.
They are two of the terms given in the literature \cite{Tim+04}.
In the context of a theory without pions,
this implies a contribution to only two of five possible
$S$--$P$ transitions.
When pions are included explicitly,
the short-range terms are expected to be of the same size as TPE,
and thus their strengths depend on the renormalization
scale. They subsume short-range dynamics that includes
the effects of heavier mesons, but whether such effects are
sufficient to saturate them is unknown.

\end{itemize}
The structure of the resulting potential is therefore significantly
different than the phenomenological potential used in the literature,
due to the specific way in which the $\bar \theta$ term breaks 
chiral symmetry.
If consideration of short-range dynamics or $I=1$ OPE
is necessary, one should also include the TPE potential 
and OPE corrections calculated here.

For the dimension-6 sources, the $I=1$ pion-nucleon coupling
appears already in leading order, 
${\cal O}(M_{QCD}/Q)$ for qCEDM and chiral-invariant sources
(gCEDM and TV FQ) and ${\cal O}(Q^2/M_{QCD}^2)$ for qEDM.
The structure of the TV potential is thus different
from that of $\bar \theta$, and depends on the source:
 
\begin{itemize}
\item For qCEDM, it contains only OPE with the well-known $I=0,1$
non-derivative pion-nucleon couplings.

\item For CI sources, there are, additionally, two 
contact interactions of the same type as found at next-to-next-to-leading order for $\bar\theta$.

\item For qEDM, OPE from the three ($I=0,1,2$)
non-derivative pion-nucleon couplings is accompanied by
one-photon exchange from the nucleon EDM.
\end{itemize}

Therefore, for all sources we have considered ($\bar\theta$,
qEDM, qCEDM, gCEDM, and TV FQ), the TV nuclear potential
presents more structure (spin, isospin,
and/or distance profile) than assumed in the usual
phenomenological approach. 
EFT offers a framework where 
the calculation of nuclear TV observables 
can be carried out in a model independent way
and characterize the low-energy manifestations of
possible TV sources.

\vspace{0.5cm}
\noindent
{\bf Acknowledgments.}
We thank  R. Timmermans for useful discussions.
UvK acknowledges the hospitality of the Kavli Institute for Theoretical
Physics China and of the Kernfysisch Versneller Instituut
at Rijksuniversiteit Groningen during the writing of this paper.
This research was supported in part by 
an APS Forum of International Physics travel grant (CMM, UvK),
by FAPERGS-Brazil under contract PROADE 2 02/1266.6 (CMM),
by the Brazilian CNPq under contract 474123/2009 (CMM),
by the Dutch Stichting voor Fundamenteel Onderzoek der Materie 
under programme 104 (JdV),
and by the US Department of Energy under grants DE-FG02-04ER41338 (EM, UvK)
and DE-FG02-06ER41449 (EM).
\appendix

\section*{Appendix: Fourier Transformation to Configuration Space}
\label{appendixdimensFt}

In general a potential obtained in EFT depends not only
on the transferred momentum $\vec q$ but also on $\vec K$,
and the 
CM momentum $\vec P$, $V(\vec q, \vec K, \vec P)$.
The Fourier transform of 
such a potential 
is defined as
\begin{equation}
V(\vec r, \vec r^{\,\prime}, \vec X, \vec X^{\,\prime}) = 
\int \frac{d^3 K}{(2\pi)^3}
\int \frac{d^3 P}{(2\pi)^3} 
\int \frac{d^3 q}{(2\pi)^3} 
e^{-i \vec P \cdot (\vec X - \vec X^{\,\prime})}  
e^{-i \vec K \cdot (\vec r - \vec r^{\,\prime})} 
e^{-\frac{i}{2} \vec q \cdot (\vec r + \vec r^{\, \prime})} 
V(\vec q, \vec K, \vec P),
\label{ftmom}
\end{equation}
where, if $\vec x_1$ and $\vec x_2$ are the positions of the incoming nucleons 
and $\vec x_1^{\,\prime}$ and $\vec x_2^{\,\prime}$ the positions of the 
outgoing nucleons, the relative coordinates are 
$\vec r = \vec x_1 - \vec x_2$ and 
$\vec r^{\,\prime} = \vec x^{\,\prime}_1 - \vec x^{\,\prime}_2$, 
while the 
CM position of the incoming and outgoing pairs are 
$2 \vec X = \vec x_1 + \vec x_2$ and 
$2 \vec X^{\prime} = \vec x^{\,\prime}_1 + \vec x^{\,\prime}_2$.
The potential in Eq. \eqref{ftmom}  has to be used in a two-nucleon 
Schr\"odinger equation of the form
\begin{equation}\label{schro}
i \frac{\partial}{\partial t} \psi(\vec r^{\,\prime}, \vec X^{\,\prime}) = 
-\left(\frac{\vec \nabla^{\,2}_{X^{\prime}}}{4 m_N} 
+ \frac{\vec \nabla^2_{r^{\prime}}}{m_N} \right) 
\psi(\vec r^{\,\prime}, \vec X^{\,\prime}) 
+ \int d^3 \vec r \int d^3 \vec X \, 
V(\vec r, \vec r^{\,\prime}, \vec X, \vec X^{\,\prime})\, \psi(\vec r, \vec X).
\end{equation}
For potentials that, like the ones in Sect. \ref{qspace}, are polynomials 
in $\vec K$ and $\vec P$,
\begin{equation}
V(\vec q, \vec K, \vec P) \propto \vec K^m \vec P^n f(\vec q),
\label{poly}
\end{equation}
 $V(\vec r, \vec r^{\,\prime}, \vec X, \vec X^{\,\prime})$ assumes the form
\begin{equation}
V(\vec r, \vec r^{\,\prime}, \vec X, \vec X^{\,\prime}) \propto  
\left(\vec\nabla^n_X \delta^{(3)}(\vec X - \vec X^{\prime} ) \right)  \, 
\left( \vec \nabla^{m}_r  \delta^{(3)}(\vec r - \vec r^{\,\prime}) \right) 
f\left(\frac{\vec r + \vec r^{\,\prime}}{2}\right),
\label{schematic}
\end{equation}
where
\begin{equation}
f(\vec r\,) =
\int \frac{d^{3}q}{\left( 2\pi \right)^{3}}e^{- i\vec{q}\cdot \vec{r}} 
\; f(\vec q \,).  
\label{TransfFourier0}
\end{equation}
Plugging Eq. \eqref{schematic} in Eq. \eqref{schro}, and integrating by parts, 
the derivatives acting on the delta functions can be turned into 
derivatives acting on $f$ and on the wave function $\psi(\vec r, \vec X)$. 
The integrals in Eq. \eqref{schro} then become trivial, and the 
Schr\"odinger equation assumes the form
\begin{equation}
i \frac{\partial}{\partial t} \psi(\vec r^{\,\prime}, \vec X^{\,\prime}) = 
-\left(\frac{\vec \nabla^{\,2}_{X^{\prime}}}{4 m_N} 
+ \frac{\vec \nabla^2_{r^{\prime}}}{m_N} \right) 
\psi(\vec r^{\,\prime}, \vec X^{\,\prime}) 
+  V(\vec r^{\, \prime}, \vec \nabla_{r^{\prime}}, \vec\nabla_{X^{\prime}}) \,
\psi(\vec r^{\, \prime}, \vec X^{\, \prime}),
\label{schro2}
\end{equation}
where the two potentials in Eqs. \eqref{schro} and \eqref{schro2} are 
related by integrations by parts. For a potential of the form \eqref{poly}, 
schematically we would have
\begin{equation}\label{newpot}
V(\vec r^{\, \prime}, \vec \nabla_{r^{\prime}}, \vec\nabla_{X^{\prime}}) 
\propto (-)^n  (-)^m   
\left\{ \frac{\nabla_{r^{\prime} \, i_1}}{2},   ... 
\left\{ \frac{\nabla_{r^{\prime} \, i_m}}{2} , f(\vec r^{\, \prime}) \right\} 
\right\}\, \nabla^n_{X^{\prime}},
\end{equation}
where the indices $i_i,\ldots,i_m$ are appropriately contracted.

In order to obtain the potential in configuration space for functions that 
diverge as the momentum transfer $|\vec q|$ goes to infinity,  
one has to define a regularization scheme. 
Here, following Ref. \cite{Friar96},
we find it convenient to extend the definition of the 
Fourier transform \eqref{TransfFourier0} to a space-time of $d=n+1$ dimensions:
\begin{equation}
V_{n}(\vec r\,) =
\int \frac{d^{n}q}{\left( 2\pi \right)^{n}}e^{- i\vec{q}\cdot \vec{r}} 
\; V(\vec q \,).  
\label{TransfFourier1}
\end{equation}
The amplitude $V(\vec q^{})$ is the expression in momentum
space of corresponding loop contributions right after the 
$d$-dimensional integration 
over loop momenta is performed, but before setting $d=4$ or 
performing the integration over Feynman parameters. 
Writing
\begin{equation}
d^n q= q^{n-1}dq \,
(1 -\cos^2 \theta)^{\frac{n-3}{2}} \,
d \cos\theta \, d \Omega_{n-1},
\end{equation}
the angular integrations are evaluated with the aid of the formulas 
\begin{equation}
\int d \Omega_{n-1} =  
\frac{2 \pi^{\frac{n-1}{2}}}{\Gamma\left(\frac{n-1}{2}\right)},
\end{equation}
and 
\begin{equation}
\frac{2 \pi^{\frac{n-1}{2}}}{\Gamma\left(\frac{n-1}{2}\right)} 
\int_{-1}^{1} d \cos\theta \; (1 -\cos^2 \theta)^{\frac{n-3}{2}} 
e^{-i  q r \cos\theta }  
= (2\pi)^{\frac{n}{2}} ( q r)^{1-\frac{n}{2}} J_{\frac{n}{2}-1}(q r),
\end{equation}
where $q = |\vec q\,|$, $r = |\vec r\, |$, and 
$J_{n}(x)$ denotes a Bessel function of the first kind.
For momentum integrals, a useful relation is \cite{StegumAbramov}
\begin{equation}
\int_0^\infty dq \, q^{\frac{n}{2}}\frac{J_{\frac{n}{2}-1}(qr)}
{\left(q^{2}+\beta ^{2}\right)^{\frac{m-n}{2}}} =
\left(\frac{r}{2}\right)^{\frac{m-n}{2}-1}
\frac{\beta ^{n-\frac{m}{2}}}{\Gamma \left(\frac{m-n}{2}\right)}
K_{\frac{m}{2}-n}\left(\beta r\right),
\label{Bessel}
\end{equation}
where $\beta$ is a constant and
$K_{n}(x)$ is the modified Bessel function of the second kind.

For example, in the case of the triangle diagrams discussed in 
Sect. \ref{qspace},
\begin{equation}
V_\triangle(\vec q \,) = 
- i \frac{g_A \bar g_0}{F^2_{\pi}} 
\frac{(4\pi \mu^2)^{\frac{3-n}{2}}}{(2\pi F_{\pi})^2} 
\boldtau^{(1)} \cdot \boldtau^{(2)} \, 
( \vec\sigma^{(1)} - \vec\sigma^{(2)} )\cdot \vec q \; 
\Gamma\left(  \frac{3-n}{2}\right) 
\int_0^1 dx \left[m_\pi^2 + q^{2} x (1-x)\right]^{\frac{n-3}{2}}.
\label{Fou1}
\end{equation}
For $m = 3$ and $\beta^2 = m_\pi^2/x (1-x)$, 
the result in Eq. \eqref{Bessel} 
allows one to cancel the divergent factor of 
$\Gamma ((3-n)/2)$ in Eq. \eqref{Fou1}
and get an expression that is finite for $r\neq 0$.
Now we can set $d=4$ and with the aid of the properties of modified Bessel
functions \cite{StegumAbramov}
we can write the potential in configuration space as 
\begin{equation}
V_\triangle(\vec r\,) =  
\frac{g_A \bar g_0}{F^2_{\pi}} \frac{1}{(2\pi F_{\pi})^2} 
\boldtau^{(1)} \cdot \boldtau^{(2)} \, 
( \vec\sigma^{(1)} - \vec\sigma^{(2)} )\cdot \vec \nabla 
\left[\frac{1}{2\pi r^3} \int_0^1 dx (1+ \beta r) \, e^{-\beta r}  \right].
\label{FouTriangle}
\end{equation}
The contributions from box and crossed diagrams can be obtained in a 
similar fashion, leading to the result in Eq. \eqref{unsub}.

Alternatively, we can isolate the short-range, divergent part of
the interaction with integration by parts.
In Eq. \eqref{Fou1}, for instance, we then obtain
\begin{eqnarray}
V_\triangle(\vec q \,) &=& 
- i \frac{g_A \bar g_0}{F^2_{\pi}} 
\frac{(4\pi \mu^2)^{\frac{3-n}{2}}}{(2\pi F_{\pi})^2} 
\boldtau^{(1)} \cdot \boldtau^{(2)} \, 
( \vec\sigma^{(1)} - \vec\sigma^{(2)} )\cdot \vec q \; 
\nonumber\\ 
&& 
\left[\Gamma\left(\frac{3-n}{2}\right) m_\pi^{n-3}
+ \Gamma\left(\frac{5-n}{2}\right)  
\int_0^1 dx \frac{q^2 x (1-2x)}{\left[m_\pi^2 
+ q^{2} x (1-x)\right]^{\frac{5-n}{2}}} \right]
\label{Fou2}.
\end{eqnarray}
The first, divergent piece in Eq. \eqref{Fou2} is a contribution to a 
delta-function potential. Applying the $d$-dimensional Fourier transform 
to Eq. \eqref{Fou2} and taking the $d\rightarrow 4$ limit we find
\begin{eqnarray}
V_\triangle(\vec r \,) &=&  
\frac{g_A \bar g_0}{F^2_{\pi}} \frac{1}{(2\pi F_{\pi})^2} 
\boldtau^{(1)} \cdot \boldtau^{(2)} \, 
( \vec\sigma^{(1)} - \vec\sigma^{(2)} )\cdot \vec \nabla \; 
\nonumber\\ 
&&  
\left[ \delta^{(3)}(\vec r) \left( L + \ln \frac{\mu^2}{m^2_{\pi}}\right)  
- \frac{1}{4\pi r} 
\int_0^1 dx \frac{1-2x}{1-x} \beta^2 \, e^{-\beta r} 
 \right]\label{Fou3},
\end{eqnarray}
where $L$ is given in Eq. \eqref{L}.
Proceeding in this way also for box and crossed terms, 
we find that Fourier transform of the 
TPE potential can be expressed as
\begin{eqnarray}
V^{(3)}_{TPE}(\vec r \,)&=&  
\frac{g_A \bar g_0}{F^2_{\pi}} \frac{1}{(2\pi F_{\pi})^2} 
\boldtau^{(1)} \cdot \boldtau^{(2)} \, 
( \vec\sigma^{(1)} - \vec\sigma^{(2)} )\cdot \vec \nabla \; 
\left\{-\delta^{(3)}(\vec r) 
\left[ (3g_A^2 -1)\left( L + \ln \frac{\mu^2}{m^2_{\pi}}\right) + 2g_A^2\right]
\right. \nonumber\\ 
&& \left. 
+ \frac{1}{4\pi r} \int_0^1 dx 
\left[g_A^2\left(4 -\frac{3}{2 x (1-x)}\right) - \frac{1-2x}{1-x} \right]
\beta^2 e^{-\beta r} 
 \right\}\label{Fou4}.
\end{eqnarray}
The piece proportional to the delta function can then be absorbed in a 
redefinition of $\bar C_{2}$ very similar to Eq. \eqref{C2ren}, 
the only difference residing in the finite pieces. 
Integrating by parts, it can be explicitly verified that the 
non-analytic piece of the expression
\eqref{Fou4}
gives the medium-range potential in the form of 
Eq. \eqref{unsub}.

As a further check of our results, we computed the Fourier transform of the 
triangle diagrams with a Gaussian regulator $\exp(-q^2/\Lambda^2)$, 
for different values of the cutoff $\Lambda$. 
The calculation was performed numerically with 
{\it Mathematica} \cite{mathematica} 
and we focused on the region $r > 1/m_{\pi}$. 
For 
$\Lambda \simeq m_{\rho}$, the result we get is still quite different from the 
Fourier transform obtained in dimensional regularization, 
but as we increase the cutoff to 
1--2 GeV, it approximates Eq. \eqref{FouTriangle} better and better.

As pointed out in Ref. \cite{Friar96}, in the
$d$-dimensional Fourier-transform procedure  the
infinities are ``regularized'' away because the nucleon distance 
is kept finite.
The ultraviolet divergences and the regulator dependence are now hidden 
in the singular behavior ($\sim 1/r^4$) 
of the potential for small $r$, 
which forces the reintroduction of a regulator in the calculation of 
matrix elements of $V(r)$. 
If the chosen regulator were dimensional regularization, 
then the $d\rightarrow 4$ limit of 
Eq. \eqref{Bessel}, which leads to the $1/r^3$ singularity in 
Eq. \eqref{FouTriangle},
must be taken in the sense of generalized functions; 
the singularity is then encoded by a delta function, 
proportional to the divergent factor $2/(d-4)$, 
with a plus distribution remaining \cite{distributions}.

\end{document}